\documentclass[12pt]{article} 

\usepackage[letterpaper,left=1in, right=1in, bottom=1.5in, top=1in]{geometry} 
\usepackage[margin=1cm]{caption}
\usepackage{subcaption,amsmath,amsthm,amsfonts,amssymb,graphicx,bbm,bm, enumitem,slashed,xcolor}
\numberwithin{equation}{section}
\usepackage{cite}
\usepackage{setspace}
\usepackage{mathrsfs}
\usepackage{upgreek} 

\usepackage[colorlinks]{hyperref}
\hypersetup{
    citecolor = {blue}
}

\usepackage{tikz}
\usepackage{tikz-cd}
\usetikzlibrary{arrows,shapes,snakes,automata,backgrounds,petri}
\usetikzlibrary{calc,arrows,cd,decorations.markings,snakes}
\tikzset{
->-/.style args={#1rotate#2}{decoration={markings, mark=at position #1 with {\arrow[scale=1.5,rotate = #2 ]{stealth}}}, postaction={decorate}}
}
\usetikzlibrary{shapes.geometric}
\tikzset{snake it/.style={decorate, decoration={snake, amplitude=15mm}}}

\usetikzlibrary{decorations.markings}

\tikzset{line/.style={line width=0.25mm},
curve/.style={line,smooth,tension=1},
->-/.style={decoration={
  markings,
  mark=at position #1 with {\arrow[>=stealth]{>}}},postaction={decorate}},
-<-/.style={decoration={
  markings,
  mark=at position #1 with {\arrow[>=stealth]{<}}},postaction={decorate}},
}

\tikzset{
    partial ellipse/.style args={#1:#2:#3}{
        insert path={+ (#1:#3) arc (#1:#2:#3)}
    }
}
\tikzset{bg/.style={opacity=.5}}

\usepackage[framemethod=TikZ]{mdframed} 
\usepackage{tikz-cd} 
\usetikzlibrary {shapes.geometric} 

\usetikzlibrary{arrows,snakes,shapes.arrows,decorations.markings}
     \tikzset{>=triangle 90}
     \tikzstyle{bbc}=[draw,circle,fill=black,scale=.75]
     \tikzstyle{rc}=[circle,fill=red,scale=.6]
     \tikzstyle{wc}=[draw,circle,scale=.75]
     
\usetikzlibrary{decorations.pathreplacing,calligraphy}

\tikzset{snake it/.style={decorate, decoration=snake}}

\tikzset{
	on each segment/.style={
		decorate,
		decoration={
			show path construction,
			moveto code={},
			lineto code={
				\path [#1]
				(\tikzinputsegmentfirst) -- (\tikzinputsegmentlast);
			},
			curveto code={
				\path [#1] (\tikzinputsegmentfirst)
				.. controls
				(\tikzinputsegmentsupporta) and (\tikzinputsegmentsupportb)
				..
				(\tikzinputsegmentlast);
			},
			closepath code={
				\path [#1]
				(\tikzinputsegmentfirst) -- (\tikzinputsegmentlast);
			},
		},
	},
	mid arrow/.style={postaction={decorate,decoration={
				markings,
				mark=at position .5 with {\arrow[#1]{stealth}}
	}}},
}

\usepackage{colortbl}
\definecolor{dgreen}{rgb}{0, 0.55, 0}
\definecolor{dblue}{rgb}{0.690, 0.824, 0.922}
\definecolor{llightyellow}{rgb}{1.0, 0.95, 0.7}
\definecolor{llightblue}{rgb}{0.7, 0.9, 1.0}
\definecolor{llightpink}{rgb}{1.0, 0.85, 0.95}
\definecolor{llightgreen}{rgb}{0.7, 1.0, 0.4}
\definecolor{llightpurple}{rgb}{0.941, 0.863, 0.937}
\colorlet{lightyellow}{llightyellow!50!white}
\colorlet{lightblue}{llightblue!50!white}
\colorlet{lightgreen}{llightgreen!50!white}
\colorlet{lightpink}{llightpink!50!white}
\definecolor{azure}{rgb}{0.0, 0.5, 1.0}
\definecolor{darkblue}{rgb}{0.15,0.35,0.7}
\definecolor{reddish}{rgb}{0.65, 0.2, 0.2}
\definecolor{brandeisblue}{rgb}{0.0, 0.44, 1.0}
\definecolor{ceruleanblue}{rgb}{0.16, 0.32, 0.75}
\definecolor{indigo(dye)}{rgb}{0.0, 0.25, 0.42}
\definecolor{grey}{rgb}{0.9,0.9,0.9}
\definecolor{dgrey}{rgb}{0.3,0.3,0.3}
\definecolor{dgreen}{rgb}{0, 0.55, 0}
\usetikzlibrary{shadings}

\usetikzlibrary {decorations.markings}

\newcommand{\goesto}{\quad\longrightarrow\quad}
\def\RR{\mathbb R}

\def\Tr{\text{Tr}}

\def\x{\mathbf{x}}

\def\q{\mathbf{q}}

\def\id{\mathbbm{1}}

\def\0{{(0)}}
\def\1{{(1)}}
\def\2{{(2)}}
\def\3{{(3)}}
\def\4{{(4)}}
\def\m1{{(\text{-}1)}}

\def\calA{\mathcal{A}}
\def\calB{\mathcal{B}}

\def\calH{\mathcal{H}}
\def\calI{\mathcal{I}}
\def\calO{\mathcal{O}}
\def\calM{\mathcal{M}}
\def\calU{\mathcal{U}}
\def\scrL{\mathscr{L}}
\def\DD{\mathbb{D}}

\newcommand{\ZZ}{\mathbb{Z}}

\newcommand{\bfg}{\mathbf{g}}
\newcommand{\bfh}{\mathbf{h}}

\usepackage[margin=1.5cm,font=small,labelfont=bf]{caption}

\title{Tilts from 2-Groups}
\author{Sven Harder,$^1$ \ Theodore Jacobson,$^{1,2}$ \ and Zhengdi Sun$^1$  \\ 
{$^1$\it\small Mani L. Bhaumik Institute for Theoretical Physics, Department of Physics and Astronomy,}\\
{\it\small University of California, Los Angeles, CA 90095, USA}  \\
{$^2$\it\small Department of Physics, Hamilton College, Clinton, NY 13323, USA}
}
\date{\today}							

\begin{document}
\maketitle
\thispagestyle{empty}

\begin{abstract}
2-group global symmetries intertwine 0-form and 1-form symmetries in an interesting way. We analyze universal constraints on lines which are charged under the 1-form subgroup of a 2-group, and show that they generically break the 0-form symmetry explicitly. This gives rise to a family of line defects parameterized by the broken symmetry generators, whose existence is invariant under the renormalization group flow of the bulk-defect system. The symmetry breaking is enforced by a `family anomaly,' which is a topological obstruction to a symmetric defect. Our main tools are the Wess-Zumino consistency condition and a generalized anomaly inflow formalism for defects and boundaries. For nonabelian continuous 2-groups the family anomaly stems from a higher Berry connection on the moduli space of defects, and constrains response functions that probe the local action of the broken symmetry.  In the continuous abelian case we apply differential cohomology to the 2-group background fields to uncover a subtle generalization of the Wess-Zumino condition, while in the discrete case we recast it in terms of the associativity of symmetry defects. We give numerous illustrative examples, computing when possible explicit forms of the tilt operator (which probes the linear response of the defect to the broken symmetry) and its higher analogs, which are crucial for matching the family anomaly. We also discuss universal features such as how symmetry violation by charged line defects is related to symmetry breaking hierarchies in the bulk, and highlight a number of subtle points including the distinction between simple and non-simple lines, and Postnikov class resolution. 
\end{abstract} 


\newpage
\begingroup
	
	\hypersetup{linkcolor=.,linktoc=all}
	\tableofcontents

\endgroup

\section{Introduction and Summary}

A basic consequence of conventional global symmetries in Quantum Field Theory (QFT) is that local operators are organized into multiplets, or representations, of the symmetry group. In this paper we address how this statement generalizes to line operators charged under 2-groups~\cite{Cordova:2018cvg,Benini:2018reh}, in which 0-form and 1-form symmetries~\cite{Gaiotto:2014kfa} intertwine in a non-trivial way.\footnote{See~\cite{Sharpe:2015mja, McGreevy:2022oyu, Cordova:2022ruw,Schafer-Nameki:2023jdn, brennan2023introduction,Iqbal:2024pee, Costa:2024wks, Kaidi:2026urc } for reviews of categorical global symmetries.} While the majority of the literature on higher-groups focuses on their background gauge fields and associated topological symmetry operators (i.e. the charges), we turn our attention to the operators charged under the symmetry, namely the line operators charged under the 1-form subgroup of a 2-group. The main goal of the paper is to highlight and explore the fact that for a 2-group consisting of 0-form symmetry $G^\0$ and 1-form symmetry $\calA^\1$, it is generically the case that:
\begin{quote}\centering
Line operators charged under $\calA^\1$ must \emph{explicitly break} $G^\0$.\footnote{Explicit symmetry breaking by a line $\scrL$ means that $G^\0$ transforms $\scrL$ to a distinct line $^g\!\scrL \not=\scrL$. In terms of symmetry defects, a symmetry-breaking line does not admit a topological junction with itself and the $G^\0$ symmetry operators. } 
\end{quote}
This fact has already been observed in several places in the literature. For instance,~\cite{vanBeest:2023dbu,Aharony:2023amq,Choi2026BerryPhaseBoundaryConformalManifolds} discuss chiral symmetry violation by 't Hooft lines in four-dimensional massless quantum electrodynamics (QED) with 2-group symmetry,~\cite{Jacobson:2024muj} studies examples of symmetry breaking on defects enforced by discrete higher-groups in $O(2)$ gauge theory, and~\cite{Choi:2022fgx,Sehayek:2026pvu} observe symmetry breaking on line and surface defects in axion models. The purpose of this paper is to understand in a more systematic and uniform way why symmetry breaking occurs on line operators charged under 2-group symmetry. We will consider both continuous and finite discrete 2-groups, in each case spelling out the conditions under which symmetry breaking occurs, and when it can be avoided. 

A key feature of 2-groups is that they constrain the emergence pattern of $G^\0$ and $\calA^\1$ in renormalization group (RG) flows from ultraviolet (UV) theories where these symmetries are not present~\cite{Cordova:2018cvg}. Namely, the 1-form symmetry must emerge before the 0-form symmetry --- it is inconsistent for the 0-form symmetry to be present without the 1-form symmetry. This is best understood in the case of continuous 2-groups where it follows from current algebra. Our analysis of charged lines gives a complementary, intuitive (though not fully rigorous) perspective on the emergence hierarchy. Suppose we start with a QFT with 2-group symmetry. We can explicitly break the 1-form symmetry by summing over the charged line operators in the path integral with some weight. But since the charged lines themselves explicitly break the 0-form symmetry, summing over them in the path integral spreads the 0-form symmetry breaking into the bulk. Therefore, as long as the 1-form symmetry is explicitly broken, the 0-form symmetry will also be broken.\footnote{In Sec.~\ref{sec:discrete} we point out that certain discrete 2-groups fail to lead to symmetry breaking on lines. In these cases the 0-form symmetry is not explicitly broken, but rather \emph{extended} when we sum over lines. }  

The simple fact that charged lines break symmetries already leads to interesting consequences, but the mechanism through which this breaking is enforced leaves additional imprints on the line. When symmetry breaking occurs, e.g. in the case of a continuous 2-group with a quantized Postnikov class,\footnote{The Postnikov class is an important characteristic of the mixing between the 0-form symmetry and 1-form symmetries. In the continuous case, it appears e.g. in the operator product expansion (OPE) coefficient of the 2-form current $J^{\mu\nu}$ in the fusion of two 1-form currents $j^\mu$~\cite{Cordova:2018cvg}, while in the discrete case it controls how the 1-form symmetry generators can terminate on the threefold junctions of 0-form symmetry generators~\cite{Benini:2018reh}.  } charged line operators necessarily come in \emph{families} generated by the $G^\0$ action. While explicit symmetry breaking on line defects is known to lead to a family of lines parameterized by the broken group~\cite{Drukker:2022pxk,Herzog:2023dop} (really the coset $G/H$ where $H$ is an unbroken subgroup), families of line operators charged under 2-groups have certain special features which distinguish them from garden-variety families of symmetry-breaking defects. We will highlight these features in various explicit examples, some of which are known, and some of which have not been explored in existing literature.

Before explaining what makes 2-group-enforced symmetry breaking on line defects special, let us give a sketch of why 2-groups may require symmetry breaking at all. One argument combines the following observations:
\begin{enumerate}
\item A line operator charged under $\calA^\1$ lives at the boundary of a Wilson surface built from the 2-form background gauge field $B^\2$ for $\calA^\1$.
\item The Wilson surface transforms like a $2d$ QFT with anomalous $G^\0$ symmetry, in a way dictated by the Postnikov class of the 2-group. 
\item Anomalous global symmetries are explicitly broken at boundaries (see e.g.~\cite{Jensen:2017eof, Thorngren:2020yht} for a comprehensive recent treatment, and references therein). 
\end{enumerate}
Note that this argument is completely kinematic and is independent of whether the line (viewed as an operator acting at a fixed time) creates a dynamical string excitation.\footnote{Such a dynamical string worldsheet carries the same 2d anomaly as the Wilson surface~\cite{Cordova:2018cvg,Hsin:2020nts}. The fact that this leads to symmetry breaking by the lines that create them was already pointed out by T.~Dumitrescu in~\cite{ClayThomastalk}. } In other words the argument is agnostic to how $\calA^\1$ is realized, i.e. whether or not it is spontaneously broken. It neither requires the bulk QFT to be a conformal field theory (CFT), nor the line to furnish a conformal defect. It also applies equally well to discrete and continuous (internal) 2-groups.

Alternatively, we can easily show that it is inconsistent for a line charged under $\calA^\1$ to be symmetric under $G^\0$ as follows. For the purposes of illustration we consider a continuous 2-group with background gauge fields $A^\1,B^\2$. While the background gauge transformation properties of $A^\1$ are standard, the 2-group requires $B^\2$ to transform under $G^\0$ as 
\begin{equation} \label{eq:ABtransf}
    A^\1 \,\to\, A^\1  + d\lambda^\0 + i[A^\1,\lambda^\0]\,, \quad B^\2 \,\to\, B^\2 + \alpha^\2(\lambda,A)\,,
\end{equation}
where $\alpha^\2(\lambda,A)$ is a 2d anomaly density for $G^\0$. Now, we perform repeated 0-form background gauge transformations with infinitesimal parameters $\lambda_1^\0, \lambda_2^\0$. One can easily verify that 
\begin{equation} \label{eq:deltaWZA}
  \left(\delta_{\lambda_1}\, \delta_{\lambda_2} - \delta_{\lambda_2}\, \delta_{\lambda_1} - \delta_{i [\lambda_1,\lambda_2]} \right)  A^\1 = 0\,,
\end{equation}
which follows from the composition rule for background gauge transformations acting on $A^\1$.\footnote{In principle one can probe this composition rule by comparing the action of $g_1 = e^{i \lambda_1}$ followed by $g_2 = e^{i\lambda_2}$ to the action of $g_1g_2$, but this requires computing gauge variations to $\calO(\lambda^2)$.  On the other hand the \emph{commutator} of the $\lambda_1,\lambda_2$ action only requires the $\calO(\lambda)$ gauge transformations in Eq.~\eqref{eq:ABtransf}.} This is the same sequence of variations used by Wess and Zumino~\cite{Wess:1971yu,Zumino:1983ew} to investigate self-consistent anomalies. They showed that the above equation should hold for any functional of $A^\1$, which implies that any 2d anomaly \emph{density} must satisfy
\begin{equation} 
 \delta_{\text{WZ}}\, \alpha^\2(\lambda,A) \equiv    \delta_{\lambda_1}\, \alpha^\2(\lambda_2,A) - \delta_{\lambda_2} \, \alpha^\2(\lambda_1,A) - \alpha^\2(i[\lambda_1,\lambda_2],A)  \, = \,  d\alpha^\1(\lambda_1,\lambda_2,A)\,,
\end{equation}
for some 1-form $\alpha^\1$. Consequently, the action of 0-form background gauge transformations on the 2-form background gauge field $B^\2$ can in general only be composed \emph{up to} a 1-form background gauge transformation $B^\2 \, \to \, B^\2 + d\Lambda^\1$ with $\Lambda^\1 = \alpha^\1(\lambda_1,\lambda_2,A)$, leading to
\begin{equation} \label{eq:deltaWZB}
\left(\delta_{\lambda_1}\, \delta_{\lambda_2} - \delta_{\lambda_2}\, \delta_{\lambda_1} - \delta_{i [\lambda_1,\lambda_2]} \right)  B^\2   \, =\, \delta_{\Lambda^\1 = \alpha^\1(\lambda_1,\lambda_2,A)} \,  B^\2 \,.
\end{equation}
We now impose a generalization of Wess-Zumino consistency --- namely, that any functional $\mathcal{F}$ of 2-group background fields must obey:
\begin{equation} \label{eq:WZModified}
\phantom{\Bigg[} \left(\delta_{\lambda_1}\, \delta_{\lambda_2} - \delta_{\lambda_2}\, \delta_{\lambda_1} - \delta_{i [\lambda_1,\lambda_2]}- \delta_{\Lambda^\1 = \alpha^\1(\lambda_1,\lambda_2,A)} \right) \, \mathcal{F}[A^\1, B^\2] \, = \, 0\,.  \phantom{\Bigg[}
\end{equation}
This follows from Eqs.~\eqref{eq:deltaWZA},~\eqref{eq:deltaWZB} and the fact that $\delta_{\Lambda^\1}\, A^\1 = 0$. See Eqs.~\eqref{eq:WZModified_abelian},~\eqref{eq:F-move} for the analogs of this consistency condition for continuous abelian and finite discrete 2-groups.

Now let $\mathcal{F}[A^\1,B^\2] = Z_\scrL[A^\1,B^\2]$ be the partition function coupled to background fields, with the insertion of a charged line operator $\scrL$ on a curve $\gamma$. If we assume that the line is invariant under $G^\0$, we reach a contradiction, because the left-hand side of the equation is non-trivial due to the action of the 1-form symmetry on the line. We can cancel this 1-form symmetry transformation by attaching the line to a Wilson surface $\int_\Sigma B^\2$ with $\partial\Sigma = \gamma$. But we again arrive at a contradiction, since the left-hand side is now non-trivial due to the anomalous properties of the Wilson surface. This is to be expected, since the Wilson surface itself satisfies Eq.~\eqref{eq:WZModified}. We conclude that a $G^\0$-symmetric line defect is inconsistent if it is charged under $\calA^\1$. 

To restore consistency, we must give up on the assumption that the line operator is invariant under the 0-form symmetry. This alone is however not enough -- the $G^\0$ action on the line must only hold up to an anomalous phase which is capable of compensating Eq.~\eqref{eq:WZModified}. Much of the paper is devoted to exploring the consequences of this anomalous phase (which we refer to as the \emph{family anomaly}) for the structure of the moduli space of line defects.\footnote{The notion of a family anomaly, or anomaly in the space of couplings, is discussed in~\cite{Gaiotto:2017yup,Tanizaki:2017bam,Kikuchi:2017pcp,Cordova:2019uob,Cordova:2019jnf,Brennan:2026ira,Debray:2023ior,Brennan:2024tlw,Copetti:2025sym,Komargodski:2025jbu,Brennan:2026ira}.} 

The same general logic applies (with different details) uniformly across the three types of 2-groups we consider: continuous nonabelian (Sec.~\ref{sec:nonabelian}), continuous abelian (Sec.~\ref{sec:abelian}), and discrete (Sec.~\ref{sec:discrete}). Each of the remaining sections is organized as follows: we first derive an inconsistency for symmetric lines, then explain how symmetry-breaking lines restore consistency. We then examine the consequences of symmetry breaking (and the ensuing family anomaly) and turn to examples which illustrate the general methods and cover the various sub-cases. Additional technical material is included in the appendices. The remainder of the introduction summarizes the main points of the paper in more detail, and concludes with a list of open questions for possible future investigation.

\subsubsection*{Continuous $\mathcal{A}^\1$ and Continuous Nonabelian $G^\0$}

For continuous nonabelian $G^\0$, we develop a novel anomaly inflow formalism for anomalous theories on manifolds with boundary similar to~\cite{Copetti:2025sym}. This formalism is more general than its application to the problem at hand. The partition function $Z_{\mathscr{L}}$ with line $\mathscr{L}$ inserted is inconsistent with Eq.~\eqref{eq:WZModified} unless we introduce a $G^\0$-valued source $U(t)$ (depending on the worldline coordinate $t$) on the line operator $\scrL$ which transforms under the natural group action $U \to g^{-1}\, U$. The full partition function now transforms under 0-form symmetry transformations as:
\begin{equation} \label{eq:correct_anomalyIntro} 
Z_{\scrL}[A^\1+d_A\lambda^\0,B^\2+\alpha^\2(\lambda,A),U + \delta_\lambda U] \\
\,=\,  Z_{\scrL}[A^\1,B^\2,U]\, e^{ -i \int_\gamma \nu^\1(\lambda,U,A)  }\,,
\end{equation}
where $\nu^\1(\lambda,U,A)$ characterizes the family anomaly on the space of line operators related by the broken $G^\0$ symmetry (see Eqs.~\eqref{eq:nu1_full},~\eqref{eq:nu1_expanded} for explicit forms of the family anomaly for $G^\0 = SU(N)$).

The family anomaly imposes strong constraints on RG flows, as it originates from a higher Berry connection~\cite{Cordova:2019uob,Cordova:2019jnf,Hsin:2020cgg,Kapustin:2020eby} with a non-trivial topological flux. While the gauge-invariant higher Berry curvature can receive arbitrary quantum corrections, its flux on the line defect moduli space is determined by the Postnikov class of the 2-group, and hence rigid. Without breaking the 2-group, it is impossible to lift the moduli space of charged lines. As a result, the flux of the higher Berry connection prevents the moduli space of line defects from collapsing to a point.

To further quantify the breaking of a continuous symmetry $G^\0$, one can consider the \textit{tilt operator} which encodes the local, linear response to the broken symmetry action~\cite{Drukker:2022pxk,Herzog:2023dop}. The tilt operator appears in the non-conservation equation for the 0-form symmetry current in the presence of a line defect,\footnote{The tilt operator is well-studied in the context of conformal field theories with boundaries or defects (BCFT/DCFT), where it furnishes an example of an exactly marginal defect operator, see e.g.~\cite{AJBray_1977,Metlitski:2020cqy,SciPostPhys.12.6.190,Drukker:2022pxk,Gimenez_Grau_2022,Herzog:2023dop,Cuomo:2023qvp,sakkas2024inversionintegralidentitiesdcfts,belton2025againbulktodefectwardidentities,Kravchuk:2025evf,girault2026consequencessymmetrybreakingconformaldefect,Belton:2026xaw}.} 
\begin{equation} \label{eq:dmujmu_tau}
d \star j^\1 = \delta^{(d-1)}(\gamma) \wedge \tau^\1 \,. 
\end{equation}
Due to the non-linear nature of the $G^\0$ action, $\tau^{(1)}$ is not enough to completely capture the response to turning on a source $U(t)$. This requires a tower of \textit{higher tilt operators} encoding the non-linear $G^\0$ action on the tilt operator itself. Such higher tilts have been explored in a DCFT context in~\cite{Drukker:2025dfm}, and play an essential role in reproducing the family anomaly.  
On the other hand, the source $U(t)$ couples to the (higher) tilt operators in a complicated way. Instead of studying the correlation functions of tilt operators themselves, we will focus on the connected response functions generated by varying the partition function with respect to the source $U(t)$. One might expect these response functions to be independent of the position on the moduli space since any two points are related by the $G^\0$ action. Instead, the family anomaly implies that the response functions in general differ by contact terms which cannot be removed by the addition of local counterterms. For instance, the 2-point response function at different points on the moduli space differ by a contact term which precisely captures the Berry connection on the space of line defects.

Importantly, the family anomaly has imprints beyond just pure contact terms. For example, the 3-point response function $G^{abc}(t_1,t_2,t_3)$ is constrained by an integrated Ward identity derived from the family anomaly:
\begin{equation} \label{eq:contracted_ward_idStart}
    \int dt\, f^{abc}\, G^{abc}(t_1,t_2,t) \, = \, - \frac{\kappa\, C_{\rm{adj}}\, \text{dim}(G)}{24\pi} \, \partial_{t_1} \delta(t_1-t_2)\,~,
\end{equation}
where $C_{\text{adj}}$ is the quadratic Casimir in the adjoint representation of $G^\0$, $t_i$ are positions along the line defect, and $a,b,c$ are Lie algebra indices. This anomalous Ward identity is known to determine the 3-point functions at separated points in BCFTs~\cite{Drukker:2025dfm, Copetti:2025sym}. We extend the analysis to general QFTs, and show that it enforces \textit{at least} a partial contact term in the 3-point function. As another example, we generalize the 2-connection defined from the 3-point functions of boundary-condition-changing operators studied in 2d BCFT \cite{Choi:2025ebk} to non-conformal line defects, and show that the topological flux of the corresponding 3-form field strength is protected and determined by the family anomaly. 

Finally, we consider examples such as 't Hooft lines in massless $\text{QED}_4$ and the pion Lagrangian with gauged baryon number (also known as the nonabelian Goldstone Maxwell model for spontaneous 2-group symmetry breaking). We explicitly compute the tilt and (in the latter case) higher tilt operators and demonstrate how they match the family anomaly. We also consider the fate of line defects when we promote the spurionic source $U(t)$ to a dynamical field. Intuitively, integrating over the group orbit should lead to a symmetrized line defect.\footnote{Note that we do not simply integrating over the global (or constant) $G^\0$ orbit, but really promote $U(t)$ to a full-fledged dynamical field. Just integrating over the zero-mode yields a line which is invariant under global, but not local, $G^\0$ transformations. We make further comments about the related distinction between simple and non-simple lines in Sec.~\ref{sec:intricacies}. } On the contrary, the family anomaly \eqref{eq:correct_anomalyIntro} of the original line defect turns into an operator-valued violation of $G^\0$, giving rise to a new tilt operator built from the worldline non-linear sigma model (NLSM) fields. The form of this tilt is completely universal, in the sense that it only depends on the family anomaly of the original defect. We highlight how the family anomaly for the defect coupled to a worldline $G^\0$ NLSM reappears due to the presence of a universal higher tilt operator.

\subsubsection*{Continuous $\mathcal{A}^\1$ and Continuous Abelian $G^\0$}

When $G^\0$ is abelian, the symmetry breaking is sensitive to the \textit{global structure} of $G^\0$. This is because 2d anomalies for abelian symmetries are always WZ consistent under infinitesimal gauge transformations. In the case that $G^\0$ is compact (i.e. $U(1)$), we find non-trivial constraints when large gauge transformations are taken into account. We develop a differential cohomology formalism for abelian 2-group bundles, which allows us to demonstrate the violation of the WZ consistency condition under small \emph{and} large gauge transformations. In this framework, we describe background gauge fields patch-wise and glue them together using transition functions and higher data. This makes it suitable for realizing symmetry defect networks for discrete abelian 2-groups, thereby bridging the methods of Sec.~\ref{sec:nonabelian} to those of Sec.~\ref{sec:discrete}. If $G^\0$ is non-compact (i.e. $\mathbb{R}$), there are no large gauge transformations, and no obstruction to a symmetric line defect.\footnote{Correspondingly, for $G^\0 = \RR$ there is no quantized, RG-invariant Postnikov class.} 

Like in the continuous nonabelian case, the inconsistency is resolved by coupling the line defect to a source for the tilt operator. In the presence of background gauge fields, $G^\0$ transformations move the line defect along the moduli space \emph{up to} anomalous phases from the family anomaly. In the case of $G^\0 = U(1)^\0$, the moduli space is a circle and the spurion is a compact background scalar field $\theta \sim \theta + 2\pi$. In a given choice of scheme, the family anomaly implies that in a background where $\theta$ winds around the line operator $\gamma$, performing a constant $U(1)^\0$ transformation yields a phase
\begin{equation} \label{eq:correct_abelian_anomalyIntro}
Z_{\scrL}[A^\1, B^\2+ \tfrac{\kappa}{2\pi}\lambda\, dA^\1, \theta^\0 + \lambda] \,=\, Z_{\scrL}[A^\1,B^\2,\theta^\0]\,e^{i \kappa \lambda\oint_\gamma \frac{d\theta^\0}{2\pi} }\,. 
\end{equation}
Due to its global nature, there is no local density that characterizes the family anomaly. As a result, we do not find any clean scheme-independent constraints on the response functions as in the nonabelian case. 

As our main example, we study in detail the universal infrared (IR) theory of spontaneous abelian 2-group breaking, the abelian Goldstone-Maxwell model. This model has a photon and a compact scalar whose spontaneously broken symmetries mix in a 2-group.\footnote{More precisely, a 3-group~\cite{Cordova:2018cvg}. For our purposes we can focus on the 2-subgroup.} We utilize electric-magnetic duality to explicitly derive the tilt operator and show how the family anomaly is matched. We explore the same defect coupled to a dynamical source, which has the effect of coupling the compact boson to a dynamical rotor on the line. This again fails to restore the 0-form symmetry but leads to a new tilt operator built from the rotor. We show that the rotor generates a phase shift, proportional to the 2-group structure constant, for bulk Goldstone modes scattering off of the defect. Finally, we discuss the realization of the tilt operator in various 2-group-preserving deformations of this model, including massless $\text{QED}_4$ and the abelian Higgs model. In the latter case, the 't Hooft line creates Abrikosov-Nielsen-Olesen (ANO) vortex strings whose worldsheet hosts chiral fermion zero modes which carry the same 2d 't Hooft anomaly as the Wilson surface $\int B^\2$.

\subsubsection*{Discrete $\mathcal{A}^\1$}

When the 1-form symmetry is discrete, the Postnikov class is captured by a class in the group cohomology $\beta \in H^3_\rho(G^\0, \mathcal{A}^\1)$, which naturally induces a cohomology class in $H^3(G^\0, U(1))$. If (and only if) this induced cohomology class is non-trivial, the Postnikov class leads to a non-trivial anomaly on the Wilson surface bounded by the line defect.  In this case a symmetric, \textit{simple}\footnote{A simple line cannot be decomposed into a sum of other lines. Relatedly, the only topological local operator on a simple line is the identity operator (see e.g.~\cite{Bhardwaj:2017xup,Chang:2018iay,Bhardwaj:2023wzd}).} charged line defect violates the WZ consistency condition, which in the discrete case makes use of the fact that the associativity of $G^\0$ symmetry defects labeled by $\mathbf{g},\mathbf{h},\mathbf{k}$ only holds up to a $\calA^\1$ symmetry defect labeled by $\beta(\mathbf{g},\mathbf{h},\mathbf{k})$~\cite{Benini:2018reh}.\footnote{To compare to the WZ consistency condition in the continuous case, we can very roughly think of the triple $\lambda_1,\lambda_2,A$ as being replaced by the three group elements $\mathbf{g},\mathbf{h},\mathbf{k}$.} 

Symmetry-breaking lines come in (discrete) families labeled by an index $i$ on which $G^\0$ acts via a permutation $\sigma_{\bfg}(i)$. Fusing symmetry operators $\bfg$ and $\bfh$ in the presence of an incoming line defect $\scrL_i$ results in a c-number phase $\nu_i(\bfg,\bfh)$. This is the discrete version of the family anomaly introduced above. The WZ consistency condition requires 
\begin{equation}\label{eq:F-moveIntro}
  \frac{\nu_{\sigma_{\bfg}(i)}(\mathbf{h},\mathbf{k})\, \nu_i(\mathbf{g},\mathbf{hk})}{\nu_i(\mathbf{gh},\mathbf{k})\nu_i(\mathbf{g},\mathbf{h})}\, \chi_{\mathbf{q}}(\beta(\mathbf{g},\mathbf{h},\mathbf{k})) =  1 ~,
\end{equation}
where $\mathbf{q}$ is the charge of the line and $\chi_{\mathbf{q}}(\beta)$ is the action of the 1-form symmetry operator $\beta$. If we assume there is only single line in the family, the above equation states that the induced anomaly class in $H^3(G^\0,U(1))$ is cohomologically trivial. 

We study a variety of interesting examples with and without symmetry breaking constraints based on our general analysis, and address several related subtleties that appear when $G^\0$ is discrete. For instance, if $G^\0$ is continuous and connected, we show that the induced anomaly is always trivial and there will never be symmetry-breaking constraints on charged lines. An instance of this is QED with two charge-2 scalars, where the faithfully acting $SO(3)^\0$ flavor symmetry forms a 2-group with the $\mathbb{Z}_2^\1$ 1-form symmetry. The standard Wilson line is indeed invariant under $SO(3)^\0$. On the other hand, the induced anomaly can be non-trivial if $G^\0$ is continuous but disconnected. This occurs in 4d $Spin(4)$ gauge theory with two fermions in the vector representation where $G^\0 = SO(3)^\0 \times \mathbb{Z}_{2,\mathcal{C}}^{\0}$ and $\mathcal{A}^\1 = \mathbb{Z}^\1_2$. The charge conjugation symmetry $\mathbb{Z}_{2,\mathcal{C}}^{\0}$ renders $G^\0$ disconnected and leads to non-trivial induced mixed anomaly with $SO(3)^\0$ -- indeed, all Wilson lines charged under $\mathcal{A}^\1$ break $\mathbb{Z}_{2,\mathcal{C}}^\0$ in this theory.

When $G^\0$ is discrete, there is no tilt operator corresponding to a linearized, local $G^\0$ action. It is therefore possible to build a symmetric \emph{non-simple} line even when the induced anomaly is non-trivial. However, it is important to emphasize the symmetry algebra in the presence of such a non-simple line must be extended from $G^\0$ to some $\widetilde{G}^\0$ for which the corresponding Postnikov class trivializes. As an explicit example, in Sec.~\ref{sec:nsl} we explore this phenomenon in the abelian Goldstone-Maxwell theory by breaking the $0$-form symmetry to be a $\mathbb{Z}_N^{(0)}$ subgroup of $U(1)^\0$. There, we find that we can construct a non-simple Wilson line which is invariant under global $\ZZ_N^\0$ transformations. But in the presence of the line, fusing $N$ symmetry operators for $\mathbb{Z}_N^{(0)}$ does not yield the identity operator, but rather a topological (on the line) vertex operator $e^{i\chi}$ built from the bulk Goldstone boson $\chi$. This is an example where bulk symmetries are extended by defect symmetries. 

A closely related subtlety is what we refer to as \textit{Postnikov class resolution}. When $G^\0$ is discrete, it is always possible to trivialize the Postnikov class by introducing a trivially-acting kernel which extends $G^\0$ to a symmetry group $\widetilde{G}^\0$. This setup often naturally occurs along the RG flow: a subgroup $H^{\0} \subset G^{\0}_{\text{UV}}$ of the UV symmetry may become trivially acting at long distances and the infrared faithful symmetry $G_{\text{IR}}^\0 = G_{\text{UV}}^\0/H^\0$ can develop an emergent Postnikov class.\footnote{This scenario can also resolve 2-groups with compact abelian 0-form symmetry, if the UV symmetry is non-compact $\mathbb{R}$.} This leads to a refinement of the usual 2-group emergence theorem, which states that if both $G^{\0}$ and $\mathcal{A}^\1$ emerge in the IR with a non-trivial Postnikov class, then the 0-form symmetry cannot emerge before the 1-form symmetry. Namely, our analysis indicates that there is another possibility where the 1-form symmetry $\mathcal{A}^\1$ is explicitly broken, and $G^\0_{\text{IR}}$ is not broken but extended to $G^\0_{\text{UV}}$ by some $H^{\0}$ in the UV-completion. Along the RG flow, $H^{\0}$ cannot become trivially-acting before $\mathcal{A}^\1$ emerges. In such a UV completion, in general, we should be able to construct a simple $G_{\text{UV}}^\0$-symmetric line operator that flows to a non-simple line, which is invariant under global $G^\0_{\text{IR}}$ action in the infrared.

Finally, we consider scenarios where a Postnikov class can be trivialized by enlarging the 1-form symmetry to $\widetilde{\calA}^\1 \supset \calA^\1$. In these situations, the Postnikov class for $\calA^\1$ is superseded by a non-trivial automorphism action $\rho$ of $G^\0$ on the larger 1-form symmetry $\widetilde{\calA}^\1$, in such a way that the symmetry-breaking constraints are explained by the action of $G^\0$ on the charges of lines. We discuss an example of this phenomenon which arises when we try to engineer a Postnikov class via symmetry fractionalization.

\subsubsection*{In Lieu of Conclusion}

We have attempted to make our analysis comprehensive, emphasizing the main idea (2-groups imply symmetry breaking by lines) while highlighting subtleties in its various manifestations. But there are still a number of remaining open questions and future directions which would be interesting to pursue:
\begin{itemize}
    \item \textbf{Constraints on spontaneous symmetry breaking patterns:} We mentioned how our analysis of charged line operators sheds light on the hierarchy between the emergence scales of the 1-form and 0-form symmetries. We currently do not have further insights into the corresponding constraints on \emph{spontaneous} symmetry breaking, i.e. the expectation that if $\calA^\1$ is spontaneously broken then $G^\0$ is as well.\footnote{This was proven for the continuous case in \cite{Cordova:2018cvg}. We comment on some genuine and near counterexamples to this hierarchy in the discrete case in Sec.~\ref{sec:intricacies}.} 

    \item \textbf{Poincar\'e 2-groups:} In this paper we only considered internal symmetries. We expect that much of what we say here generalizes to 2-groups involving spacetime symmetries~\cite{Cordova:2018cvg,Cordova:2020tij}. This requires an understanding of the boundaries of gravitationally anomalous theories~\cite{Hellerman:2021fla}. 

    \item \textbf{Dimensional reduction:} The fate of 2-groups under dimensional reduction  was studied in \cite{Damia:2022seq,Nardoni:2024sos}. In the context of circle compactifications (or theories with 2-group symmetries at finite temperature~\cite{Iqbal:2020lrt}), it would be interesting to study the local operators obtained by wrapping symmetry-breaking lines on the circle. 

    \item \textbf{Higher ($n>2$)-groups:} We focused on 2-groups where lines are the highest-dimensional charged objects. Our techniques generalize straightforwardly to e.g. 3-groups, where surface operators may be forced to break 0-form or 1-form symmetries (see~\cite{Jacobson:2024muj} for an example). 3-group symmetries occur in theories with axions \cite{Cordova:2018cvg,Seiberg:2018ntt,Brennan:2020ehu,Hidaka:2020iaz,Hidaka:2020izy,Nakajima:2022feg,Choi:2022fgx,Anber:2024gis,Sehayek:2026pvu} and in gauge theories with disconnected gauge groups~\cite{Barkeshli:2022wuz,Barkeshli:2023bta,Jacobson:2024muj}. For work on even higher groups see~\cite{Tanizaki:2019rbk,Cherman:2020cvw, Hidaka:2021mml, Hidaka:2021kkf,Kan:2023yhz}.
\end{itemize}

\section{Continuous Nonabelian 2-Groups}
\label{sec:nonabelian} 

In this section, we consider a continuous nonabelian 0-form symmetry $G^\0$ which forms a 2-group with a $U(1)^\1$ 1-form symmetry (extending the discussion to more general continuous 1-form symmetries is straightforward). We work in Euclidean signature where the 2-group background gauge fields $A^\1$ and $B^\2$ couple to currents $j_\mu$ and $J_{\mu\nu}$ as 
\begin{equation}
S \, \supset \,  i \int  A^{\1\, a} \wedge \star j^a    + B^\2 \wedge \star J\,,
\end{equation}
where $a$ is a Lie algebra index. The fundamental feature of the 2-group background fields is the modified gauge transformation rule~\cite{Cordova:2018cvg}: 
\begin{equation} \label{eq:cont_2group_trans}
A^\1 \,\to\, A^\1 + d\lambda^\0 + i [A^\1,\lambda^\0] \,, \quad B^\2 \,\to\, B^\2 + d\Lambda^\1 + \alpha^\2(\lambda,A)\,.
\end{equation}
Crucially, $\alpha^\2(\lambda,A)$ is the anomaly density of a $2d$ QFT with anomalous $G^\0$ symmetry. This 2d anomaly is in turn controlled by the Postnikov class, which we can intuitively think of as the inflow action $\calI^\3(A)$ (or the anomaly polynomial $d\calI^\3(A)$) giving rise to the anomaly $\alpha^\2(\lambda,A)$. 

We start in Sec.~\ref{sec:WZconsistency} by analyzing the implications of the modified gauge transformation rule for the partition function with a charged line operator insertion. This partition function is inconsistent unless it depends on a spurion reflecting the explicit breaking of $G^\0$ by the line operator. We develop a novel inflow/descent formalism in the presence of boundaries to show that the partition function, which depends on background gauge fields \emph{and} the spurion parameterizing the family of line defects, must transform anomalously under $G^\0$. The contribution of the line operator to this anomaly depends on the spurion and, borrowing language from~\cite{Cordova:2019jnf,Cordova:2019uob,Debray:2023ior}, we refer to it as the `family anomaly.'\footnote{This is distinct from the notion of a `defect anomaly' for symmetric defects~\cite{Delmastro:2022pfo,Brennan:2022tyl,Antinucci:2024izg,Brennan:2025acl,Komargodski:2025jbu}. } The family anomaly descends from a non-trivial higher Berry curvature in the space of defect couplings. 

Next, in Sec.~\ref{sec:family_anomaly} we analyze the implications of the family anomaly for correlation functions of tilt operators which probe the defect's response to the action of the broken symmetry. In particular, it forces the three-point response function of the tilt operator to include a separated points contribution (and/or a \emph{partial} contact term) which is dictated by the Postnikov class. We also we explain the relation of these signatures to the higher Berry phase in the space of 1d conformal boundary conditions~\cite{Choi:2025ebk,Wen:2025xka}. Finally, we study the manifestation of symmetry breaking in massless $\text{QED}_4$, and in the effective field theory for spontaneously broken nonabelian 2-groups (which describes e.g. the long-distance behavior of $U(N)$ quantum chromodynamics (QCD)). We also explicitly compute a universal form for the tilt operator obtained by taking any charged line defect and promoting its spurion to a dynamical field. 

\subsection{Wess-Zumino Consistency for Line Defects}
\label{sec:WZconsistency}

We study partition functions with a line operator of charge $q$ inserted and turn on 2-group background fields. To preserve invariance under 1-form symmetry transformations, we attach a Wilson surface to the line:\footnote{Attaching the $B^\2$ surface is convenient for the analysis, but not necessary. As in the introduction, we could keep the bare line and would arrive at the same symmetry breaking conclusion due to Eq.~\eqref{eq:WZModified}. The Wilson surface helps us go further and quantify how symmetry breaking (and in particular the family anomaly) helps restore Wess-Zumino consistency. }
\begin{equation} \label{eq:wilson_surface}
\scrL_{q,\gamma} \, e^{-iq \int_\Sigma B^\2}\,, \quad \partial\Sigma = \gamma\,. 
\end{equation}
This is analogous to the coupling of a background Wilson line to a charged local operator --- the combination above is invariant under 1-form background gauge transformations \eqref{eq:cont_2group_trans}. In the following we set $q=1$, drop the charge subscript, and denote
\begin{equation}
Z_{\scrL,\Sigma}[A^\1,B^\2] = \int \mathscr{D}\text{fields}  \, \scrL_{\gamma}\, e^{-i\int_\Sigma B^\2} \, e^{-S[A^\1,B^\2]}\,.    
\end{equation}
\emph{If} we assume that the line $\scrL$ is symmetric under $G^\0$, then the partition function with the line and Wilson surface inserted transforms by an anomalous phase:
\begin{equation} \label{eq:naive_anomaly} 
Z_{\scrL,\Sigma}[A^\1+d_A\lambda^\0,B^\2 + \alpha^\2(\lambda,A)] \ \stackrel{?}{=} \ Z_{\scrL,\Sigma}[A^\1,B^\2]\, e^{-i \int_\Sigma \alpha^\2(\lambda,A)}\,. 
\end{equation}
where we ignored possible bulk 't Hooft anomalies which play no role in what follows. The Wilson surface gives rise to an anomalous transformation equivalent to that of a $2d$ QFT on a surface $\Sigma$ with boundary. Partition functions on manifolds with boundary are known to violate the Wess-Zumino consistency condition~\cite{Wess:1971yu}. For a healthy anomalous $G^\0$-action, the anomalous variation $\calA(\lambda,A) = \int_\Sigma \alpha(\lambda,A)$ must satisfy  
\begin{equation}
\delta_{\rm{WZ}}\, \calA(\lambda,A) \, \equiv \,   \delta_{\lambda_1}\, \calA(\lambda_2,A) - \delta_{\lambda_2}\, \calA(\lambda_1,A) - \calA(i[\lambda_1,\lambda_2],A) = 0 \,. 
\end{equation}
An anomaly which can be written as the integral of a local density $\alpha^\2$ over a manifold \emph{without} boundary will be Wess-Zumino consistent as long as 
\begin{equation} \label{eq:omega1}
\delta_{\rm{WZ}} \, \alpha^\2(\lambda,A) =  d\alpha^\1(\lambda_1,\lambda_2,A)\,. 
\end{equation}
On the other hand, this is not strong enough to ensure consistency if the anomaly is obtained by integrating $\alpha^\2$ on a manifold \emph{with} boundary, since 
\begin{equation}
\delta_{\rm{WZ}}\, \calA(\lambda,A) =   \int_{\partial\Sigma} \alpha^\1(\lambda_1,\lambda_2,A) \,. 
\end{equation}
If the right-hand side of the equation is non-vanishing, the line operator coupled to background fields is inconsistent~\cite{Jensen:2017eof,Thorngren:2020yht}. To resolve the inconsistency, we must abandon the assumption that the line operator is $G^\0$-symmetric. Equivalently, the line operator must explicitly break the 0-form component of the 2-group. 

Let us take a step further and quantify how explicit symmetry breaking restores WZ consistency. Solutions to the bulk consistency condition are given by the Stora-Zumino descent procedure (see e.g.~\cite{Zumino:1983ew,Alvarez-Gaume:1984zlq,Manes:1985df,Thorngren:2020yht} for a review). Equivalently, it is believed that all consistent anomalies arise via \emph{anomaly inflow}~\cite{Callan:1984sa}, i.e. as the boundary-localized variation of a gapped invertible phase in one dimension higher (for a modern viewpoint see e.g.~\cite{Kapustin:2014tfa,Freed:2014iua,Freed:2016rqq,Monnier:2019ytc}). For the purposes of our discussion, such inflow theories are symmetry-protected topological (SPT) phases for $G^\0$. On the other hand, anomalies on manifolds with boundaries are seemingly incompatible with anomaly inflow, because boundaries of boundaries are empty. To fix this, we follow~\cite{Copetti:2025sym} and develop a generalized inflow formalism that incorporates boundaries. In addition to the familiar background gauge fields extended into a bulk SPT phase, this generalized inflow formalism involves a spacetime-dependent spurion which encodes the explicit symmetry breaking on the boundary. In App.~\ref{app:descent} we rephrase this inflow formalism in terms of the descent procedure. Our discussion there is more general, and can be applied to boundaries and higher-codimension defects in spacetime dimension $d$.  

\begin{figure}[h!]
    \centering
\includegraphics[width=0.5\linewidth]{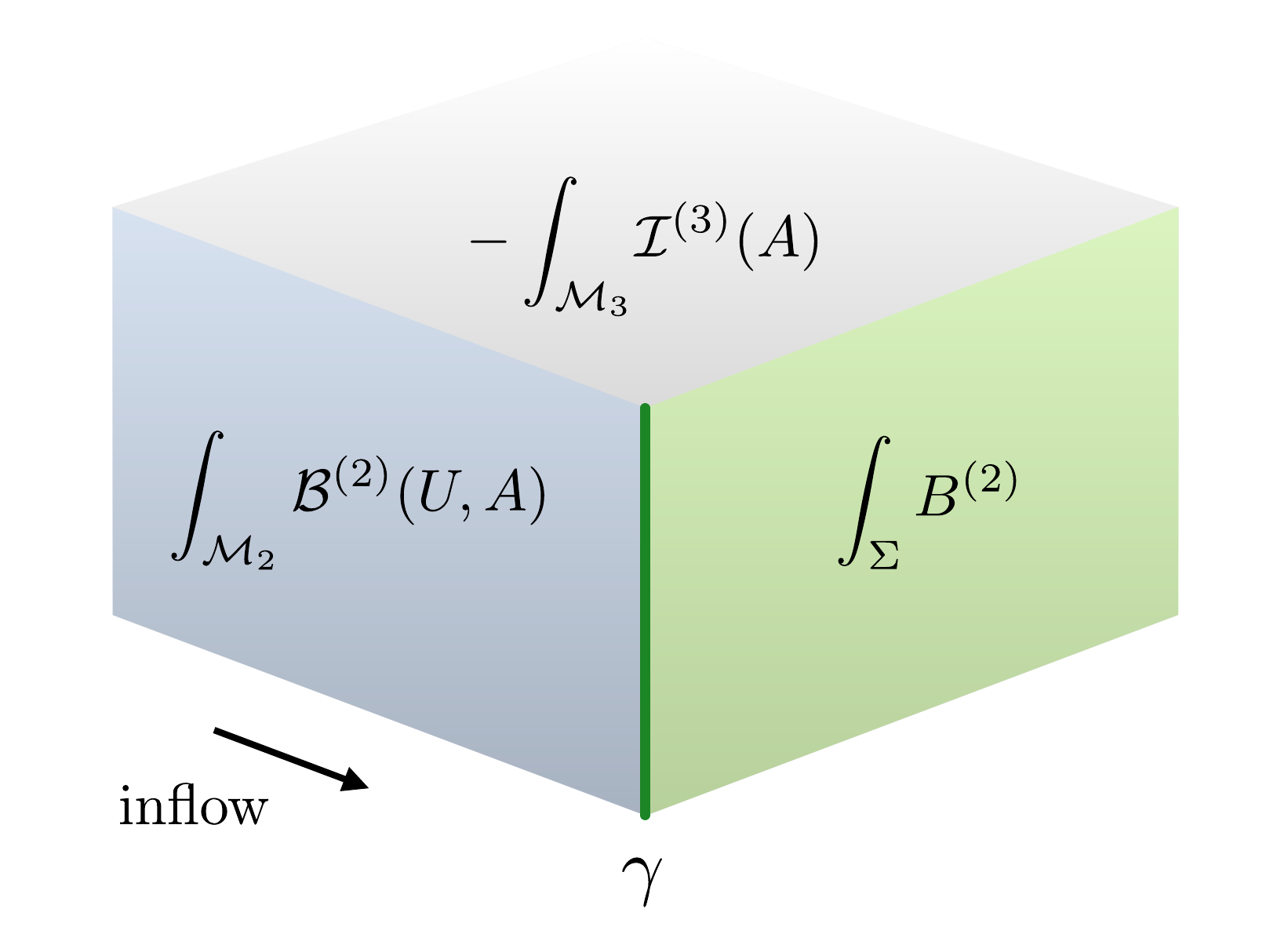}
        \caption{The anomalous transformation of a charged line operator on $\gamma$ attached to a Wilson surface $\Sigma$ is Wess-Zumino consistent if it can be cancelled by anomaly inflow. This requires a second trivially gapped topological boundary theory on $\calM_2$ for the 3d SPT phase $\calI^\3$.   \label{fig:inflow}}
\end{figure}

To formulate the inflow picture, we imagine extending the Wilson surface $\Sigma$ into a bulk 3-manifold $\calM_3$.\footnote{For theories in spacetime dimension $d>2$, $\calM_3$ can be chosen to be embedded in the physical spacetime, or stretch along a genuine auxiliary dimension.} The manifold $\calM_3$ must have an additional boundary component $\partial \calM_3 = \Sigma \cup \calM_2$ with $\partial\Sigma = - \partial \calM_2 = \gamma$ required by $\partial(\partial \calM_3) = 0$. See Fig.~\ref{fig:inflow}. The action of the SPT phase satisfies
\begin{equation}
\delta \calI^\3(A) = d\alpha^\2(\lambda,A)\,,
\end{equation}
which we can use to cancel the anomaly on $\Sigma$, at the expense of leaving an anomaly on $\calM_2$. This in turn must be canceled without using dynamical degrees of freedom. By definition, there is no local effective action involving just the gauge field which can do this. To do this, we introduce a spurion $U(x)$ on $\calM_2$ which transforms under $G^\0$ and look for local terms $\calB^\2$ built from the classical background fields $U$ and $A$ which can saturate the anomaly $\alpha^\2$, 
\begin{equation}  \label{eq:naiveBvar}
\delta \calB^\2(U,A) = \alpha^\2(\lambda,A) \, + \, d\nu^\1(\lambda,U,A) \,,
\end{equation}
where we allowed for the possibility that the anomaly is only matched up to an exact term. We will see in a moment that this exact term plays a crucial role. Before that, let us provide a simple physical way to think about $\calB^\2$. Any symmetry-preserving boundary of a Chern-Simons-SPT for continuous $G^\0$ must be gapless. However, if we explicitly break the symmetry at the boundary, there is no obstruction to gapping it. The background field $U$ should be thought of as the spurion for this explicit symmetry breaking, and the term $\calB^\2(U,A)$ as the long-distance effective action obtained by integrating out the gapped degrees of freedom. If we turn off the background gauge field, $U$ simply parameterizes the space of $G^\0$-breaking gapped theories, and $\calB^\2(U)$ is the \emph{higher Berry phase} over this parameter space~\cite{Hsin:2020cgg,Thorngren:2020yht,Kapustin:2020eby}. As pointed out in \cite{Copetti:2025sym}, for perturbative $G^\0$ anomalies we can take $U \in G^\0$ and $\calB^\2(U,A)$ to be the $2d$ Wess-Zumino term~\cite{Wess:1971yu,novikov,POLYAKOV1983121,D'Adda:141193,cmp/1103940923} coupled to a background gauge field. 

In a slight abuse of language we will also refer to the background-gauge-field-dependent quantity $\calB^\2(U,A)$ as the higher Berry connection. It is the $G^\0$-breaking analog of the topological action for a $G^\0$-preserving SPT phase. In fact, there is a symmetry-breaking boundary analog for each layer of the standard inflow mechanism: 
\renewcommand{\arraystretch}{1.5}
\begin{center}
\begin{tabular}{ c | c}
   Bulk & Boundary \\
  \hline 
Anomaly polynomial $\calI^{(d+2)}(A)$ & Higher Berry curvature $\calH^{(d+1)}(U,A)$ \\
Anomaly inflow $\calI^{(d+1)}(A)$ & Higher Berry phase $\calB^{(d)}(U,A)$ \\
't Hooft anomaly $\alpha^{(d)}(\lambda,A)$  & Family anomaly $\nu^{(d-1)}(\lambda,U,A)$
\end{tabular}
\end{center}
\bigskip
So far we have successfully canceled the anomalies on the surfaces $\Sigma$ and $\calM_2$, at least away from their common boundary. The variation of the 2-form $\calB^\2$ contains an extra term $\nu^\1(\lambda,U,A)$ which localizes to the defect worldline $\gamma$. To complete the inflow picture, this phase must be canceled by the line operator itself. For this to be possible the line must depend on $U$ (and possibly $A$), so we write 
\begin{equation}
\scrL_{\gamma}[U,A]\,. 
\end{equation}
We will show momentarily that this dependence cannot just be due to local counterterms built from $U$ and $A$. 

To see this, we revisit the Wess-Zumino consistency condition. Inflow tells us that the partition function with the line inserted (and the Wilson surface attached) transforms as
\begin{multline} \label{eq:correct_anomaly} 
Z_{\scrL,\Sigma}[A^\1+d_A\lambda^\0,B^\2+\alpha^\2(\lambda,A),U + \delta_\lambda U] \\
=  Z_{\scrL,\Sigma}[A^\1,B^\2,U]\, e^{-i \int_\Sigma \alpha^\2(\lambda,A) -i \int_\gamma \nu^\1(\lambda,U,A)  }\,,
\end{multline}
in contrast to Eq.~\eqref{eq:naive_anomaly}. We refer to the line-localized contribution $\nu^\1(\lambda,U,A)$ as the \emph{family anomaly} associated to the line. The spurion-dependent partition function is Wess-Zumino consistent provided
\begin{equation}
\delta_{\rm{WZ}} \left( \alpha^\2(\lambda,A) + d\nu^\1(\lambda,U,A) \right) = 0 \,,
\end{equation}
which recalling Eq.~\eqref{eq:omega1} requires
\begin{equation} \label{eq:cancel_WZ}
\delta_{\rm{WZ}} \, \nu^\1(\lambda,U,A) = - \alpha^\1(\lambda_1,\lambda_2,A) + d\nu^\0(\lambda_1,\lambda_2,U)\,. 
\end{equation}
Hence, the role of the family anomaly is to cancel the Wess-Zumino inconsistency of the Wilson surface with boundary. It follows that the dependence of the line on $U$ and $A$ cannot be solely through local $1d$ counterterms $\beta^\1(U,A)$, since the variation of any such term is WZ consistent on its own, $\delta_{\rm{WZ}} \, \delta\beta^\1(U,A) = 0$. In other words, the spurion $U$ must couple to a non-trivial operator on the line. At the linearized level, this operator is known as the \emph{tilt operator}. More precisely, if we parameterize $U(t) = e^{i\omega(t)}$, then 
\begin{equation}
    \scrL_\gamma[e^{i \omega(t)}] = \scrL_\gamma[\id]\, \exp\left( i \int dt\,  \omega^a(t) \, \tau^a(t) + \cdots \right) \,,
\end{equation}
where $\tau$ is the tilt operator and the dots refer to terms higher order in $\omega$.

We end with some comments about the freedom in the above inflow formalism:
\begin{itemize}
\item We can change the presentation of the anomaly by adding a local counterterm $\mu^\2(A)$ to the Wilson surface attached to the line and to the topological boundary of the SPT. This will shift $\alpha^\2(\lambda,A) \to \alpha^\2(\lambda,A) + \delta \mu^\2(A)$ but $\alpha^\1$ and $\nu^\1$ are unaffected. 
\item We can shift $\alpha^\2(\lambda,A) \to \alpha^\2(\lambda,A) + d\Lambda^\1(\lambda,A)$, which corresponds to activating a 1-form background gauge transformation that depends on the 0-form gauge field and gauge parameter (but not on the spurion). This will shift $\nu^\1(\lambda,U,A) \to \nu^\1(\lambda,U,A) - \Lambda^\1(\lambda,A)$ and $\alpha^\1(\lambda_1,\lambda_2,A) \to \alpha^\1(\lambda_1,\lambda_2,A)  + \delta_{\rm{WZ}}\,   \Lambda^\1(\lambda,A)$. 
\item We can shift $\calB^\2(U,A) \to \calB^\2(U,A) + \eta^\2(U,A)$ as long as $\delta \eta^\2(U,A) = d\eta^\1(\lambda,U,A)$ (this implies $\delta_{\rm{WZ}}\, \eta^\1(\lambda,U,A) = d\eta^\0(\lambda_1,\lambda_2,U,A)$). This can be absorbed by $\nu^\1(\lambda,U,A) \to \nu^\1(\lambda,U,A) + \eta^\1(\lambda,U,A)$ and $\nu^\0(\lambda_1,\lambda_2,A) \to \nu^\0(\lambda_1,\lambda_2,A) + \eta^\0(\lambda_1,\lambda_2,A)$. The only example we are aware of is when $\eta^\2(U,A) = d\rho^\1(U,A)$ is exact. This is equivalent to adding a local counterterm to the line, which simply changes $\nu^\1(\lambda,U,A) \to \nu^\1(\lambda,U,A) + \delta \rho^\1(U,A)$ but does not modify the WZ consistency condition.
\end{itemize}
\bigskip

To ground the above discussion in an example, let us consider the case where $G^\0 = SU(N)^\0$ such that the 2-group transformation rule \eqref{eq:cont_2group_trans} takes the form 
\begin{equation}
    B^\2 \ \to \  B^\2 + \frac{\kappa}{4\pi}\Tr\, (\lambda^\0 dA^\1)\,,
\end{equation}
where the trace is taken in the fundamental representation with generators normalized to $\Tr(T^aT^b) = \frac{1}{2}\delta^{ab}$. The parameter $\kappa \in \ZZ$ characterizes the Postnikov class. In other words, the 2-form background gauge field shifts by the perturbative $G^\0$ anomaly with anomaly density
\begin{equation} \label{eq:anomaly_choice}
\alpha^\2(\lambda,A) = \frac{\kappa}{4\pi}\Tr\, (\lambda^\0 dA^\1) 
\end{equation}
and inflow action 
\begin{equation}
  \int_{\calM_3}  \mathcal{I}^{(3)}(A) = \frac{\kappa}{4\pi}\int_{\calM_3} \mathrm{Tr}\left(A^\1\wedge dA^\1 + \frac{2i}{3} A^\1\wedge A^\1\wedge A^\1\right) ~.
\end{equation}
Given this concrete anomaly, we can work out the full finite gauge transformation of $\mathcal{I}^{(3)}$. Denoting $^g\!A^\1 = g^{-1}(A^\1 - i \, d)g$, 
\begin{multline}\label{eq:I3gaugevar}
    \mathcal{I}^{\3}(^g\!A)-\mathcal{I}^{\3}(A) =  \frac{\kappa}{12\pi} \mathrm{Tr}\left( g^{-1}dg\wedge g^{-1}dg \wedge g^{-1}dg \right) + \frac{i\kappa}{4\pi}d\left[\mathrm{Tr}\left((dg)g^{-1}\wedge A^\1\right)\right] \\
  = d\left[ \kappa\, \calB^\2_{\rm{WZ}}(g) + \frac{i\kappa}{4\pi}\Tr\left((dg)g^{-1}\wedge A^\1\right) \right]   = d\left[ \frac{\kappa}{4\pi}\Tr (\lambda^\0 dA^\1) + \calO(\lambda^2) \right] \,,
\end{multline}
where $\calB_{\rm{WZ}}^{(2)}(g)$ is the $2d$ Wess-Zumino-Witten term~\cite{Wess:1971yu,novikov,POLYAKOV1983121,D'Adda:141193,cmp/1103940923} and we restricted to infinitesimal $g \approx 1 + i \lambda$. The anomaly is inconsistent on a manifold with boundary, since
\begin{equation}
\delta_{\rm{WZ}}\, \alpha^\2(\lambda,A) = d\alpha^\1(\lambda_1,\lambda_2,A)\,, \quad \alpha^\1(\lambda_1,\lambda_2,A) =  \frac{i\kappa}{4\pi}\Tr\, ([\lambda^\0_1,\lambda^\0_2]A^\1)\,. 
\end{equation}
To restore WZ consistency, we introduce the spurion $U \in G^\0$ which transforms as $U \to g^{-1}\, U$. Observe that the gauge variation of the closed 3-form 
\begin{equation}
\calH^\3(U,A) = - \frac{\kappa}{12\pi}\Tr\left( U^{-1}dU \wedge U^{-1}dU \wedge U^{-1}dU \right) - \frac{i\kappa}{4\pi}d\left[  \Tr\left( (dU)U^{-1}\wedge A^\1 \right) \right]
\end{equation}
is exactly Eq.~\eqref{eq:I3gaugevar}. Since $\calH^\3$ is closed, we can write it locally as a total derivative, 
\begin{equation}
\calH^\3(U,A) = d\calB^\2(U,A) \,, \quad \calB^{(2)}(U,A) = - \kappa\, \calB_{\rm{WZ}}^{(2)}(U) - \frac{i\kappa}{4\pi} \Tr\left( (dU)U^{-1}\wedge A^\1 \right)\,.
\end{equation}
A local expression of the $2d$ WZ term $\calB_{\rm{WZ}}^{(2)}(U)$ can be found using the Poincar\'e lemma: we take $U = e^{i\omega^a T^a}$ and use the local tangent space coordinates near the identity to write
\begin{equation}\label{eq:BWZ}
    \calB_{\rm{WZ}}^{(2)}(U) \equiv \frac{1}{2} B_{ab}(\omega)\, d\omega^a \wedge d\omega^b = - \frac{1}{4\pi} \left[\frac{\sinh X(\omega)-X(\omega)}{X^2(\omega)}\right]_{ab}\, d\omega^a \wedge d\omega^b \,,
\end{equation}
where $X^{ab}(\omega) \equiv -f^{abc}\,\omega^c$ as in \cite{Choi:2025ebk}. To continue the inflow procedure, we need to compute the gauge variation of $\calB^\2$ relative to the $2d$ anomaly $\alpha^\2$. In App.~\ref{app:Bvariations} we show that 
\begin{equation}
\delta \calB^\2(U,A) - \alpha^\2(\lambda,A) \, =\,  d\nu^\1(\lambda,U,A)\,,
\end{equation}
where 
\begin{equation} \label{eq:nu1_full}
    \nu^\1(\lambda,U,A) = - \frac{\kappa}{2\pi} \left(\frac{1}{X(\omega)}+\frac{1}{1-e^{X(\omega)}}\right)^{ab} \lambda^a \, d\omega^b - \frac{\kappa}{4\pi}\lambda^a A^a\,.
\end{equation}
This holds to all orders in $\omega$. Expanding $U$ near the identity we get
\begin{equation}  \label{eq:nu1_expanded}
 \nu^\1(\lambda,U,A) =  - \frac{\kappa}{4\pi} \lambda^a (d\omega^a + A^a) + \frac{\kappa}{24\pi} f^{abc} \lambda^a \omega^b d\omega^c \,+\, \calO(\omega^3) \,.
\end{equation}
We note that the leading term is scheme-dependent, since the local counterterm $iq\int_\gamma \frac{\kappa}{4\pi}\Tr(\omega\, A)$ can remove it in favor of a tower of terms starting with $\sim f^{abc} \lambda^a \omega^b A^c$. In App.~\ref{app:counterterms} we show that the term $f^{abc}\lambda^a \omega^b d\omega^c$ is scheme-\emph{independent}. We discuss the physical implications of this term in the next section.

\subsection{Imprints of the Family Anomaly}
\label{sec:family_anomaly}

Suppose we have a family of line operators $\scrL_\gamma[U]$ depending on a spurion $U$. As discussed in the previous section, we can couple the line defect to a modulated source $U(t)$, where $t$ is the coordinate along $\gamma$. We define
\begin{equation}
Z_\scrL[U(t)]  = \int \mathscr{D}\textrm{fields} \ \scrL_\gamma[U(t)]\ e^{-S} ~.
\end{equation}
For now, we have set the 2-group background fields to zero. Let us decompose $U(t) = e^{i\omega(t)}\, U_0$ as a modulation on top of some reference point $U_0$ on the group manifold. We can write the effect of the modulation as\footnote{We expect such a formula to make sense only in some patch of the moduli space, since the coordinates $\omega^a$ are not globally well-defined. }  
\begin{equation} \label{eq:source_coupling}
\scrL_\gamma[e^{i\omega(t)}\, U_0] =  \scrL_\gamma[ U_0 ] \, \exp\left( i \int dt\, \omega^a(t)\, \tau^a(U_0;t) + \cdots \right)\,,
\end{equation}
where $\tau^a(U_0;t)$ is the tilt operator defined at $U_0$ and the $\cdots$ refer to terms which are higher order in the modulation $\omega(t)$. The tilt operator appears on the right-hand side of the non-conservation equation 
\begin{equation}
\partial^\mu \, j_\mu^a(x) = \delta^\3(\x) \, \tau^a(U_0;t)\,,
\end{equation}
where $x^\mu = (t,\x)$. On the other hand, the terms higher order in $\omega(t)$ encode the nonlinear transformation of the tilt operator itself (for instance they appear in the Ward identity for $\partial^\mu j_\mu^a(x)\, \tau^b(t')$). These terms are cumbersome to parameterize, but are crucially important for matching the family anomaly. For an example of such `higher' tilts, see the discussion around Eq.~\eqref{eq:higher_tilt}.

The central objects we will study are the connected $n$-point response functions around a given reference point $U_0$ in the space of line defects,
\begin{equation} \label{eq:connected_n_pt}
G^{a_1 \cdots a_n}(U_0; t_1, \ldots, t_n)  := \left. (-i)^n \frac{\delta}{\delta \omega^{a_1}(t_1)} \cdots \frac{\delta}{\delta \omega^{a_n}(t_n)} \log Z_\gamma[e^{i\omega(t)}\, U_0]  \right|_{\omega = 0} ~.
\end{equation}
The 1-point function is simply the expectation value of the tilt operator at the point $U_0$,  
\begin{equation}
G^a(U_0;t) \ = \ \langle \tau^a(U_0;t) \rangle_{U_0}\,.
\end{equation}
If we are only interested in correlators at separated points, the $n$-point response function is equivalent to the $n$-point connected correlation function of the tilt operator which appears on the right-hand side of the current non-conservation equation~\cite{Drukker:2025dfm}. In general, however, the $n$-point response function will also contain (partial) contact terms including the higher tilts. As we will see, the most direct signature of the family anomaly is in contact terms, so it is more appropriate to study the response functions instead of correlators of the tilt itself.

We will analyze how the 2-group structure, and in particular the Wess-Zumino-enforced anomaly in Eq.~\eqref{eq:correct_anomaly}, constrains correlation functions of tilt operators. This was recently investigated in~\cite{Copetti:2025sym, Drukker:2025dfm} in the context of symmetry breaking on conformal boundaries and defects. Here we revisit the question in the 2-group context, without assuming conformal symmetry (neither in the bulk nor on the line). To see the effect of the 2-group, we consider how the response functions depend on the position $U_0$ on the group manifold. Since different line defects are related by the action of the symmetry, one might expect the response functions $G^{a_1\cdots a_n}$ to be identical for different choices of $U_0$ (up to corresponding covariant transformations of the indices $a_i$). However, the non-trivial anomaly indicates that this expectation only holds at separate points, and is violated by contact terms. The statement of the anomaly is that there is no choice of counterterms which can remove these contact terms at all points on the group manifold.

Recall that the partition function with the line defect inserted, coupled to a modulated source for tilts, satisfies
\begin{equation}
    Z_\scrL[g^{-1}\, U(t)] = Z_\scrL[U(t)]\, \exp\left(-i\int_\gamma \nu^{(1)}(g,U(t))\right) ~,
\end{equation} 
where $g$ is a constant gauge transformation. Let us choose $U(t) = e^{i\omega(t)}$ to be a small modulation and set $g^{-1} = U_0 \equiv e^{iu_0}$ equal to a fixed position on the group manifold. The anomalous transformation implies the following relation between a modulation at $U_0$ and at the identity, 
\begin{equation}
    \log Z_\scrL[e^{i \omega^a(t) (e^{X(u_0)})^{ab}\, T^b}\, U_0] = \log Z_\scrL[e^{i\omega(t)}] - i \int_\gamma \nu^{(1)}(U_0^{-1},e^{i\omega(t)}) ~,
\end{equation}
where $X^{ab}(u_0) = -f^{abc}\, u_0^c$, and we used $e^{iu_0}\,  T^a \, e^{-iu_0} = (e^{X(u_0)})^{ab}\,T^b$. Taking $n$ functional derivatives leads to the following relation on the $n$-point response function defined in Eq.~\eqref{eq:connected_n_pt} at the origin of the moduli space and at the point $U_0$, 
\begin{multline}
     (e^{X(u_0)})^{a_1}{}_{b_1} \cdots (e^{X(u_0)})^{a_n}{}_{b_n}\, G^{b_1 \cdots b_n}(U_0;t_1,\cdots,t_n)  \\
   =\  G^{a_1 \cdots a_n}(\mathbbm{1};t_1,\cdots,t_n) \ +\  (-i)^n \frac{\delta}{\delta \omega^{a_1}(t_1)} \cdots \frac{\delta}{\delta \omega^{a_n}(t_n)} \left(-i\int_\gamma \nu^{(1)}(U_0^{-1},e^{i\omega(t)})\right)\Bigg|_{\omega = 0} ~.
\end{multline}
Response functions at different points on the moduli space of line defects differ by contact terms due to the anomaly. Extracting the contact term for all $n$-point functions with finite $U_0$ is tedious but straightforward. 

\subsubsection{Two-Point Response Functions}
\label{sec:2point_res}

The anomaly first contributes at $n  = 2$.  To see this we compute the finite global transformation of the $2d$ WZ term expanded to second order in the modulation. This computation is performed in App.~\ref{app:finite_trans}. Setting $\lambda_0 = -u_0$ in Eq.~\eqref{eq:nu_finite_1} yields
\begin{multline}
\nu^\1(U_0^{-1}, e^{i\omega(t)}) = - \frac{\kappa}{2}\left( M(-u_0)\, B(u_0)\, M(u_0) - B(0) \right)_{ab} \, \omega^a \wedge d\omega^b   + \calO(\omega^3) \\
=  \frac{\kappa}{8\pi} \left( \frac{\sinh X(u_0)- X(u_0)}{\cosh X(u_0) - 1} \right)^{ab}\, \omega^a \wedge d\omega^b + \calO(\omega^3)\,,
\end{multline}
where
\begin{equation}
    M^{ab}(u_0) = \left(\frac{X(u_0)}{e^{X(u_0)}-1}\right)^{ab}\,, \quad X^{ab}(u_0) = -f^{abc}\, u_0^c\,. 
\end{equation}
From this we can acquire the following relation for the 2-point response functions on the moduli space of boundary conditions 
\begin{equation} \label{eq:2pt_response_final}
\begin{split}
  (e^{X(u_0)})^{ac} (e^{X(u_0)})^{be}  & \ G^{ce}  (U_0;t_1,t_2)  \\
   = & \ G^{a b}(\mathbbm{1};t_1,t_2)\ + \ \frac{i\kappa}{8\pi}\left( \frac{\sinh X(u_0)- X(u_0)}{\cosh X(u_0) - 1 } \right)^{a b} \partial_{t_1}\delta(t_1 - t_2) \\ 
   = & \ G^{a b}(\mathbbm{1};t_1,t_2)\ - \ \frac{i\kappa}{24\pi}(f^{a b c} \, u_0^c \, +\, \cdots)\, \partial_{t_1}\delta(t_1 - t_2) \,. 
\end{split}
\end{equation}
To summarize, we have shown that the family anomaly contributes to a mismatch in contact terms in the 2-point response function at two distinct points on the moduli space of line defects. Remarkably, the difference in contact terms is essentially given by the WZ 2-form itself. 

Before moving on, we note that (as is always the case with anomalies), the precise contact term on the right-hand side of Eq.~\eqref{eq:2pt_response_final} is scheme-dependent and subject to counterterm ambiguities. The most general boundary counterterm of $\omega^a(t)$ that could change the above $\nu^{(1)}$ takes the form:
\begin{equation}
Z_\scrL[e^{i\omega(t)}]  \ \longrightarrow \  Z_\scrL[e^{i\omega(t)}] \, e^{i \kappa \int_\gamma \Lambda_a(\omega) d\omega^a}\,, 
\end{equation}
Adding this counterterm changes $\nu^{(1)}$ to 
\begin{multline}
    \nu^{(1)}(U_0^{-1},e^{i\omega(t)}) \\ \goesto 
    - \frac{\kappa}{2}\left( M(-u_0)\, (B(u_0)+d\Lambda(u_0))\, M(u_0) - (B(0)+d\Lambda(0)) \right)_{ab} \, \omega^a \wedge d\omega^b   + \calO(\omega^3)\,,
\end{multline}
the result of which is to gauge transform the WZ 2-form $B_{ab}$. 

\

To conclude the discussion, we point out another signature of the family anomaly via the 2-pt response function. Consider a line operator $\scrL$ with the following modulated boundary condition
\begin{equation}
    \begin{tikzpicture}[baseline=(current bounding box.center), scale = 0.8]
        \draw[dgreen, line width = 0.5mm, ->-=0.5] (2,0) arc (0:120:2);
        \draw[blue, line width = 0.5mm, ->-=0.5] (-1,1.7320508076) arc (120:240:2);
        \draw[red, line width = 0.5mm, ->-=0.5] (-1,-1.7320508076) arc (240:360:2);
        \filldraw (-1,1.7320508076) circle (2pt);
        \filldraw (-1,-1.7320508076) circle (2pt);
        \filldraw (2,0) circle (2pt);
        \node at (0,0) {$\scrL$};
        \node[dgreen, above] at (1,1.75) {\footnotesize $e^{i\alpha}$};
        \node[blue, left] at (-2,0) {\footnotesize $e^{i\beta}$};
        \node[red, below] at (1,-1.9) {\footnotesize $e^{i\gamma}$};
        \node[right] at (2,0) {\footnotesize $t_1$};
        \node[left] at (-0.95,1.9) {\footnotesize $t_2$};
        \node[left] at (-0.95,-1.9) {\footnotesize $t_3$};
    \end{tikzpicture} ~, \quad \quad \quad {\color{dgreen} e^{i\alpha}}, {\color{blue} e^{i\beta}}, {\color{red} e^{i\gamma}} \in G ~,
\end{equation}
where $\omega(t)$ takes constant values labeled by $\alpha,\beta,\gamma$ in three segments. As shown in detail in App.~\ref{app:hbp}, from the correlation function $Z_{\scrL}[e^{i\omega(t)}]$ of this modulated loop operator, one can define a 2-form $\mathcal{B}$ on the group manifold
\begin{equation}
\begin{aligned}
    \mathcal{B}^\2 &= - \frac{i}{2}\left(\left(\frac{\partial^2}{\partial \beta^a \partial \gamma^b} - \frac{\partial^2}{\partial \beta^b \partial \gamma^a}\right)_{\beta=\gamma = \alpha} \log Z_{\scrL}[e^{i\omega(t)}]\right) d\alpha^a \wedge d\alpha^b \\
    &=  \kappa \, \mathcal{B}^\2_{\mathrm{WZ}} \, + \, \Omega_{ab}(t_1,t_2,t_3) \theta^a \wedge \theta^b ~,
\end{aligned}
\end{equation}
where $\theta^a$ is the left-invariant 1-form on $G^\0$. In the calculation, the anomalous transformation of the 2-pt response function \eqref{eq:2pt_response_final} allows us to separate the anomaly contribution $\kappa\,\mathcal{B}^\2_{WZ}$ from the second term $\Omega_{ab}(t_1,t_2,t_3) \theta^a \wedge \theta^b$ which is controlled by the QFT dynamics via the 2-pt response function (at a given point of $G^\0$). Furthermore, $\Omega_{ab}(t_1,t_2,t_3)$ is a constant along $G^\0$ which implies that $\Omega_{ab} \theta^a \wedge \theta^b$ is a globally well-defined 2-form on $G^\0$ and leads to a \textit{topologically trivial} field strength; in contrast, the anomaly contribution $\kappa\, \mathcal{B}^\2_{\mathrm{WZ}}$ leads to a \textit{topologically non-trivial} field strength with $\kappa$ units of flux. This generalizes a similar result on the higher Berry curvature on $2d$ boundary conformal manifolds acquired in ~\cite{Choi2026BerryPhaseBoundaryConformalManifolds}. Further discussion comparing the two approaches is in App.~\ref{app:hbp}.

\subsubsection{Three-Point Response Function}
\label{sec:3point_res}

In the section above we observed that the $n$-point response functions at distinct points on the moduli space of defects differ by contact terms coming from the anomaly. Instead of comparing contact terms at two points on the moduli space, we can track the infinitesimal change of the contact term at a given point. This gives a local characterization of the anomaly on the moduli space, and allows us to find the effect of the anomaly on correlation functions beyond pure contact terms. 

We start with the Ward identity for the 2-point response function at two points in the moduli space, 
\begin{equation} \label{eq:2pt_response_again}
  (e^{X(u_0)})^{ac} (e^{X(u_0)})^{be}\, G^{ce}  (U_0;t_1,t_2) = G^{ab}(\mathbbm{1};t_1,t_2) - \frac{i\kappa}{24\pi}\left( f^{abc} \, u_0^c \, + \cdots \right)  \partial_{t_1}\delta(t_1 - t_2) \,,
\end{equation}
and expand around $u_0 =0$ ($U_0 = \id$) by taking the $u_0$-derivative on both sides. On the right-hand-side we get a non-trivial contact term. Taking the derivative of the left-hand-side will bring down an insertion of the integrated tilt operator in the correlator of two tilts. The true situation is more complicated because the modulated source itself depends on $u_0$. Relegating the details to App.~\ref{app:2to3pt}, we find 
\begin{multline} \label{eq:integrated_3pt}
 \int dt\, G^{a b c}(\id; t_1,t_2,t) =  - \frac{\kappa}{24\pi}f^{a b c} \, \partial_{t_1}\delta(t_1 - t_2)
 \\
 +\frac{i}{2}  \left( f^{c h b} \,G^{a h}(\id;t_1,t_2) +  f^{c h a} \,G^{b h}(\id;t_2,t_1)\right) \\
  +\frac{1}{12}(f^{c e a}f^{e h b}+f^{c e b}f^{e h a}) \delta(t_1-t_2)\,  G^h(\id;t_1) \,. 
\end{multline}
This can be compared to Eq.~(13) in \cite{Drukker:2025dfm}. At face value, this equation implies that a combination of 1-, 2-, and 3-point functions must saturate the anomaly-induced contact term. By contracting both sides of the equation with $f^{abc}$, the 1-point and 2-point functions drop out, and we obtain a Ward identity for the integrated 3-point function:
\begin{equation} \label{eq:contracted_ward_id}
    \int dt\, f^{abc}\, G^{abc}(\id;t_1,t_2,t) \, = \, - \frac{\kappa\, C_{\rm{adj}}\, \text{dim}(G)}{24\pi} \, \partial_{t_1} \delta(t_1-t_2)\,,
\end{equation}
where $C_{\rm{adj}}$ is the quadratic Casimir in the adjoint representation. The contact term must be matched by the 3-point response function
\begin{equation} \label{eq:3pt_ward} 
    \int dt\,  G^{abc}(\id;t_1,t_2,t) \, \supset \, - \frac{\kappa}{24\pi} \,f^{abc}\, \partial_{t_1} \delta(t_1-t_2)\,.
\end{equation}
The remaining question is what kind of structures in the \emph{un}integrated 3-point function are capable of matching the contact term in the integrated 3-point function. 

Bose symmetry implies that $G^{a_1a_2a_3}(\id; t_1,t_2,t_3)$ is invariant under simultaneous exchange of pairs $(a_i,t_i)$. To match the contact term, we only have to consider terms of the form $G^{a_1a_2a_3}(\id; t_1,t_2,t_3) \, \supset \, f^{a_1a_2a_3} \, F(t_1,t_2,t_3)$,
where $F$ is completely antisymmetric. Translation invariance implies that $F(t_1,t_2,t_3) = \mathsf{F}(t_{12},t_{23})$ with $t_{ij} = t_i - t_j$. The Fourier transform is  
\begin{equation}
\widehat{F}(E_1,E_2,E_3) = 2\pi \delta(E_1+E_2+E_3) \, \widehat{\mathsf{F}}(E_1,E_2,E_3)\,,
\end{equation}
which is also completely antisymmetric in its arguments. The Ward identity is simply
\begin{equation}
\widehat{\mathsf{F}}(E,-E,0) = \frac{i\kappa}{24\pi} \, E\,. 
\end{equation}
In principle, the right-hand-side of the Ward identity can receive contributions from pure contact terms, partial contact (also known as semi-local) terms,\footnote{See e.g.~\cite{Bzowski:2014qja,Dymarsky:2014zja,Nakayama:2018dig,Schwimmer:2018hdl} for examples where partial contact terms play an important role in anomaly-matching.} and genuine separated points pieces. We observed in Sec.~\ref{sec:family_anomaly} that adding a local counterterm has the effect of changing the representative of the WZ 2-form $B_{ab}$. This cannot remove the contact term mismatch of 2-point functions in Eq.~\eqref{eq:2pt_response_final}, nor the contact term in the 3-point function in Eq.~\eqref{eq:3pt_ward}. We can see this more explicitly: A pure contact term in $F(t_1,t_2,t_3)$ is an antisymmetric polynomial in momentum space. But any antisymmetric polynomial in $E_1,E_2,E_3$ is proportional to the Vandermonde determinant, 
\begin{equation}
\widehat{\mathsf{F}}(E_1,E_2,E_3) = (E_1-E_2)(E_2-E_3)(E_1-E_3)\, \mathsf{H}(E_1,E_2,E_3)\,,
\end{equation}  
where $\mathsf{H}$ is a symmetric polynomial. The Ward identity yields 
\begin{equation}
\mathsf{H}(E,-E,0) = \frac{i\kappa}{48\pi}\frac{1}{E^2}
\end{equation}
which is inconsistent with $\mathsf{H}$ being a polynomial. We conclude that \emph{pure contact terms cannot saturate the anomalous Ward identity.}

This leaves us with partial contact terms and separated points contributions. In App.~\ref{app:partial_contact} we argue that the partial contact term
\begin{equation}
\mathsf{F}(t_{12},t_{23}) = \frac{i\kappa}{48\pi} \delta^{''}\!(t_{12}) \, \Theta(t_{23}) \, e^{-\epsilon\, t_{23}} + \text{signed permutations}.
\end{equation}
can match the anomalous Ward identity if it is understood as a distribution where we send $\epsilon \to 0$ only after performing all integrals. Note that this partial contact term is not consistent with conformal symmetry. 

Finally, the anomaly can be matched by terms which are non-trivial at separated points $t_1 \not = t_2 \not= t_3$. An example is the conformally invariant 3-point function discussed in~\cite{Drukker:2025dfm,Copetti:2025sym},\footnote{This expression should be defined with an $i\epsilon$ prescription, and then appropriately antisymmetrized. The Ward identity is matched regardless of the choice of prescription.  }
\begin{equation}
\mathsf{F}(t_1,t_2,t_3) =  \frac{\kappa}{16\pi^3} \frac{1}{(t_1-t_2)(t_2-t_3)(t_1-t_3)}\,.
\end{equation}
Along a general RG flow where neither the bulk nor defect are conformal, we can only say that some combination of separated points contributions and partial contact terms will saturate the anomaly. If both the bulk and the defect are conformal, we can rule out the partial contact terms and we are left with just the separated points piece. 

To summarize, the family anomaly (or higher Berry phase) leads to an anomalous Ward identity for the integrated 3-point response function \eqref{eq:contracted_ward_id} which cannot be saturated by pure contact terms, i.e. it is completely scheme-independent.

\subsection{Examples}\label{sec:ExNonAbel}

We now turn to known examples of theories with continuous nonabelian 2-group global symmetries and analyze their line operators. In each case we demonstrate how charged lines explicitly break the 0-form part of the 2-group, and when possible give explicit expressions for tilt operators. Given a QFT with 2-group symmetry, there is neither a unique family of line defects nor a notion of a `minimal' line defect, as we can dress any consistent line with additional degrees of freedom which also contribute to symmetry violation. To emphasize this point, in Sec.~\ref{sec:nlsm_on_line} we take a generic symmetry-breaking line and couple it to a $G^\0$ nonlinear sigma model. This leads to a universal form for the tilt operator which does not depend on the details of the original line defect.

\subsubsection{QED-Like Models}
\label{sec:qed}

A quintessential model with 2-group symmetry is given by abelian gauge theory coupled to $N_f$ Dirac fermions $\Psi^I$ of charge $q$ under a compact $U(1)$ gauge group in four dimensions~\cite{Cordova:2018cvg}. This model has been studied in the context of monopole scattering, where the 't Hooft line defect models an infinitely heavy probe monopole off of which the massless fermions scatter~\cite{vanBeest:2023dbu,Aharony:2023amq}. The Lagrangian of the theory is given by:
\begin{equation}
    \mathscr{L} = - \frac{1}{4e^2}f_{\mu\nu}f^{\mu\nu} - i \Bar{\psi}_i\Bar{\sigma}^\mu D_\mu \psi^i - i\overline{\widetilde{\psi}}_{\widetilde{j}}\Bar{\sigma}^\mu D_\mu \widetilde{\psi}^{\widetilde{j}}~,
\end{equation}
where we have written the Dirac fermions in terms of Weyl fermions $\Psi^I = (\psi_\alpha^I, \overline{\widetilde{\psi}}^{\Dot{\alpha}, I})$ 
with gauge charges:
\begin{equation}
    q[\psi_\alpha^i] = q, \quad q[\widetilde{\psi}_\alpha^{\widetilde{i}}] = -q ~.
\end{equation}
We emphasize that the fermions transform under different flavor symmetry groups: 
\begin{equation}
    \begin{split}
        SU(N_f)^\0_L&:\psi^i \mapsto (g_L)^i_{\,j}\, \psi^j\\
        SU(N_f)^\0_R&:\widetilde{\psi}^{\widetilde{j}}\mapsto (g_R)^{\widetilde{j}}_{\,\widetilde{i}} \, \widetilde{\psi}^{\widetilde{i}}~.
    \end{split}
\end{equation}
We ignore the anomalous $U(1)_A^\0$ axial symmetry, which is broken by the ABJ-anomaly. The nonabelian flavor symmetries form a 2-group with the magnetic 1-form symmetry and their Postnikov classes can be read off from the mixed gauge-flavor-flavor triangle diagram:
\begin{equation}
    \kappa_L = q \quad \kappa_R=-q~.
\end{equation}
The diagonal combination of the flavor symmetries does not participate in the 2-group.

Commonly, one defines the 't Hooft line as a disorder operator by excising a small ball along the line $\gamma$ in spacetime and imposing boundary conditions for the gauge field:
\begin{equation}
    \frac{1}{2\pi}\int_{S^2_\epsilon} f^\2 = q_M\in \mathbb{Z}
\end{equation}
inside the path integral. In the late 70's, it was understood that it is necessary to add boundary conditions for the fermions to ensure self-adjointness of the fermion Hamiltonian~\cite{Kazama:1976fm}. Modern expositions of this problem can be found in~\cite{Aharony:2023amq, vanBeest:2023dbu} and we review the argument in App.~\ref{app:SymmetryBreakingMasslessQED}. The problem of self-adjointness is limited to the $s$-wave sector, which can be reduced to a system of two-dimensional fermions. From there, it is clear that the boundary must break the flavor symmetry due to the 't Hooft anomaly of the effective two-dimensional theory.

The upshot is that the 't Hooft lines of this model are labeled by a modulus $U\in SU(N_f)$ which characterizes the fermion boundary conditions. It transforms as:
\begin{equation}
    SU(N_f)_L \times SU(N_f)_R: \quad U \ \to \  g_L\, U \, g^{-1}_R\,.
\end{equation}
At a given modulus, the line preserves a twisted $SU(N_f)$ subgroup of the full flavor symmetry with $g_L = U g_R U^{-1}$.

In many UV-complete theories, the 2-group is explicitly broken in the UV but emerges as a symmetry of the low-energy effective theory. The simplest UV-completion for the above model is discussed in \cite{Cordova:2018cvg}. Consider a $SU(2)$ Georgi-Glashow model with an adjoint scalar $\Phi$ and $N_f$ massless Weyl fermion doublets $\psi_\alpha^{I,a}$ where $I$ is the flavor index and $a$ the gauge index. We add a potential $V(\Phi)$ that Higgses the gauge group to $U(1)$, and forbid any Yukawa couplings that could gap out the fermions. The Weyl fermion doublets become Dirac fermions $\Psi^I$ with charge $q=1$ under the $U(1)$ gauge group in the IR, leading to massless QED with $N_f$ flavors. If we Higgs the UV gauge theory far above the strong coupling scale $v\gg\Lambda$, we can construct finite energy monopole saddles, which are not accessible in the IR effective theory. We expect the worldline of the lightest magnetic monopole to be represented by a 't Hooft line in the low-energy theory. Scattering fermions off of the 't Hooft-Polyakov monopole corresponds to scattering in the background of the 't Hooft line defect in the low energy theory. Following this prescription, we can in principle determine the 't Hooft line modulus $U$ in the IR theory by studying the worldline interactions of scattering states in the 't Hooft-Polyakov background. In the model above this is not necessary, since the UV gauge theory preserves the standard, untwisted $SU(N_f)_V$-symmetry. The monopole and all of its interactions preserve this symmetry and we find $U = \id$. To realize an IR 't Hooft line at a generic point on the moduli space as a bona-fide UV monopole, we would have to seek out UV completions which break the $SU(N_f)_L \times SU(N_f)_R$ flavor symmetry entirely. We leave the interesting problem of how to realize a general IR 't Hooft line in a UV complete model for future work.

\subsubsection{Nonabelian Goldstone-Maxwell Model}

Next we consider a model in which we can directly compute the tilt operator. This is the nonabelian Goldstone-Maxwell model, i.e. the long-distance effective field theory for spontaneous nonabelian 2-group symmetry breaking~\cite{Antinucci:2024bcm,Berean-Dutcher:2025ohp}. For simplicity, we will stick to the QCD-like symmetry breaking pattern with $SU(N_f)_L^\0 \times SU(N_f)_R^\0 \to SU(N_f)_V^\0$, but the discussion can easily be generalized to generic Lie groups $G^\0 \to H^\0$. 
The Goldstone field of this nonlinear sigma model (NLSM) is a matrix $u\in SU(N_f)$ which transforms as
\begin{equation}
        SU(N_f)_L \times SU(N_f)_R: \ u\mapsto g_L^\dag\,  u\, g_R\,. 
\end{equation}
The standard `pions' are given by $u= e^{i\uppi^a T^a}$. We will couple the theory to a background gauge field $A_L^\1$ for $SU(N_f)_L^\0$ and denote the $SU(N_f)_L \times SU(N_f)_R$ currents as:
\begin{equation}
    L = u^\dag du, \quad R = u du^\dag ~.
\end{equation}
The action is then given by:\footnote{For $N_f > 2$ one can also include a 4d Wess-Zumino term for the pions (or for $N_f = 2$ its discrete mod-2 analog~\cite{Witten:1983tx}), which is famously required for anomaly matching in QCD~\cite{Witten:1983tw}. It plays no role in what follows so we omit it. See e.g.~\cite{Davighi:2018inx,Lee:2020ojw} for modern treatments of topological terms in sigma models.  }
\begin{equation}
    S[u,A^\1_L] = \frac{f_\uppi^2}{4}\int \text{Tr}\left((L-iA^\1_L)\wedge \star(L-iA^\1_L)\right)~.
\end{equation}
The non-trivial third homotopy group of the target space $\pi_3(SU(N_f)) = \ZZ$ leads to a $U(1)_B^\0$ `Baryon-number' symmetry with current:
\begin{equation}
    \star J_B^\1 = \frac{1}{24\pi^2}\text{Tr}\left(L\wedge L \wedge L\right) = \frac{1}{2\pi}d\calB^\2_{\rm{WZ}}(u)\,.
\end{equation}
This is just the Skyrmion current. It is not invariant under $SU(N_f)_L$ background gauge-transformations that map:
\begin{equation}
L  \ \to\  L - i u^\dag d\lambda_L^\0 u\,, \quad 
                \star J_B \ \to \ \star J_B - \frac{i}{8\pi^2}d\left(\text{Tr}(d\lambda_L^\0\wedge R)\right)\,. 
\end{equation}
The current transforms by a non-trivial operator under background gauge transformations, but we can add an improvement term to reduce the anomalous transformation to a c-number. The coupling to a background $U(1)_B^\0$ gauge field is:
\begin{equation}
    S \ \supset \ i\int\,A_B^\1\wedge \left(\star J_B + \frac{i}{8\pi^2}d\left(\text{Tr}(A^\1_L\wedge R)\right)\right) \,.
\end{equation}
Under an $SU(N_f)_L$ transformation this term shifts by
\begin{equation}
   \delta S =  -\frac{i}{8\pi^2} \int dA_B^\1 \wedge \text{Tr}(\lambda_L^\0 \wedge dA^\1_L)\,,
\end{equation}
indicating a mixed anomaly between $U(1)_B^\0$ and $SU(N_f)_L$ (the same computation with $SU(N_f)_R$ yields a mixed anomaly with the opposite coefficient so that $SU(N_f)_V$ does not have a mixed anomaly with $U(1)_B^\0$).   

To obtain the nonabelian Goldstone Maxwell model we gauge $U(1)_B^\0$, replacing $A_B^\1 \to a^\1$ and path integrating over $a^\1$ with a standard Maxwell term. If we also introduce a background gauge field $B^\2$ for the resulting magnetic 1-form symmetry $U(1)_m^\1$ of the Maxwell sector, we have:
\begin{equation}
    \begin{split}
        S[u,a^\1,A^\1_L,B^\2] =\,\,& \frac{f_\uppi^2}{4}\int \text{Tr}\left((L-iA^\1_L)\wedge \star(L-iA^\1_L)\right) + \frac{1}{2e^2}\int da^\1\wedge\star da^\1 \\
        &+i\int\,a^\1\wedge \star J_B + \frac{i}{2\pi}\int \,da^\1\wedge\left(B^\2 + \frac{i}{4\pi}\text{Tr}(A^\1_L\wedge R)\right)\,.
    \end{split}
\end{equation}
What was previously a c-number violation of $SU(N_f)_L$ becomes operator-valued and can happily be canceled by imposing the 2-group transformation rule
\begin{equation}
    B^\2 \ \to\  B^\2 + \frac{1}{4\pi}\text{Tr}(\lambda_L^\0 dA^\1_L)\,. 
\end{equation}
In particular, this model has a 2-group symmetry mixing $SU(N_f)_L$ and $U(1)_m^\1$ with level $\kappa=1$. Gauging $U(1)_B^\0$ does not spoil the fact that the $SU(N_f)_L$ symmetry is spontaneously broken --- at low energies there are no $U(1)_B^\0$-charged excitations at all, and we get a nearly decoupled Maxwell sector with a massless photon.\footnote{Ref.~\cite{Davighi:2024zjp} studies the fate of the 2-group when the Maxwell sector is completely Higgsed by a pair of scalar fields. } 

The line defects charged under the magnetic $U(1)_m^\1$ symmetry are again 't Hooft lines. Since the gauge field appears in the action only through its field strength, we can easily define the 't Hooft line by cutting open an electric surface. The electric 1-form symmetry is not conserved due to disorder operators that represent Skyrmions, but the gauge field equation of motion takes the familiar form:
\begin{equation}
    d \left[\frac{1}{e^2}\star f^\2 +\frac{i}{2\pi}\left(\calB^\2_{\text{WZ}} + \frac{i}{4\pi}\text{Tr}\left(A_L^\1\wedge R\right) +B^\2\right)\right]=0 ~.
\end{equation}
Though conserved, this current is not well-defined due to the appearance of the trivialization $\calB^\2_{\text{WZ}}$ which is itself not globally well-defined. We can try to define the 't Hooft line naively as:
\begin{equation}
    H_{q,\gamma} \stackrel{?}{=} \exp \left[\frac{2\pi q}{e^2}\int_\Sigma \star f^\2 + iq\int_\Sigma \calB_{\text{WZ}}^\2 + \frac{i}{4\pi}\text{Tr}\left( A_L^\1 \wedge R \right) + B^\2\right]\,,
\end{equation} 
where $q \in \ZZ$ and $\partial\Sigma = \gamma$. But this is similarly ambiguous due to the integral of $\mathcal{B}^\2_{\text{WZ}}$ on an open surface. There are different paths to render it well-defined:
\begin{itemize}
    \item We restrict our path integral to small fluctuations $\uppi^{\,a}$ around the identity, such that we can always pick a local trivialization on that target space patch and $\mathcal{B}_{\text{WZ}}^\2$ is well-defined on the restricted field space. In this case, we have an electric 1-form symmetry, because the disorder operators representing Skyrmions are disallowed.

    \item Even if we restrict to low momenta with $|u^\dag\partial u|\ll f_\uppi$, we need to be more cautious. The field can explore the entire target space and it is necessary to introduce additional line degrees of freedom or impose a boundary condition on the group-valued field~\cite{Elitzur:2001qd, Gawedzki:2002se}. These will generically contribute additional pieces to the tilt operator.
\end{itemize}
For our purposes, it suffices to understand the 't Hooft lines locally around a point of the moduli space (which we take to be the identity). The $SU(N_f)_L$ violation occurs due to the fact that the 't Hooft line is defined with a 2d Wess-Zumino-Witten term on a surface with boundary. In this example, the computation of the tilt operator follows the same techniques we used to compute the family anomaly (see App.~\ref{app:Bvariations}). Namely, we have the relation
\begin{equation}
    \nu^\1(\lambda,u,A) = \lambda^a\, \tau^a(u,A)\,,
\end{equation}
where we replaced the source $U$ with the pion field $u$. From Eq.~\eqref{eq:nu1_full} we can read off the tilt-operator as:
\begin{equation}
    \tau^a = -\frac{1}{2\pi}\left(\frac{1}{X(\uppi)}+ \frac{1}{\id - e^{X(\uppi)}}\right)^{ab}d\uppi^{\,b} - \frac{1}{4\pi}A^a_L~,
\end{equation}
which can be expanded in leading order as:
\begin{equation}
    \tau^a = -\frac{1}{4\pi}(d\uppi^{\,a} +A_L^a) + \frac{1}{24\pi}f^{abc}\, \uppi^{\,b} d\uppi^{\,c} + \mathcal{O}(\uppi^4)\,.
\end{equation}

As mentioned in Sec.~\ref{sec:family_anomaly}, the linear coupling of the source to the tilt operator is not enough to reproduce the family anomaly. Fortunately, we can make progress by studying how the tilt operator itself transforms under $SU(N_f)_L$. The linear coupling
\begin{equation}
H_{q,\gamma}\, \exp\left( i q \int \omega^a\, \tau^a(u,A) \right)
\end{equation}
is not enough to ensure $SU(N_f)_L$-covariance (even to $\calO(\lambda,\omega,\uppi)$), since $\tau$ itself transforms by an operator. To linear order in $\uppi$ we can fix this by including an $\calO(\omega^2)$ term coupled to a higher tilt\footnote{Alternatively we can use the transgression formula to compute the finite transformation of the 2d WZ term as in App.~\ref{app:finite_transgression}.}
\begin{multline} \label{eq:higher_tilt}
    H_{q,\gamma}[e^{i\omega(t)}] = H_{q,\gamma}\, \exp\Bigg[ - \frac{i q}{4\pi} \int \omega^a\, \left(d\uppi^{\,a}+A^{\1 a}_L - \frac{1}{6}f^{abc}\, \uppi^{\, b}d\uppi^{\, c} + \cdots \right) \\
    + \frac{iq}{24\pi} \int f^{abc}\,  \omega^a d\omega^b \, \uppi^c + \cdots  \Bigg] \,. 
\end{multline}
This combination is invariant under $SU(N_f)_L$ transformations up to $\calO(\lambda,\omega,\uppi)$, but there is a c-number violation at $\calO(\lambda,\omega^2)$ which comes precisely from the higher tilt and matches the scheme-independent $f^{abc}\lambda^a\omega^b d\omega^c$ term in the family anomaly displayed in Eq.~\eqref{eq:nu1_expanded}.

\subsubsection{Nonlinear Sigma Models on Lines: A Universal Tilt Operator}
\label{sec:nlsm_on_line}

Consider a generic line operator charged under the 1-form subgroup of a continuous nonabelian 2-group, with tilt operator $\tau$. The non-invariance of the line under $G^\0$ transformations is absorbed into the spurious transformation of the source $U$. One might expect to be able to `symmetrize' the line defect by promoting the spurion to a dynamical field which is integrated over in the path integral,\footnote{Such couplings of symmetry-breaking defects and boundaries to nonlinear sigma models have been studied in~\cite{Metlitski:2020cqy,Cuomo:2023qvp}.} 
\begin{equation}
    \widetilde{\scrL}_\gamma \, = \,  \int \mathscr{D}u\, \scrL_\gamma[u]\, e^{-S_\gamma[u]}\,,
\end{equation}
with some worldline action $S_\gamma$. This, however, is too naive, and the resulting line defect still violates the $G^\0$ symmetry. When $U$ is a classical background field, the family anomaly is a c-number violation of the spurious symmetry. But once $U \to u$ is made dynamical, the family anomaly turns into an \emph{operator-valued} violation of the symmetry, 
\begin{equation}
 \int \mathscr{D}u\, \scrL_\gamma[u]\, e^{-S_\gamma[u]} \goesto \int \mathscr{D}u\, \scrL_\gamma[u]\, e^{-i\int_\gamma \nu^\1(\lambda,u)}\, e^{-S_\gamma[u]} \,.
\end{equation}
In other words, by coupling the original line defect to a $G^\0$ NLSM (or particle on the group manifold), the original family anomaly turns into a new tilt operator. This universal tilt has exactly the same form as the tilt in the nonabelian Goldstone-Maxwell model, 
\begin{equation}
    \tau^a = -\frac{1}{2\pi}\left(\frac{1}{X(\uppi)}+ \frac{1}{\id - e^{X(\uppi)}}\right)^{ab}d\uppi^{\,b} \ =\   -\frac{1}{4\pi}d\uppi^{\,a}  + \frac{1}{24\pi}f^{abc}\, \uppi^{\,b} d\uppi^{\,c} + \mathcal{O}(\uppi^4)\,,
\end{equation}
where now the `pions' $u = e^{i\uppi}$ just live on the worldline $\gamma$. The family anomaly is matched by the transformation of the higher tilt as discussed around Eq.~\eqref{eq:higher_tilt}.

\section{Continuous Abelian 2-Groups}
\label{sec:abelian} 

In the previous section we argued for symmetry breaking on line defects by analyzing the Wess-Zumino consistency condition for infinitesimal background gauge transformations. This analysis does not lead to any constraints when the 2-group (and in particular the 0-form symmetry $G^\0$) is abelian. In particular, the 2d abelian anomaly $\alpha^\2(\lambda,A) = \frac{\kappa}{2\pi}\lambda^\0 dA^\1$ appears to be Wess-Zumino consistent~\cite{Jensen:2017eof,Thorngren:2020yht}. 

The obstruction is instead global. A compact $U(1)$ symmetry allows for large gauge transformations, and in an abelian 2-group these large transformations do not commute with small gauge transformations. To see this effect we formulate the abelian 2-group background fields using differential cohomology. This allows us to keep track not only of the local 1-form and 2-form gauge fields but also their transition functions and integer cocycles. We formulate the obstruction to a symmetric line defect as a Wess-Zumino consistency condition for the large and small gauge transformations. 

The main example we consider is the abelian Goldstone-Maxwell theory. The theory is free, and we explicitly compute the tilt operator and show how it reproduces the family anomaly and restores Wess-Zumino consistency. We also couple the line to a dynamical rotor degree of freedom, recompute the tilt operator, and find that the resulting defect gives a phase shift for s-wave scattering determined by the Postnikov class. We finish the section by embedding the model in a larger theory with 2-group symmetry and exploring how the tilt operator appears in other IR phases.

\subsection{Differential Cohomology for Abelian 2-Group Bundles}

Differential cohomology is a streamlined but concrete way to capture the global structure of fields (see e.g.~\cite{Alvarez:1984es, Bauer_2005, Freed:2006ya, Freed:2006yc, MooreDiffCoh, Kapustin:2014gua, Cordova:2019jnf,Davighi:2020vcm} for a physics-oriented exposition of the formalism). The formalism works particularly well for (generalized) abelian gauge fields, and can be used define topological terms in a way that makes their invariance under small gauge transformations (which act on local gauge fields) as well as large ones (which act on their transition functions and other higher data). For our purposes, we will use differential cohomology to carefully analyze the global consistency of 2-group background fields, and the Wilson surfaces built from them. 

\begin{figure}[h!]
    \centering
\includegraphics[width=0.75\linewidth]{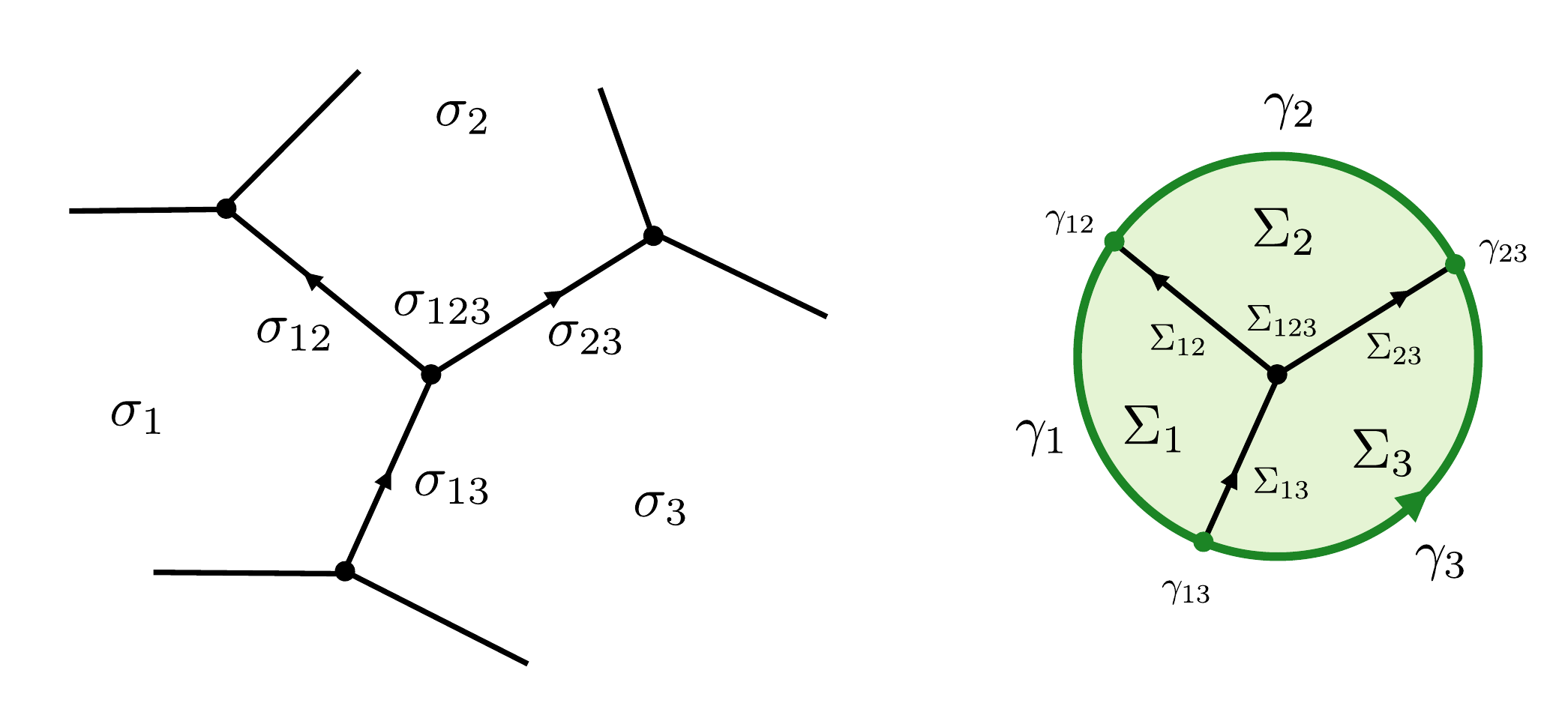}
        \caption{An example of a triple intersection of patches. The orientation of each region is such that e.g. $\partial \sigma_1 \supset \sigma_{12}$ and $\partial\sigma_{12} \supset - \sigma_{123}$. On the right, we show the associated partitioning of the Wilson surface $\Sigma$ with boundary $\gamma$. \label{fig:patches}}
\end{figure}

We cover our spacetime with open patches $\calU_i$ labeled by an ordered list $i_1< i_2 < \cdots$ and choose an associated partition of spacetime into closed, oriented regions $\sigma_i$ where $\sigma_i \subset \calU_i$. The overlap regions $\calU_{i_1} \cap \calU_{i_2}\cap \cdots \cap \calU_{i_k}$ contain the common boundaries of the partition $\sigma_{i_1} \cap \sigma_{i_2} \cap \cdots \cap \sigma_{i_k}$, which are codimension-$(k-1)$, which we take to be connected (see Fig.~\ref{fig:patches}). We define $\sigma_{i_1 \cdots i_k}$ to be the oriented common boundary such that $\partial\sigma_{i_1 \cdots i_{k-1}} =(-1)^k \cup_{i_k} \,  \sigma_{i_1 \cdots i_{k-1} i_k}$. These are totally antisymmetric in their indices. The 2-group background field is described by a set of differential forms and integers which live on patches and their overlaps. A given piece of the background field data carries some number of patch indices. The differential acting on these indices is $(\delta X)_{i_1 \cdots i_{k+1}} = \sum_j (-1)^{k+j} X_{i_1 \cdots \hat{i}_{j} \cdots i_{k+1}}$ where $\hat{i}$ means the index is omitted.\footnote{For our purposes we will just need 
\begin{equation}
    (\delta X)_{ij} = X_j - X_i ~, \quad (\delta Y)_{ijk} = Y_{ij} - Y_{ik} + Y_{jk} ~, \quad (\delta Z)_{ijkl} = - Z_{ijk} + Z_{ijl} - Z_{ikl} + Z_{jkl} ~.
\end{equation}
}

Locally, the 2-group background gauge fields consist of a $\RR$-valued 1-form $A^\1$ and 2-form $B^\2$ in each patch. The gauge fields in neighboring patches are related on double overlaps by 
\begin{equation} \label{eq:connection} 
\begin{split}
(\delta A^\1)_{ij} &= d\Phi^\0_{ij} \,, \quad\quad (\delta B^\2)_{ij} = d\Omega^\1_{ij} + \frac{\kappa}{2\pi} \Phi^\0_{ij} \, dA^\1_j \,. 
\end{split}
\end{equation}
Here $\Phi^\0_{ij}, \Omega^\1_{ij}$ are $\RR$-valued 0- and 1-form transition functions respectively. The appearance of $U(1)^\0$ gauge field data in the connection formula for the gauge field for $U(1)^\1$ is the signature of the 2-group bundle. Taking $\delta$ of the connection formulas yields the cocycle conditions on triple overlaps, 
\begin{equation}\label{eq:firstcocycle} 
\begin{split}
(\delta\Phi^\0)_{ijk} &= 2\pi\, C_{ijk}\,, \quad\quad (\delta\Omega^\1)_{ijk} =-\kappa \, C_{ijk}\, A^\1_{k} +  d\Psi^\0_{ijk}\,, 
\end{split}
\end{equation}
where we introduced integers $C_{ijk}$ and an $\mathbb{R}$-valued scalar $\Psi^\0_{ijk}$ on the triple overlaps. The integers $C_{ijk}$ capture the first Chern number of the $U(1)$ gauge field, i.e. its quantized flux through closed 2-cycles. Repeatedly taking $\delta$ gives further cocycle conditions on higher overlaps, 
\begin{equation}\label{eq:secondcocycle}
(\delta C)_{ijkl} = 0 \,, \quad \quad (\delta\Psi^\0)_{ijkl} = \kappa \, C_{ijk}\, \Phi^\0_{kl} + 2\pi Y_{ijkl}\,, \quad\quad (\delta Y)_{ijklm} = - \kappa\, C_{ijk}\, C_{klm}\,, 
\end{equation}
where we introduced an integer $Y_{ijkl}$ on the quadruple overlaps which characterizes the quantized flux of the 2-form connection through closed 3-cycles. The last condition is the statement that the second Chern class of the $U(1)^\0$ connection gets trivialized by the flux of the $U(1)^\1$ connection (compare to the discussion near Eq.~(7.6) in \cite{Cordova:2018cvg}). All in all, the 2-group data is captured by $(A^\1_i, B^\2_i;\, \Phi^\0_{ij}, \Omega^\1_{ij};\, C_{ijk}, \Psi^\0_{ijk};\, Y_{ijkl})$ and the associated cocycle conditions.\footnote{We can describe networks of topological defects by setting $A^\1_i = 0$ and $B^\2_i = 0$ on each patch, $\Omega^\1_{ij} =0$ on the double overlaps, and set $\Phi^\0_{ij}$ and $\Psi^\0_{ijk}$ to be constants. Then we just have $\delta\Phi^\0 = 2\pi C$ and the relations in Eq.~\eqref{eq:secondcocycle}.}

Background gauge transformations for $U(1)^\1$ are split into small gauge transformations parameterized by an $\RR$-valued 1-form $\Lambda^\1_i$ on each patch, and large gauge transformations parameterized by an $\RR$-valued 0-form $\Pi^\0_{ij}$ on each double overlap and integers $Z_{ijk}$ on triple overlaps, 
\begin{equation} \label{eq:one_form_small_gauge_transform} 
\begin{split}
B^\2_i & \goesto B^\2_i + d\Lambda^\1_i\,, \\ 
 \Omega^\1_{ij} &\goesto \Omega^\1_{ij} + (\delta\Lambda^\1)_{ij} - d\Pi^\0_{ij} \,, \\
\Psi^\0_{ijk} &\goesto \Psi^\0_{ijk} -(\delta \Pi^\0)_{ijk} + 2\pi Z_{ijk} \,, \\
Y_{ijkl} & \goesto Y_{ijkl} + (\delta Z)_{ijkl} \,. 
\end{split}
\end{equation}
One can check that on a closed surface $S$, the Wilson surface (here we use the shorthand $S_{i_1\cdots i_k} \equiv S \cap \sigma_{i_1 \cdots i_k}$)
\begin{equation}
\exp\left( i q \int_S B^\2 \right) =\exp\left(i q \sum_i \int_{S_i} B^\2_i - iq \sum_{i<j} \int_{S_{ij}} \Omega^\1_{ij} - i q\sum_{i < j < k} \left. \Psi^\0_{ijk} \right|_{S_{ijk}}  \right)
\end{equation}
is invariant under the above background gauge transformations provided $q \in \ZZ$.\footnote{It is useful to use the following identities
\begin{equation}
    \sum_i \int_{\sigma_i } dX^\1_i = \sum_{i<j} \int_{\sigma_{ij}} (\delta X^\1)_{ij}\,, \quad \sum_{i<j} \int_{\sigma_{ij} } dX^\0_{ij} = - \sum_{i<j<k} \left. (\delta X^\0)_{ijk}\right|_{\sigma_{ijk}} 
\end{equation}
for 1-forms $X^\1_i$ living on $\calU_i$ and 0-forms $X^\0_{ij}$ living on double-overlaps $\calU_{ij}$. } 

Now we turn to $U(1)^\0$ gauge transformations. Small gauge transformations are parameterized by $\RR$-valued $\lambda^\0_i$ in each patch, and act as  
\begin{equation} \label{eq:small_gauge_transform} 
\begin{split}
A^\1_i & \goesto A^\1_i + d\lambda^\0_i \,, \\ 
\Phi^\0_{ij} &\goesto \Phi^\0_{ij} + (\delta\lambda^\0)_{ij}\,, \\
B^\2_i & \goesto B^\2_i +\frac{\kappa}{2\pi} \lambda^\0_i \, dA^\1_i \,, \\
\Psi^\0_{ijk} & \goesto \Psi^\0_{ijk} + \kappa\, C_{ijk}\, \lambda^\0_k\,.
\end{split}
\end{equation}
Large background gauge transformations for the $U(1)^\0$ are parameterized by integers $K_{ij}$ on double overlaps, and act on the transition functions and higher data as
\begin{equation} \label{eq:large_gauge_transform} 
\begin{split}
\Phi^\0_{ij} & \goesto \Phi^\0_{ij} + 2\pi K_{ij}\,, \\
C_{ijk} &\goesto C_{ijk} + (\delta K)_{ijk} \, , \\
\Omega^\1_{ij} & \goesto \Omega^\1_{ij}  -\kappa\, K_{ij}\, A^\1_j \,, \\
\Psi^\0_{ijk} &\goesto \Psi^\0_{ijk}  + \kappa \, K_{ij}\, \Phi^\0_{jk} \,, \\
Y_{ijkl} & \goesto Y_{ijkl} -\kappa\, \left((C+\delta K)_{ijk}\, K_{kl} + K_{ij}\, C_{jkl}\right)  \,. 
\end{split}
\end{equation}

A crucial feature of the gauge transformation rules is that the small and large gauge transformations parameterized by $\lambda^\0_i$ and $K_{ij}$ only commute \emph{up to} a large gauge transformation for the 1-form symmetry. This is the abelian analog of the discussion around Eq.~\eqref{eq:deltaWZB}. In particular, 
\begin{equation}
   ( \delta_\lambda\, \delta_K - \delta_K\, \delta_\lambda )\, B^\2 \, = \, \delta_{\Pi_{ij} = \kappa\, K_{ij}\, \lambda_j}\, B^\2\,.
\end{equation}
The 2-group-modified WZ consistency condition for functionals of the background gauge fields is
\begin{equation} \label{eq:WZModified_abelian}
       ( \delta_\lambda\, \delta_K - \delta_K\, \delta_\lambda - \delta_{\Pi_{ij} = \kappa\, K_{ij}\, \lambda_j} )\,\mathcal{F}[A^\1,B^\2] \, = \, 0\,. 
\end{equation}
Just as we did in the nonabelian case, we can quickly derive an inconsistency for $U(1)^\0$-symmetric lines charged under $U(1)^\1$. Consider the partition function $Z_\scrL[A^\1,B^\2]$ in the presence of a charge-1 line $\scrL$, which for the sake of contradiction we assume is symmetric under $U(1)^\0$. This object satisfies (as always ignoring possible bulk anomalies which do not affect the argument)
\begin{equation}
\begin{split}
       ( \delta_\lambda\, \delta_K - \delta_K\, \delta_\lambda - \delta_{\Pi_{ij} = \kappa\, K_{ij}\, \lambda_j} )\,Z_\scrL[A^\1,B^\2] &= - \delta_{\Pi_{ij} = \kappa\, K_{ij}\, \lambda_j} \,Z_\scrL[A^\1,B^\2] \\
       &= \left(1-e^{i\kappa \sum_{i<j} \left. K_{ij}\, \lambda_j \right|_{\gamma_{ij}}} \right) \,Z_\scrL[A^\1,B^\2] \not = 0\,. 
\end{split}
\end{equation}
Just as in the nonabelian case, the WZ consistency condition is violated by any $U(1)^\0$-symmetric, $U(1)^\1$-charged line operator.

To proceed, we allow the line to break $U(1)^\0$ and couple it to a spurion $\theta$ on $\gamma$, which is just a compact scalar background field which shifts under the $U(1)^\0$ action. In the patch formalism such a background field is described by a collection of real fields $\theta^\0_i$ satisfying 
\begin{equation}
(\delta \theta^\0)_{ij} = 2\pi \, W_{ij} + \Phi^\0_{ij}\,, \quad  (\delta W)_{ijk} = - C_{ijk}\,,
\end{equation}
with the integers $W_{ij}$ on double overlaps capturing the winding of the compact scalar $\theta$. Compactness is reflected in a (background) gauge redundancy whereby we shift the spurion by an independent multiple of $2\pi$ in each patch, 
\begin{equation} \label{eq:theta_2pi}
\theta^\0_i \to \theta^\0_i + 2\pi L_i \,, \quad W_{ij} \to W_{ij} + (\delta L)_{ij}\,. 
\end{equation}
Background $U(1)^\0$ gauge transformations act as shifts of the spurion,  
\begin{equation} \label{eq:theta_transformation} 
\theta^\0_i \to \theta^\0_i + \lambda^\0_i \,, \quad W_{ij} \to W_{ij} - K_{ij}\,. 
\end{equation}

Now we can repeat the procedure from Sec.~\ref{sec:nonabelian}. Namely, we introduce the Wilson surface on $\Sigma$ to cancel the 1-form transformation of the line, and extend the background gauge fields and the spurion into an auxiliary bulk $\calM_3$ with boundary $\partial\calM_3 = \Sigma \cup \calM_2$ as in Fig.~\ref{fig:inflow}. We place an abelian analog of the higher Berry phase on $\calM_2$. The combination of the inflow action on $\calM_3$, the Wilson surface on $\Sigma$, and the higher Berry connection on $\calM_2$ have a consistent anomalous transformation localized on $\Sigma$ and $\gamma$ (the latter being the family anomaly). 

To extract the family anomaly, it suffices to compute the gauge variation of the higher Berry phase modulo the 2d anomaly coming from the Wilson surface. The anomalous transformation of the Wilson surface can be computed using Eqs.~\eqref{eq:small_gauge_transform}, \eqref{eq:large_gauge_transform}\,, 
\begin{equation} \label{eq:B_anomaly} 
\begin{split}
\int_\Sigma \alpha^\2_{\rm{small}}(\lambda,A) \equiv \delta_\lambda \int_\Sigma B^\2  &= \frac{\kappa}{2\pi} \sum_i \int_{\Sigma_i} \lambda_i^\0 dA^\1_i + \kappa \sum_{i<j<k} \left. C_{ijk}\, \lambda_k^\0\right|_{\Sigma_{ijk}} \, ,\\
\int_\Sigma \alpha^\2_{\rm{large}}(K,A) \equiv \delta_K \int_\Sigma B^\2 &= -\kappa \sum_{i<j} \int_{\Sigma_{ij}} K_{ij}  \, A^\1_j  + \kappa \sum_{i < j < k} \left. K_{ij}\, \Phi^\0_{jk} \right|_{\Sigma_{ijk}} \,.
\end{split}
\end{equation}
The analog of the higher Berry phase in the abelian case can be written as~\cite{Cordova:2019jnf}
\begin{equation}
\frac{\kappa}{2\pi} \int_{\Sigma} \theta\, dA = \frac{\kappa}{2\pi} \sum_i \int_{\Sigma_i} \theta_i^\0 \, dA^\1_i + \kappa \sum_{i<j} \int_{\Sigma_{ij}} W_{ij}\, A^\1_j - \kappa \sum_{i<j<k} \left. W_{ij}\, \Phi^\0_{jk}\right|_{\Sigma_{ijk}}\,.
\end{equation}
Under small and large gauge transformations it transforms as (taking $\calM_2$ to be $\Sigma$ itself) 
\begin{equation}
\begin{split}
\delta_\lambda \, \frac{\kappa}{2\pi} \int_{\Sigma} \theta\, dA &= \frac{\kappa}{2\pi} \sum_i \int_{\Sigma_i} \lambda_i^\0 dA^\1_i + \kappa \sum_{i<j} W_{ij} \int_{\Sigma_{ij}} d\lambda^\0_j  - \kappa \sum_{i < j < k} \left. W_{ij}\, (\delta\lambda^\0)_{jk} \right|_{\Sigma_{ijk}} \,, \\
\delta_K \, \frac{\kappa}{2\pi} \int_{\calM_2} \theta\, dA &=  - \kappa \sum_{i<j} \int_{\Sigma_{ij}}  K_{ij}\, A^\1_j  + \kappa \sum_{i < j < k} \left. K_{ij}\, \Phi^\0_{jk} \right|_{\Sigma_{ijk}}  \text{ mod } 2\pi \,. 
\end{split}
\end{equation}
Comparing to Eq.~\eqref{eq:B_anomaly}, we see that $e^{i \int_\Sigma (B^\2 - \frac{\kappa}{2\pi}\theta\, dA)}$ is invariant under large gauge transformations, so that $\nu^\1_{\rm{large}}(K,\theta,A) = 0$, but not invariant under small gauge transformations, 
\begin{multline} \label{eq:abelian_fam_anomaly} 
\delta_\lambda\left( \int_\Sigma B^\2 - \frac{\kappa}{2\pi} \int_{\Sigma} \theta\, dA \right)  = - \int_\gamma \nu^\1_{\rm{small}}(\lambda,\theta,A) \\
=  \kappa \sum_{i<j<k} \left. C_{ijk}\, \lambda_k^\0\right|_{\Sigma_{ijk}} - \kappa \sum_{i<j} W_{ij} \int_{\Sigma_{ij}} d\lambda^\0_j + \kappa \sum_{i < j < k} \left. W_{ij}\, (\delta\lambda^\0)_{jk} \right|_{\Sigma_{ijk}}  \\
= \kappa \sum_{i<j} W_{ij} \left. \lambda^\0_j \right|_{\gamma_{ij}} \, ,
\end{multline}
where we used the fact that $C_{ijk} = -(\delta W)_{ijk}$. Finally, we can use the above equation to read off the Wess-Zumino consistent anomalous transformation for the line defect coupled to the Wilson surface,
\begin{multline} \label{eq:correct_abelian_anomaly}
Z_{\scrL,\Sigma}[A^\1+d\lambda^\0, B^\2+ \tfrac{\kappa}{2\pi}\lambda^\0\, dA^\1, \theta^\0 + \lambda^\0]\\
= Z_{\scrL,\Sigma}[A^\1,B^\2,\theta^\0]\,e^{-i\frac{\kappa}{2\pi}\sum_i \int_{\Sigma_i} \lambda^\0_i dA^\1_i }\,  e^{i \kappa \sum_{i<j} W_{ij} \left. \lambda^\0_j \right|_{\gamma_{ij}} } \,. 
\end{multline}
In particular, since small and large gauge transformations commute in their action on the $\theta, A$ background fields, no local counterterm built from them can reproduce the family anomaly $\sim W_{ij}\, \lambda_j$ which is not Wess-Zumino consistent on its own. We conclude that the line operator $\scrL_\gamma$ must depend on the spurion through a coupling to a non-trivial operator on the line, which is nothing but the tilt:
\begin{equation}
    \scrL_\gamma[\theta] = \scrL_\gamma[0]\, \exp\left( -i\int dt \, \theta(t) \, \tau(t) \right) \,.
\end{equation}

In the given scheme, the family anomaly says that if $\theta(t)$ winds by $2\pi$ along the time direction (characterized by some $W_{ij} = 1$), applying a global symmetry transformation yields the defect at $\theta(t)+\lambda$ times a phase $e^{i\kappa \lambda}$. It is tempting to say that winding configurations of $\theta$ carry charge, but this is misleading because the charge is violated on the line. Relatedly, we can compute what happens to the line operator insertion when we perform the $2\pi$ shift of $\theta$ in Eq.~\eqref{eq:theta_2pi}. Such a shift is equivalent to performing a set of 0-form and 1-form symmetry transformations with
\begin{equation}
\begin{split}
\lambda^\0_i &= 2\pi L_i\,, \ K_{ij} = -(\delta L)_{ij} \,, \\
\Lambda^\1_i &= -\kappa \, L_i \, A_i\,, \ \Pi^\0_{ij} = - \kappa\,  L_i\,  \Phi^\0_{ij} \,, \ Z_{ijk} = -\kappa\,  C_{ijk} (L_i+ L_k)\,.
\end{split}
\end{equation}
Note that $A^\1_i, \Phi^\0_{ij}, C_{ijk}$ are unaffected. In our current choice of scheme, this sequence of transformations does not activate the family anomaly (since $\kappa \in \ZZ$).
If we consider just the line operator inserted with $B^\2$ turned off and the Wilson surface absent, we just pick up the 1-form symmetry transformation:
\begin{equation}
Z_\scrL[A^\1,\theta^\0_i+2\pi L_i ,W_{ij} \to W_{ij} + (\delta L)_{ij} ] = Z_\scrL[A^\1,\theta^\0]\, e^{-i \kappa\left(  \sum_i \int_{\gamma_i } L_i\, A_i^\1 +  \sum_{i<j} \left. L_i \Phi^\0_{ij} \right|_{\gamma_{ij}} \right) } \,. 
\end{equation}
In particular if we set $L_i = L$ to be the same constant in each patch we get the simple relation
\begin{equation} \label{eq:notpump} 
Z_\scrL[A^\1,\theta^\0+2\pi L ] = Z_\scrL[A^\1,\theta^\0]\, e^{-i \kappa L\int_\gamma A^\1 } \,. 
\end{equation}
This looks very similar to a `charge pump,' whereby sending $\theta^\0 \to \theta^\0+2\pi$ deposits $\kappa$ units of charge on the line defect. This interpretation is however \underline{not} available here, as the line is not invariant under the $U(1)^\0$ symmetry so the charge has no meaning. In particular, unlike the examples discussed in~\cite{Debray:2023ior,Hsin:2020cgg,Komargodski:2025jbu}, nothing stops us from adding a $U(1)^\0$-violating counterterm $\sim \int_\gamma \theta^\0 A^\1$ which removes the pump while further complicating the family anomaly in Eq.~\eqref{eq:correct_abelian_anomaly}. 

\subsection{Example: Abelian Goldstone-Maxwell Model}
\label{sec:abelian_GMM}

In \cite{Cordova:2018cvg}, the authors introduced the abelian Goldstone-Maxwell model, a universal low-energy effective theory describing the spontaneous breaking of a continuous abelian 2-group in 4d. To be self-contained, let us describe a novel path to this model starting from a compact scalar $\chi \sim \chi + 2\pi$, 
\begin{equation}
S_\chi[A^\1,\Theta^\3] = v^2 \int\,\left(d\chi - A^\1\right)\wedge \star\left(d\chi - A^\1\right) + \frac{i}{2\pi}\int d\chi \wedge \Theta^\3\,.
\end{equation} 
Here we coupled the action to background gauge fields $A^\1$ for the $U(1)^\0$ shift symmetry and $\Theta^\3$ for the $U(1)^\2$ winding symmetry of the compact scalar. To generate a theory with a continuous abelian 2-group, we seek a QFT with a mixed 't Hooft anomaly between two 0-form symmetries --- then we will gauge one of them. Naively, the free compact scalar in 4d just has a mixed anomaly between the two aforementioned symmetries, with inflow action
\begin{equation}
\calI^{(5)}(A^\1,\Theta^\3) = \frac{1}{2\pi} A^\1 \wedge d\Theta^\3\,. 
\end{equation}
But we can enrich the theory with a trivially-acting 0-form symmetry $U(1)_C^\0$ (with background gauge field $C^\1$) by coupling the winding symmetry current to a prescribed background field
\begin{equation}
\Theta^\3 = \frac{\kappa}{2\pi} A^\1 \wedge dC^\1\,, \quad \kappa \in \ZZ\,. 
\end{equation}
This induces the anomaly 
\begin{equation}
\calI^{(5)}(A^\1,C^\1) = \frac{\kappa}{(2\pi)^2} \, A^\1 \wedge dA^\1 \wedge dC^\1 \,. 
\end{equation}
In other words, we can use the winding symmetry to choose a fractionalization class of a trivially-acting 0-form symmetry and induce a mixed anomaly. This mechanism for inducing anomalies via symmetry fractionalization is common and has been discussed in e.g.~\cite{Hsin:2019fhf,Hsin:2019gvb,Brennan:2022tyl,Delmastro:2022pfo,Brennan:2025acl,Seiberg:2025bqy,Lu:2026jnq}. 

Now we gauge the trivially-acting $U(1)_C^\0$ symmetry by replacing $C^\1 \to a^\1$ and path-integrating over $a^\1$. Since $U(1)_C^\0$ was trivially-acting, the result of gauging is Maxwell theory. Denoting its magnetic 1-form symmetry background gauge field by $B^\2$, we arrive at\footnote{The model actually has a 3-group symmetry involving the \emph{electric} 1-form symmetry, $U(1)^\0$, and the 2-form winding symmetry~\cite{Cordova:2018cvg}. It plays no role in our analysis of the 2-group (for instance, we can explicitly break the electric 1-form symmetry by coupling $a^\1$ to charged matter fields).}
\begin{multline} \label{eq:abelian_GM}
S[A^\1, B^\2] = v^2 \int\,\left(d\chi - A^\1\right)\wedge \star\left(d\chi - A^\1\right) + \frac{1}{2e^2}\int da^\1 \wedge \star da^\1 \\
- \frac{i\kappa}{(2\pi)^2}\int \chi \wedge dA^\1 \wedge da^\1 + \frac{i}{2\pi}\int da^\1 \wedge B^\2\,. 
\end{multline}
The 2-group symmetry shows up in the modified transformation rule of the 2-form background gauge field under $U(1)^\0$ transformations which shift $\chi\to\chi+\lambda^\0$,  
\begin{equation}
A^\1 \to A^\1 +d\lambda^\0\,, \quad B^\2 \to B^\2 + \frac{\kappa}{2\pi}\lambda^\0\, dA^\1\,. 
\end{equation}
The associated current is
\begin{equation} \label{eq:almost_improved_current} 
j^\1 = 2iv^2\left(d\chi - A^\1\right)-\frac{\kappa}{(2\pi)^2}\star\left(d\chi \wedge da^\1 \right)\,. 
\end{equation}
Note that this is \emph{almost} an improvement of the ordinary shift symmetry current $j^\1 = 2iv^2 d\chi$, but neither $\chi$ nor $a$ are globally well-defined. 

Now we insert a 't Hooft line with charge $q$, which sets boundary conditions for the dynamical gauge field such that $d^2a = 2\pi q\, \delta^\3(\gamma)$. We can quickly get a sense for the symmetry breaking by looking at the classical conservation equation for the current --- setting $A^\1 = 0$ and using the equations of motion we find 
\begin{equation}
d\star j^\1 = 0 + \frac{\kappa}{(2\pi)^2}\,  d\chi \wedge d^2a^\1  = \frac{\kappa q}{2\pi}\, d\chi \wedge \delta^\3(\gamma)\,.
\end{equation}
In the presence of the monopole defect the current is no longer conserved, and we find a tilt operator $\tau = - \frac{\kappa q}{2\pi} d\chi$.\footnote{In accordance with our observations around Eq.~\eqref{eq:almost_improved_current}, if $\chi$ is non-compact the tilt operator can be absorbed into a new current which is conserved in the presence of the 't Hooft line.}  To go beyond the classical level, we need a proper definition of the 't Hooft line in the presence of background fields,\footnote{This expression differs from the definition given in~\cite{Cordova:2018cvg} by a crucial boundary term $\sim \int_\gamma \chi \, A^\1$ which ensures that it is well-defined under $2\pi$ shifts of $\chi$. } 
\begin{equation} \label{eq:abelian_H} 
    H_{q,\gamma} = \exp\left[\frac{2\pi q}{e^2}\int_{\Sigma }\star da^\1 +iq \int_{\Sigma}\left(B^\2 + \frac{\kappa}{2\pi}\, d\chi \wedge A^\1 \right)\right]
\end{equation}
where $\partial\Sigma = \gamma$. The gauge field equations of motion ensure that this definition is invariant under the choice of bounding surface $\Sigma$. It however fails to be $U(1)^\0$ invariant, transforming as
\begin{equation}
H_{q,\gamma} \goesto H_{q,\gamma}\, \exp\left(-\frac{iq\kappa}{2\pi} \int_\gamma \lambda^\0 (d\chi-A^\1) \right) ~.
\end{equation}
We can identify the tilt-operator to be:
\begin{equation} \label{eq:abelian_tilt} 
    \tau = - \frac{q\kappa}{2\pi}\left(d\chi-A^\1\right)\,,
\end{equation}
in accordance with our classical analysis above. Since the symmetry group is abelian, it is simple to write down the line at any point on the $S^1$ moduli space:
\begin{equation} \label{eq:Htheta}
    H_{q,\gamma}[\theta] = H_{q,\gamma}\, \exp\left(\frac{iq\kappa}{2\pi} \int_\gamma \theta\,  (d\chi-A^\1)\right) ~.
\end{equation}

\subsubsection{Matching the Family Anomaly}

It is not immediately obvious how the tilt in Eq.~\eqref{eq:abelian_tilt} is capable of reproducing the family anomaly in Eq.~\eqref{eq:abelian_fam_anomaly}. To see this explicitly it is easier to work in the dual magnetic frame. Since the action \eqref{eq:abelian_GM} only depends on the gauge field through its field strength $da$, we can apply the standard abelian duality procedure to obtain a dual action in terms of the magnetic gauge field $b^{(1)}$, 
\begin{multline}
S_{\text{dual}}[A^\1,B^\2] = v^2 \int \left(d\chi - A^\1 \right)\wedge\star \left(d\chi - A^\1\right)\\
+\frac{e^2}{2(2\pi)^2}\int \left(db^\1  -\frac{\kappa}{2\pi}\,  d\chi \wedge A^\1-B^\2\right) \wedge\star \left(db^\1 -\frac{\kappa}{2\pi}\,  d\chi \wedge A^\1-B^\2\right)\,.
\end{multline}
The compact scalar and the magnetic gauge field now couple through the $A^\1$-dependence in the magnetic Maxwell term. In order for this coupling to be background gauge invariant, we impose 
\begin{equation} \label{eq:b_transform} 
b^\1  \goesto b^\1 + \Lambda^\1 - \frac{\kappa}{2\pi} \lambda^\0 \, (d\chi- A^\1)\,, \\
\end{equation}
under 0-form and 1-form gauge transformations. Note the unusual feature that the magnetic gauge field shifts by an operator under background gauge transformations. Under electric-magnetic duality the 't Hooft line from Eq.~\eqref{eq:abelian_H} maps to 
\begin{equation}
H_{q,\gamma} = \exp\left( i q \int_\gamma b^\1\right)\,. 
\end{equation}
To compute the family anomaly, we have to treat the dynamical fields within the same differential cohomology formalism described above for the 2-group background fields. We have a compact scalar $\chi^\0_i$ and gauge field $b^\1_i$ in each patch, with 
\begin{equation}
\begin{split}
(\delta\chi^\0)_{ij} &= 2\pi n_{ij} + \Phi^\0_{ij}\,, \quad\quad (\delta b^\1)_{ij} = d\phi^\0_{ij} + \Omega^\1_{ij} - \frac{\kappa}{2\pi} \Phi^\0_{ij} \, (d\chi^\0_j-A^\1_{j})\,, \\
\end{split}
\end{equation}
where on double overlaps we have integers $n_{ij}$ capturing the winding of $\chi$ and transition functions $\phi^\0_{ij}$ for the gauge field. Both are summed over in the path integral. Repeatedly taking $\delta$ of the connection formulae and using the connection and cocycle conditions in Eqs.~\eqref{eq:connection},~\eqref{eq:firstcocycle},~\eqref{eq:secondcocycle} gives
\begin{equation}
\begin{split}
(\delta n)_{ijk} = - C_{ijk} \,, \quad (\delta\phi^\0)_{ijk} = -\Psi^\0_{ijk} + \kappa\, C_{ijk}\, \chi^\0_k + 2\pi m_{ijk}\,, \quad (\delta m)_{ijkl} = Y_{ijkl} -\kappa\, C_{ijk}\, n_{kl}\,.
\end{split}
\end{equation}
where $m_{ijk}$ are integers on triple overlaps which we sum over in the path integral. The total data of the dynamical fields is captured by $(\chi^\0_i,\, n_{ij})$ and  $(b^\1_i,\, \phi^\0_{ij},\, m_{ijk})$. These are subject to the gauge redundancies
\begin{equation}
\begin{split}
\chi^\0_i & \goesto \chi^\0_i + 2\pi  r_i \,, \quad\quad b^\1_i  \goesto b^\1_i + d\alpha^\0_i\,, \\
n_{ij} &\goesto n_{ij} + (\delta r)_{ij}\,, \quad\quad \phi^\0_{ij} \goesto \phi^\0_{ij} + (\delta \alpha^\0)_{ij} + 2\pi s_{ij} \,, \\
m_{ijk} & \goesto m_{ijk} -\kappa\, C_{ijk}\, r_k + (\delta s)_{ijk} \,. 
\end{split}
\end{equation}
where $r_i, s_{ij}$ are integers and $\alpha^\0_i$ real-valued `magnetic' gauge transformation parameters. Small background gauge transformations act as 
\begin{equation} \label{eq:small_gauge_dyn}
\begin{split}
\chi_i^\0 &  \goesto \chi_i^\0 + \lambda^\0_i \,, \\
b^\1_i & \goesto b^\1_i + \Lambda^\1_i - \frac{\kappa}{2\pi} \lambda^\0_i \, (d\chi^\0_i-A^\1_i)\,,
\end{split}
\end{equation}
as in Eq.~\eqref{eq:b_transform}. The advantage of the patch formalism is that we can study large gauge transformations which affect the transition functions and higher data of the dynamical fields, 
\begin{equation} \label{eq:large_gauge_dyn} 
\begin{split}
\phi^\0_{ij} &  \goesto \phi^\0_{ij} +\Pi^\0_{ij} + \kappa\, K_{ij}\,  \chi^\0_j   \,, \\
n_{ij} & \goesto n_{ij} - K_{ij} \,, \\
m_{ijk} & \goesto m_{ijk} -\kappa\, K_{ij}\, n_{jk} + Z_{ijk}\,. 
\end{split}
\end{equation}
The 't Hooft line transforms by
\begin{multline} \label{eq:abelian_trans}
H_{q,\gamma} = \exp\left(i q\sum_i \int_{\gamma_i} b^\1_i + i q\sum_{i<j} \left.\phi^\0_{ij}\right|_{\gamma_{ij}}\right) \\
\goesto H_{q,\gamma} \, \exp\left(-\frac{i q\kappa}{2\pi} \sum_i \int_{\gamma_i}\lambda^\0_i\, (d\chi^\0_i-A^\1_i) +i q\kappa \sum_{i<j} \left. K_{ij}\, \chi^\0_{j}\right|_{\gamma_{ij}} \right)\,.
\end{multline}

Finally, we can revisit the 't Hooft line coupled to a source for the tilt operator \eqref{eq:Htheta}, which in the differential cohomology formalism is written as
\begin{equation}
H_{q,\gamma}[\theta] = \exp\Bigg( i q \sum_i \int_{\gamma_i} \left( b^\1_i + \frac{\kappa}{2\pi} \theta^\0_i \,(d\chi_i^\0 -A^\1_i) \right) + i q \sum_{i<j} \left. \left( \phi^\0_{ij} + \kappa\, W_{ij}\, \chi_j^\0\right) \right|_{\gamma_{ij}} \Bigg) ~.
\end{equation}
It is then straightforward to verify from Eq.~\eqref{eq:large_gauge_dyn} and~\eqref{eq:theta_transformation} that the modulated 't Hooft line is invariant under large $U(1)^\0$ gauge transformations, but under a small background gauge transformation the shift of $\chi$ in the last term gives the family anomaly from Eq.~\eqref{eq:correct_abelian_anomaly}. 

\subsubsection{Coupling to a Rotor}

So far we have studied just one family of 't Hooft lines in the abelian Goldstone-Maxwell model. As in Sec.~\ref{sec:nlsm_on_line}, we can study a larger class of lines by starting with the 't Hooft line defined in Eq.~\eqref{eq:Htheta} and promoting the moduli parameter $\theta$ to a dynamical field $\uptheta$. The result is a quantum-mechanical rotor coupled to the bulk periodic scalar field:\footnote{Before promoting $\theta \to \uptheta$ to a dynamical rotor we added a counterterm $\sim \int_\gamma \theta A$ relative to Eq.~\eqref{eq:Htheta} so that the rotor is $2\pi$-periodic. See the discussion near Eq.~\eqref{eq:notpump}.  }
\begin{multline} \label{eq:H_QM}
    \widetilde{H}_{q,\gamma} =\int\mathscr{D}\uptheta \exp\left[\frac{2\pi q}{e^2}\int_{\Sigma }\star da^\1 +iq \int_{\Sigma}\left(B^\2 + \frac{\kappa}{2\pi}d\chi\wedge A^\1\right)\right] \\
 \times   \exp\left[- \int dt \, \left( \frac{I}{2}(\dot{\uptheta}-A^\1_t)^2 + \frac{iq\kappa}{2\pi} \chi \,\dot\uptheta  \right)\right] ~.
\end{multline}
Here $I$ is the moment of inertia of the rotor and $t$ is the coordinate along $\gamma$. Despite the fact that we are integrating over the modulus $\uptheta$, we have failed to restore the $U(1)$ symmetry --- we simply get a different tilt operator 
\begin{equation}
   \widetilde{\tau} =  - \frac{q\kappa}{2\pi}(d\uptheta-A^\1)\,.
\end{equation}
This has the universal form described in Sec.~\ref{sec:nlsm_on_line}, for the case $G = U(1)$.

Given that the bulk-defect system is still quadratic, we can solve for the dynamics induced by the defect. In the absence of background gauge fields the compact scalar and the Maxwell field are completely decoupled. Let us focus on the scalar field dynamics.\footnote{Ignoring the gauge field, our rotor defect can be regarded as a generalization of the (bosonized) fermion-rotor model of~\cite{Polchinski:1984uw}, recently revisited in~\cite{loladze2024,Loladze:2025jsq}, to four bulk dimensions. } From the point of view of the rotor, the bulk field $\chi$ looks like a time-dependent magnetic flux. From the point of view of the bulk field $\chi$, integrating out the rotor degree of freedom induces a (series of) localized mass term(s) on the line, 
\begin{equation}
\widetilde{H}_{q,\gamma} \sim  H_{q,\gamma} \, \sum_{n \in \ZZ} \exp\left( -\frac{1}{2(2\pi)^2 I}\int dt (\chi(t)-2\pi n )^2 \right)\,, 
\end{equation}
where $n$ labels the energy levels of the rotor. This indeed breaks the $U(1)^\0$ shift symmetry explicitly. In App.~\ref{app:rotor} we show that this induces a scattering length $\ell$ for s-wave scattering with
\begin{equation}
\ell = \left(\frac{q\kappa}{2\pi}\right)^2 \frac{1}{2v^2 I_r}\,,  
\end{equation}
with $I_r$ a renormalized  moment of inertia for the rotor.

\subsubsection{Deformations}

Previously, we argued that the symmetry breaking should be agnostic to the bulk-realization of the 2-group. We will consider a `UV' model which can realize the abelian Goldstone-Maxwell theory at long distances, but also other phases. 

A simple completion which does the job is the QED-like model as discussed in Sec. 6.2 of \cite{Cordova:2018cvg}. We take four Weyl fermions and complete the compact field $\chi$ into a complex scalar $\phi$ coupled to a $U(1)$ gauge field $a$ with charges:
    \begin{equation}
        \begin{array}{c|ccccc}
             & \psi_1 & \psi_2 & \psi_3 & \psi_4 & \phi_A \\
            \cline{1-6}
            \text{gauge} & +1 & +1 & -1 & -1 & 0\\
            \text{global}  & +1 & -1 & 0 & 0 & -1
        \end{array}
    \end{equation}
    This charge assignment is identical to $N_f=2$ massless QED (discussed above in Sec.~\ref{sec:qed}) where the flavor symmetry is explicitly broken to the $U(1)$-Cartan of the left action, with charges shown in the above table. We add a potential for $\phi$ and Yukawa couplings:
    \begin{equation}
        \Delta\mathscr{L} = y_1 \phi\,  \psi_1 \psi_3 + y_2 \overline{\phi}\,  \psi_2\psi_4 + \text{h.c.} -\mu^2|\phi|^2- \frac{\lambda}{4}|\phi|^4
    \end{equation}
    By dialing $\mu^2$ from large negative to large positive values, we can go from the abelian Goldstone-Maxwell model to a version of massless QED where the $U(1)^\0$ symmetry participates in a 2-group with the $U(1)^\1$ magnetic symmetry with $\kappa = 1$. In the latter phase the symmetry breaking is enforced by the choice of fermion boundary conditions as in Sec.~\ref{sec:qed}. 

 Starting with massless QED (i.e. after giving $\phi$ a large mass squared) we can introduce a charge-1 Higgs field $H$ which is neutral under $U(1)^\0$. We write Yukawa couplings
    \begin{equation}
        \Delta\mathscr{L} = \widetilde{y}_1 \overline{H}^2\, \psi_1\psi_2 + \widetilde{y}_2 H^2\, \psi_3 \psi_4 + \text{h.c.} - \mu^2_H|H|^2 - \frac{\lambda_H}{4}|H|^4
    \end{equation}
    In the $\mu_H^2\ll0$ phase, the fermions are gapped out (this is not obstructed by any perturbative anomalies) and the gauge group is Higgsed to nothing. In the Higgs phase, the 't Hooft line acting on the vacuum creates a solitonic ANO string. The string world-sheet carries a $2d$ chiral anomaly~\cite{Cordova:2018cvg} and it ends on the 't Hooft line, so the 't Hooft line must break the anomalous symmetry. The symmetry breaking appears in a way similar to the monopole background: the Higgs field $H$ winds asymptotically in the ANO vortex background and an index theorem ensures the existence of bound chiral world-sheet fermion zero modes, which need to be quantized \cite{Jackiw:1981ee, Weinberg:1981eu}. If the worldsheet has a boundary, one needs to provide a boundary condition for the fermions to render the Hamiltonian self-adjoint. This will break the symmetry.

\section{Discrete 2-Groups}
\label{sec:discrete}

In this section we consider discrete 2-groups, where only a finite discrete 1-form symmetry $\calA^\1$ participates in the 2-group. The 0-form symmetry $G^\0$ itself may be continuous or finite. Rather than phrasing the discussion in terms of background fields as in the previous sections, we primarily use the diagrammatics of topological symmetry operators to derive the obstruction to a symmetric line defect. We discuss numerous examples, including 2-groups that admit symmetric lines, and finish by pointing out some subtleties that arise in the discrete case regarding extended and enhanced symmetries. 

We start with a lightning review of discrete 2-group global symmetries following \cite{Benini:2018reh}. The 2-group data consists of:
\begin{enumerate}
    \item A 0-form symmetry $G^\0$ and a 1-form symmetry $\mathcal{A}^\1$,

    \item A group homomorphism $\rho$ that describes the action of the  0-form symmetry on 1-form symmetry defects
    \begin{equation}
        \rho: G^\0 \to \text{Aut}(\mathcal{A}^\1) \,,
    \end{equation}

    \item A Postnikov class $\beta\in H^3_\rho(G^\0, \mathcal{A}^\1)$ which is the RG-invariant data.
\end{enumerate}
When $G^\0$ and $\mathcal{A}^\1$ are discrete, the Postnikov class $\beta$ can be understood diagrammatically in terms of symmetry operators~\cite{Benini:2018reh}. Consider the triple intersection of three 0-form symmetry operators $\mathbf{g},\mathbf{h},\mathbf{k}\in G^\0$ in codimension-3. In a 2-group, such a junction is only topological if a 1-form symmetry operator $\beta(\mathbf{g},\mathbf{h},\mathbf{k})$ \emph{terminates} at the triple intersection:
\begin{equation}\label{eq:discrete}
\begin{tikzpicture}[baseline=(current bounding box.center),scale= 0.8]
    \draw[line width = 0.3mm] (3.129,2.246) -- (3.129,2.246-4); 
    \draw[line width = 0.3mm] (0,2) arc (-90:-81:20); 
    \draw[dashed, dash phase = -2, line cap=round, line width = 0.4mm] (0,-2) arc (-90:-81:20); 
    \draw[line width = 0.3mm] (-4.158234, 2.437048) -- (-4.158234, 2.437048-4); 
    \draw[line width = 0.3mm] (0,2) arc (-90:-96.8799:20); 
    \draw[line width = 0.3mm]  (-2.8111,2.1985) arc (-98.0799:-102:20); 
    \draw[line width = 0.3mm] (0,-2) arc (-90:-102:20); 
    \draw[line width = 0.3mm, dashed, line cap=round, line width = 0.3mm] (-1.423-2,-1.948+1.5) -- (-1.423,-1.948); 
    \draw[line width = 0.3mm] (1.571-2,2.063+1.5) -- (1.571,2.063); 
    \draw[line width = 0.3mm, dashed] (-0.429,3.563) -- (-3.423,-0.448); 
    \draw[line width = 0.3mm] (-0.429,3.563) -- (-1.5508, 2.0602); 

    \draw[line width=0.5mm, blue!80!black, decoration={markings, mark=at position .5 with {\arrow[scale=1.5,rotate=0]{stealth}}}, postaction=decorate] (0,0) arc (-110:-154:5); 

    \draw[line width = 0.5mm, red!90!black] (0,-2) to (0,2); 
    \draw[line width = 0.5mm, red!90!black] (0,0) arc (127.46:164.03:3.8); 
    \draw[line width = 0.5mm, red!90!black, dashed, line cap=round] (0,0) arc (-57.239:-25.77:3.8); 
    \draw[line width = 0.5mm, red!90!black] (1.3663, 1.5446) arc (-25.77:-17.85:3.8); 

    \node at (-3.9,2) {\footnotesize $\mathbf{g}$}; 
    \node at (-.4,3.3) {\footnotesize $\mathbf{h}$}; 
    \node at (2.9,2) {\footnotesize $\mathbf{k}$}; 
    \node at (2.5,-2.4) {\footnotesize $\mathbf{g}\mathbf{h}\mathbf{k}$}; 
    \node at (-0.5,-1.7) {\footnotesize $\mathbf{g}\mathbf{h}$}; \node at (1.2,1.8) {\footnotesize $\mathbf{h}\mathbf{k}$};
    \node[above, blue] at (-2.7,2.4) {\footnotesize $\beta(\mathbf{g}, \mathbf{h},\mathbf{k})$};
    \filldraw (0,0) circle [radius=.05];
    \draw[line width = 0.3mm] (0,2) to (3,1) to (3,-3) to (0,-2); 

\end{tikzpicture} 
\end{equation}
Notice that we can dress the codimension-2 fusion junction of 0-form symmetry operators $\mathbf{g},\mathbf{h}$ by a 1-form symmetry operator $t(\mathbf{g},\mathbf{h}) \in \mathcal{A}^\1$:
\begin{equation}
    \begin{tikzpicture}[baseline=(current bounding box.center), scale = 0.8]
        \draw[line width = 0.3mm] (9.5,0) to[out=-90, in=180, looseness=1] (12.25,-1) -- (13.25,-0.285) to[out=180, in=-95, looseness=1] (10.5,0.715) -- (9.5,0);
        \draw[line width = 0.3mm, dashed] (10.5,0.715) -- (9.863,0.715);
        \draw[line width = 0.3mm] (9.863,0.715) -- (7.85,0.715);
        \draw[line width = 0.3mm] (7.85,0.715) -- (6.5,0);
        \draw[line width = 0.3mm] (6.5,0) -- (9.5,0);

        \draw[line width = 0.3mm, black] (10,0.85) to[out=30, in=180, looseness=0.5] (12.5,1);
        \draw[line width = 0.3mm, black](12.5,1) -- (13.5,1.715) to[out=180, in=30, looseness=0.5] (11,1.565) -- (10,0.85);
        \draw[line width = 0.3mm] (9.5,0) to[out=95, in=210, looseness=1] (10,0.85) -- (11,1.565);
        \draw[line width = 0.3mm, dashed] (11,1.565) .. controls (10.787575,1.442357) and (10.613768,1.212493) .. (10.539225,0.967134);
        \draw[line width = 0.3mm] (10.539225,0.967134) .. controls (10.513871,0.883679) and (10.5,0.798431) .. (10.5,0.715);
        
        \draw[line width = 0.5mm, red] (9.5,0) -- (10.5,0.715);
       
        \node at (7.2,0.8) {\footnotesize $\mathbf{gh}$};
        \node at (13,-0.75) {\footnotesize $\mathbf{g}$};
        \node at (13.5,2) {\footnotesize $\mathbf{h}$};
    \end{tikzpicture} \rightarrow \begin{tikzpicture}[baseline=(current bounding box.center), scale = 0.8]
        \draw[line width = 0.3mm] (9.5,0) to[out=-90, in=180, looseness=1] (12.25,-1) -- (13.25,-0.285) to[out=180, in=-95, looseness=1] (10.5,0.715) -- (9.5,0);
        \draw[line width = 0.3mm, dashed] (10.5,0.715) -- (9.863,0.715);
        \draw[line width = 0.3mm] (9.863,0.715) -- (7.85,0.715);
        \draw[line width = 0.3mm] (7.85,0.715) -- (6.5,0);
        \draw[line width = 0.3mm] (6.5,0) -- (9.5,0);

        \draw[line width = 0.3mm, black] (10,0.85) to[out=30, in=180, looseness=0.5] (12.5,1);
        \draw[line width = 0.3mm, black](12.5,1) -- (13.5,1.715) to[out=180, in=30, looseness=0.5] (11,1.565) -- (10,0.85);
        \draw[line width = 0.3mm] (9.5,0) to[out=95, in=210, looseness=1] (10,0.85) -- (11,1.565);
        \draw[line width = 0.3mm, dashed] (11,1.565) .. controls (10.787575,1.442357) and (10.613768,1.212493) .. (10.539225,0.967134);
        \draw[line width = 0.3mm] (10.539225,0.967134) .. controls (10.513871,0.883679) and (10.5,0.798431) .. (10.5,0.715);
        
        \draw[line width = 0.5mm, blue] (9.5,0) -- (10.5,0.715);
       
        \node at (7.2,0.8) {\footnotesize $\mathbf{gh}$};
        \node at (13,-0.75) {\footnotesize $\mathbf{g}$};
        \node at (13.5,2) {\footnotesize $\mathbf{h}$};

        \node[blue] at (11.3,0.6) {\footnotesize $t(\mathbf{g},\mathbf{h})$};
    \end{tikzpicture} ~,
\end{equation}
under which the Postnikov class changes by 
\begin{equation}
    \beta(\mathbf{g},\mathbf{h},\mathbf{k}) \mapsto \beta(\mathbf{g},\mathbf{h},\mathbf{k}) \frac{{}^{\mathbf{g}}t(\mathbf{h},\mathbf{k}) t(\mathbf{g},\mathbf{hk})}{t(\mathbf{gh},\mathbf{k})t(\mathbf{g},\mathbf{h})} ~,
\end{equation}
where we use ${}^g a$ to denote the $g$-transformation of some object $a$. This is the reason why the Postnikov class $\beta$ is only defined as a cohomology class. 

Similar to the continuous case, the Postnikov class $\beta$ also shows up in the relation between the background fields $A^{\1}$ and $B^{\2}$ of $G^\0$ and $\mathcal{A}^\1$:
\begin{equation}
    \delta_{A} B^\2 = (A^{\1})^*\beta ~,
\end{equation}
where $\delta_A$ is the twisted differential encoding the $G^\0$ action on $\mathcal{A}^\1$, and $A^*\beta$ is the pull-back of $\beta \in H^3(BG^\0,\mathcal{A}^\1)$ along $A^\1: \mathcal{M}\rightarrow BG^\0$ and can be expressed in terms of $A^\1$ using cup products and Bockstein homomorphisms.

\subsection{Constraints from Symmetry Defects}

We now discuss consistency conditions for line operators charged under $\calA^\1$. This amounts to understanding the representation theory of discrete 2-groups, as studied in e.g.~\cite{Bartsch:2023pzl,Bhardwaj:2023wzd}. The question of explicit symmetry breaking on line defects amounts to asking whether the line defect furnishes a 2-representation of non-trivial dimension.  

Consider a line operator $\scrL_{\mathbf{q}}$ that transforms in an irreducible representation $\mathbf{q}$ of the 1-form symmetry $\mathcal{A}^\1$. The automorphism $\rho$ specifies the action of the 0-form symmetry on $\mathcal{A}^\1$. Acting with a 0-form-symmetry generator $\bfg \in G^\0$ can change the 1-form charge $\rho_{\bfg}(\mathbf{q})$ of the new line ${}^{\bfg} \scrL_{\mathbf{q}}$. If $\rho_{\bfg}(\mathbf{q}) \neq \mathbf{q}$ for any $\bfg \in G^\0$, then ${}^{\bfg} \scrL_{\mathbf{q}}$ cannot be the same line operator as $\scrL_{\mathbf{q}}$. In this case, the line $\scrL_{\mathbf{q}}$ breaks the 0-form symmetry, regardless of whether the Postnikov class is non-trivial.  

Now let us suppose $\mathbf{q}$ is invariant under $G^\0$. Now the Postnikov class $\beta$ becomes important. In general, a line of charge $\mathbf{q}$ may not be acted on by the full 1-form symmetry $\calA^\1$. The faithfully-acting group is a finite cyclic group and fits into a short exact sequence 
\begin{equation}
    \ker \mathbf{q} \, \rightarrow\,  \mathcal{A}^{(1)} \, \rightarrow\,  \mathcal{A}^\1/ \ker \mathbf{q} ~.
\end{equation}
This allows us to define the \textit{effective} Postnikov class $\mathbf{q}_*\beta$ for any line of charge $\mathbf{q}$ from the long exact sequence
\begin{equation}
    \cdots \, \rightarrow\,  H_\rho^3(G^\0,\mathcal{A}^\1) \, \xrightarrow{\mathbf{q}_*} \, H^3(G^\0,\mathcal{A}^\1/ \ker \mathbf{q}) \, \rightarrow \,  \cdots 
\end{equation}
to quotient out the trivially-acting subgroup of the 1-form symmetry. When the effective Postnikov class is trivial, we do not get any constraints on the $G^\0$-symmetry breaking of the charged line operators. Therefore, in the following, we restrict ourselves to the case where $\mathcal{A}^\1 = \mathbb{Z}_N^\1$ and consider lines which are faithfully acted upon by $\calA^\1$.\footnote{We can also consider 2-groups where the 0-form symmetry $G^\0$ is discrete and the 1-form symmetry is continuous. However, in this case, only a discrete subgroup of the 1-form symmetry participates non-trivially in the 2-group, so the situation is identical to the discrete-discrete case above. The same holds even if $G^\0$ is continuous as long as the Postnikov class is finite. }  

To proceed, we follow \cite{Bartsch:2023pzl}. Denote a generic \textit{simple} line operator by $\scrL_i$, where the subscript $i$ is acted upon by $G^\0$ as ${}^{\mathbf{g}}\scrL_i = \scrL_{\sigma_{\mathbf{g}}(i)}$.\footnote{We comment on non-simple lines later in Sec.~\ref{sec:nsl} when we discuss examples.} Our goal is to understand when it is consistent for the line to be a $G^\0$ singlet, i.e. when a single line label suffices. There is a corresponding topological junction associated with this symmetry action where $\scrL_i$ pierces a $\mathbf{g}$ symmetry operator and emerges as $\scrL_{\sigma_{\mathbf{g}}(i)}$.\footnote{If the line $\scrL_i$ is non-simple, then there exist multiple local junctions between $\scrL_i$ and the 0-form symmetry operator $\mathbf{g}$.} Moving a codimension-2 junction of $G^\0$ symmetry operators across $\scrL_i$ may lead to a non-trivial phase:
\begin{equation}
 \begin{tikzpicture}[baseline=(current bounding box.center), scale = 0.8]
        \draw[line width = 0.5mm, dgreen] (11.5,-1.5) -- (11.5,-1);
        \draw[line width = 0.5mm, dgreen, densely dotted] (11.5,-1) -- (11.5,-0.6);
        \draw[line width = 0.5mm, dgreen, ->-=0.5] (11.5,-0.6) -- (11.5,1.0);
        \draw[line width = 0.5mm, dgreen, densely dotted] (11.5,1.0) -- (11.5,1.35);
        \draw[line width = 0.5mm, dgreen] (11.5,1.35) -- (11.5,2.5);
        
        \draw[line width = 0.3mm] (9.5,0) to[out=-90, in=180, looseness=1] (12.25,-1);
        \draw[line width = 0.3mm] (12.25,-1) -- (13.25,-0.285);
        \draw[line width = 0.3mm] (13.25,-0.285) .. controls (12.757345,-0.285000) and (12.161375,-0.311120) .. (11.650000,-0.250132);
        \draw[line width = 0.3mm] (11.350000,-0.201144) .. controls (10.823615,-0.087700) and (10.451827,0.164381) .. (10.5,0.715);
        \draw[line width = 0.3mm, dashed] (10.5,0.715) -- (9.863,0.715);
        \draw[line width = 0.3mm] (9.863,0.715) -- (7.85,0.715);
        \draw[line width = 0.3mm] (7.85,0.715) -- (6.5,0);
        \draw[line width = 0.3mm] (6.5,0) -- (9.5,0);

        \filldraw[dgreen!70!black] (11.5,-0.6) circle (1.5pt);
        \filldraw[dgreen!70!black] (11.5,1.35) circle (1.5pt);

        \draw[line width = 0.3mm, black] (10,0.85) to[out=30, in=180, looseness=0.5] (12.5,1);
        \draw[line width = 0.3mm, black] (12.5,1) -- (13.5,1.715);
        \draw[line width = 0.3mm, black] (13.5,1.715) .. controls (13.212156,1.715000) and (12.304379,1.742917) .. (11.650000,1.691810);
        \draw[line width = 0.3mm, black] (11.350000,1.659475) .. controls (11.194361,1.636646) and (11.070975,1.605977) .. (11,1.565);
        \draw[line width = 0.3mm, black] (11,1.565) -- (10,0.85);
        \draw[line width = 0.3mm] (9.5,0) to[out=95, in=210, looseness=1] (10,0.85) -- (11,1.565);
        \draw[line width = 0.3mm, dashed] (11,1.565) .. controls (10.787575,1.442357) and (10.613768,1.212493) .. (10.539225,0.967134);
        \draw[line width = 0.3mm] (10.539225,0.967134) .. controls (10.513871,0.883679) and (10.5,0.798431) .. (10.5,0.715);
        
        \draw[line width = 0.5mm, red] (9.5,0) -- (10.5,0.715);
       
        \node at (7.2,0.8) {\footnotesize $\mathbf{gh}$};
        \node at (13,-0.75) {\footnotesize $\mathbf{g}$};
        \node at (13.5,2) {\footnotesize $\mathbf{h}$};
        \node at (11.8,-1.65) {$\color{dgreen!70!black}{\scrL}_i$};
        \node at (12.3,0.3) {$\color{dgreen!70!black}{\scrL}_{\sigma_{\mathbf{g}}(i)}$};
        \node at (12.4,2.5) {$\color{dgreen!70!black}{\scrL}_{\sigma_{\mathbf{gh}}(i)}$};
        \end{tikzpicture} 
   \ = \ \nu_i(\bfg,\bfh) \quad  
    \begin{tikzpicture}[baseline=(current bounding box.center), scale = 0.8]
        \draw[line width = 0.5mm, dgreen] (8.5,0) -- (8.5,-1.5);
        \draw[line width = 0.5mm, dgreen, densely dotted] (8.5,0) -- (8.5,0.35);
        \draw[line width = 0.5mm, dgreen, ->- = 0.5] (8.5,0.35) -- (8.5,2.5);

        \draw[line width = 0.3mm] (9.5,0) to[out=-90, in=180, looseness=1] (12.25,-1) -- (13.25,-0.285) to[out=180, in=-95, looseness=1] (10.5,0.715) -- (9.5,0);
        \draw[line width = 0.3mm, dashed] (10.5,0.715) -- (9.863,0.715);
        \draw[line width = 0.3mm] (9.863,0.715) -- (8.6,0.715);
        \draw[line width = 0.3mm] (8.4,0.715) -- (7.85,0.715);
        \draw[line width = 0.3mm] (7.85,0.715) -- (6.5,0);
        \draw[line width = 0.3mm] (6.5,0) -- (9.5,0);

        \filldraw[dgreen!70!black] (8.5,0.35) circle (1.5pt);

        \draw[line width = 0.3mm, black] (10,0.85) to[out=30, in=180, looseness=0.5] (12.5,1);
        \draw[line width = 0.3mm, black](12.5,1) -- (13.5,1.715) to[out=180, in=30, looseness=0.5] (11,1.565) -- (10,0.85);
        \draw[line width = 0.3mm] (9.5,0) to[out=95, in=210, looseness=1] (10,0.85) -- (11,1.565);
        \draw[line width = 0.3mm, dashed] (11,1.565) .. controls (10.787575,1.442357) and (10.613768,1.212493) .. (10.539225,0.967134);
        \draw[line width = 0.3mm] (10.539225,0.967134) .. controls (10.513871,0.883679) and (10.5,0.798431) .. (10.5,0.715);
        
        \draw[line width = 0.5mm, red] (9.5,0) -- (10.5,0.715);
       
        \node at (7.2,0.8) {\footnotesize $\mathbf{gh}$};
        \node at (13,-0.75) {\footnotesize $\mathbf{g}$};
        \node at (13.5,2) {\footnotesize $\mathbf{h}$};
        \node at (8.8,-1.65) {$\color{dgreen!70!black}{\scrL}_i$};
        \node at (9.4,+2.5) {$\color{dgreen!70!black}{\scrL}_{\sigma_{\mathbf{gh}}(i)}$};
    \end{tikzpicture}
\end{equation}
We can dress the junction between $\scrL_i$ and the symmetry operator $\mathbf{g}$ with a local counterterm, which is a c-number phase $\gamma_i(\bfg)$. Therefore, $\nu_i(\mathbf{g},\mathbf{h})$ is only defined up to:
\begin{equation}
    \nu_i(\bfg,\bfh) \mapsto \nu_i(\bfg,\bfh) \frac{\gamma_{\sigma_{\mathbf{g}}(i)}(\mathbf{h})\gamma_i(\mathbf{g})}{\gamma_i(\mathbf{gh})} ~.
\end{equation}

As mentioned around Eq.~\eqref{eq:discrete}, the key feature of the Postnikov class $\beta$ in discrete 2-groups is that the 1-form symmetry line $\beta(\mathbf{g},\mathbf{h},\mathbf{k}) \in \mathcal{A}^\1$ terminates at the triple junction of 0-form symmetry operators. As a consequence, the fusion of 0-form symmetry defects is associative only \emph{up to} the 1-form symmetry~\cite{Benini:2018reh}. The action of the 0-form symmetry on lines charged under the 1-form symmetry must therefore also be non-associative. By matching the non-associativity of the fusion of bulk symmetry defects with their action on lines, we obtain a consistency condition which is the discrete analog of the Wess-Zumino condition exploited in Secs.~\ref{sec:nonabelian},~\ref{sec:abelian}.\footnote{The same technique of using the F-symbol or `associator' was used in \cite{Thorngren:2020yht} to argue for symmetry breaking at the boundary of theories with discrete 't Hooft anomalies.} 

\begin{figure}[h!]
    \centering
    \includegraphics[width=0.95\linewidth]{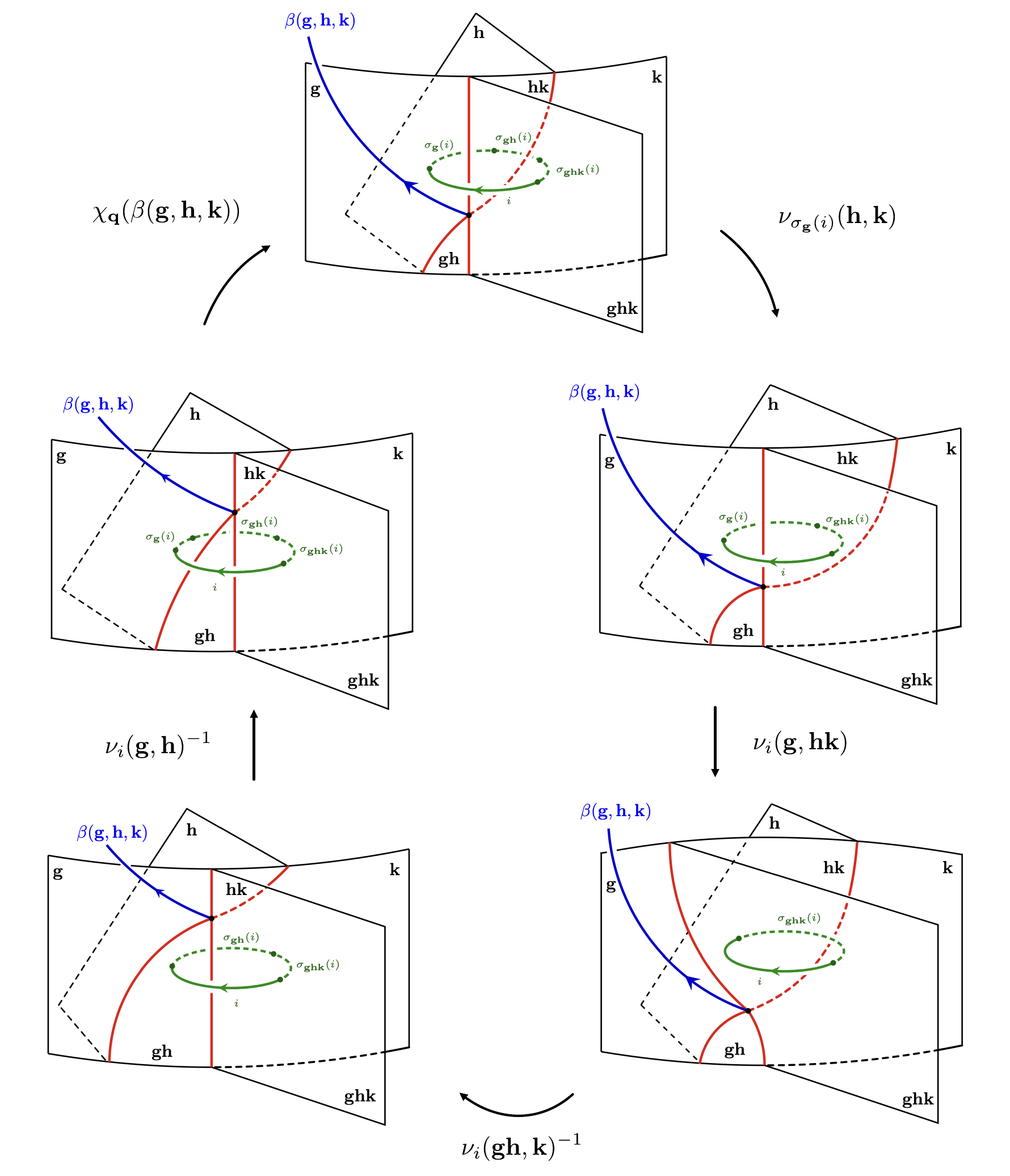}
    \caption{Derivation of the consistency condition \eqref{eq:F-move} by the above topological manipulations of the triple junction of $G^\0$ symmetry operators around a line operator charged under $\calA^\1$.}
    \label{fig:pentagon}
\end{figure}

We start with a closed loop $\scrL_{\mathbf{q},i}$ in the presence of the triple junction of 0-form symmetries. Moving the junction around $\scrL_{\mathbf{q},i}$ as in Fig.~\ref{fig:pentagon} leads to the consistency condition
\begin{equation}\label{eq:F-move}
  \frac{\nu_{\sigma_{\bfg}(i)}(\mathbf{h},\mathbf{k})\, \nu_i(\mathbf{g},\mathbf{hk})}{\nu_i(\mathbf{gh},\mathbf{k})\nu_i(\mathbf{g},\mathbf{h})}\, \chi_{\mathbf{q}}(\beta(\mathbf{g},\mathbf{h},\mathbf{k})) =  1 ~,
\end{equation}
where $\chi_{\mathbf{q}}(a)$ denotes the action of the 1-form symmetry operator $a$ on $\scrL_{\mathbf{q},i}$. This is the discrete analog of Eq.~\eqref{eq:cancel_WZ}.

For $G^\0$-invariant charge $\mathbf{q}$, the quantity $\chi_{\mathbf{q}}(\beta(\mathbf{g},\mathbf{h},\mathbf{k}))$ defines an induced cohomology class in $H^3(G^\0,U(1))$ with coefficients in $U(1)$ as opposed to $\calA^\1$. Suppose that the line $\scrL_{\mathbf{q},i}$ is a singlet. Then we can drop the $i$ label and Eq.~\eqref{eq:F-move} reads
\begin{equation} \label{eq:singlet_condition}
    \chi_{\mathbf{q}}(\beta(\mathbf{g},\mathbf{h}, \mathbf{k})) \stackrel{!}{=} \frac{\nu(\mathbf{gh},\mathbf{k})\,\nu(\mathbf{g},\mathbf{h})}{\nu(\mathbf{h},\mathbf{k})\, \nu(\mathbf{g},\mathbf{hk})} = (\delta \nu)(\mathbf{g},\mathbf{h},\mathbf{k})\,, 
\end{equation}
which says that  $\chi_{\mathbf{q}}(\beta)$ is cohomologically trivial. Conversely, 
\begin{quote}
If the induced 3-cocycle $\chi_{\mathbf{q}}(\beta) \in H^3(G^\0,U(1))$ is cohomologically non-trivial, there does not exist a $G^\0$-singlet line operator $\scrL_{\mathbf{q}}$.
\end{quote}
In more physical terms, the induced 3-cocycle $\chi_\q(\beta)$ characterizes the $G^\0$ 't Hooft anomaly on the Wilson surface bounding the line operator $\scrL_\q$ coupled to background gauge fields (as a consistency check, 2d anomalies for bosonic $G^\0$-symmetry are indeed classified by $H^3(G^\0,U(1))$). The symmetry breaking result can then be rephrased as 
\begin{quote}
If the Wilson surface bounding a line charged under $\calA^\1$ has a $G^\0$ anomaly, the line must explicitly break $G^\0$.
\end{quote}
This statement is identical to the continuous case. What is special about discrete symmetries is that \emph{not every Postnikov class leads to a non-trivial anomaly on the Wilson surface.} In the next section we illustrate the different possibilities with explicit examples.

\subsection{Examples}
\label{sec:discrete_examples}

In the following, we illustrate the different situations in simple examples. We will mainly focus on two cases:
\begin{enumerate}
    \item The Postnikov class is non-trivial, but trivializes as an anomaly class in $H^3(BG,U(1))$ and does not lead to symmetry breaking constraints.
    \item The Postnikov class induces a non-trivial anomaly class which implies symmetry breaking by the charged lines. 
\end{enumerate}
We also comment on more subtle cases in the final subsection. 

\subsubsection{Postnikov Class without Symmetry Breaking}
\label{sec:nobreaking}

We begin with two well-known examples of 2-group symmetry, and find that there is no enforced $G^\0$-symmetry breaking by charged line operators. The first example is the 4d abelian Higgs model with two charge-2 scalars. The faithfully acting 0-form symmetry is $G^\0 = SO(3)^\0$ and the theory has a $\calA^\1 = \mathbb{Z}_2^\1$ electric 1-form symmetry. Together, they form a 2-group with Postnikov class~\cite{Benini:2018reh,Bartsch:2023pzl}
\begin{equation}
 \beta = \mathrm{Bock}(w_2(SO(3)^\0))
\end{equation}
where $w_2 \in H^2(SO(3)^\0,\mathbb{Z}_2)$ is the second Stiefel-Whitney class of $SO(3)$ and $\mathrm{Bock}: H^2(SO(3)^\0,\mathbb{Z}_2) \rightarrow H^3(SO(3)^\0,\mathbb{Z}_2)$ is the Bockstein homomorphism, which can be explicitly written as
\begin{equation}
    \mathrm{Bock}(w_2) = \frac{1}{2}\delta \widetilde{w}_2 \, \in H^3(SO(3)^\0,\ZZ_2) \,,
\end{equation}
where $\widetilde{w}_2$ is a $\ZZ_4$ lift of the $w_2$ class. The 2-group structure survives as long as we preserve the $\ZZ_2\times \ZZ_2$ subgroup of $SO(3)^\0$ of $\pi$ rotations around any two orthogonal axes. Being a torsion class, the Bockstein of the Stiefel-Whitney class becomes exact when embedded in $U(1)$, where we can write
\begin{equation}
    \mathrm{Bock}(w_2) = \delta\left( \frac{1}{2}\widetilde{w}_2 \right) \, \in H^3(SO(3)^\0,U(1)) \,.
\end{equation}
Indeed, there is nothing inconsistent about the standard charge-1 Wilson line in this theory which does not explicitly break the $SO(3)^\0$ flavor symmetry.\footnote{It does, however, contribute the non-trivial phases $c(\mathbf{g},\mathbf{h}) = e^{\frac{i\pi}{2} \widetilde{\omega}_2(\mathbf{g},\mathbf{h})}$ whose role is to match the induced Postnikov class in Eq.~\eqref{eq:singlet_condition}. This is the square-root of the phase $e^{i\pi w_2(\mathbf{g},\mathbf{h})}$ generated by the junction passing through a charge-2 Wilson line, which can terminate on the charge-2 Higgs fields transforming in a projective representation of $SO(3)^\0$. } 

This example turns out to be part of a more general story when the 0-form symmetry $G^\0$ is continuous and $\mathcal{A}^\1$ is discrete. First, recall that on a given line operator, the faithfully acting 1-form symmetry must be isomorphic to some $\mathbb{Z}_N^{(1)}$. Therefore, the effective Postnikov class $\beta \in H^3(BG^\0,\mathbb{Z}_N^\1)$ must satisfy $\beta^N = 1$, and the induced class $\chi_\q(\beta) \in H^3(BG^\0,U(1))$ also satisfies $\chi_q(\beta)^N = 1$. On the other hand, if $G^\0$ is \emph{connected}, the possible 2d anomaly coefficients (classified by $H^3(BG^\0,U(1)) \simeq H^4(BG^\0,\mathbb{Z})$) always take values in $\ZZ$ rather than in a finite cyclic group.\footnote{Here, the coefficient $U(1)$ has the topology of $S^1$.  For a proof of this statement, see \href{https://mathoverflow.net/questions/180173/h4bg-mathbb-z-torsion-free-for-g-a-connected-lie-group}{here}.}
This means that the Postnikov class will always induce a trivial anomaly $H^3(BG^\0,U(1))$,  and there will never be a constraint on symmetry breaking. 
This conclusion can be evaded if $G^{\0}$ is \emph{disconnected}~\cite{Bhardwaj:2022scy}, in which case some anomaly coefficients may be mod $N$. We will provide an explicit example later. 

\bigskip

The second example without symmetry breaking is 3d $\mathbb{Z}_3$-gauge theory with a twist, with an action that can be written in terms of two $U(1)$ gauge fields $u,v$ as
\begin{equation}
    S = \frac{3i}{2\pi} \int u \wedge dv + \frac{i}{2\pi} \int u \wedge du ~.
\end{equation}
This theory contains only topological line operators $\displaystyle e^{i k \oint u}$ ($k\in \mathbb{Z}_9$) generating an anomalous $\mathbb{Z}^\1_9$ 1-form symmetry. It is shown in~\cite{Benini:2018reh,Thorngren:2015gtw}, that one can couple this theory to a 2-group background involving \emph{any} discrete trivially-acting 0-form symmetry $G^\0$ and the 1-form symmetry $\mathbb{Z}^\1_3 \subset \mathbb{Z}^\1_9$. Its Postnikov class is given by: 
\begin{equation}
    \beta = \mathrm{Bock}(w) ~, \quad w \in H^2(BG,\mathbb{Z}_3) ~,
\end{equation}
where $\mathrm{Bock}:H^2(BG,\mathbb{Z}_3) \rightarrow H^3(BG,\mathbb{Z}_3)$ is the Bockstein homomorphism associated with the extension $\mathbb{Z}_3 \rightarrow \mathbb{Z}_9 \rightarrow \mathbb{Z}_3$. However, for the same reason described above, $\mathrm{Bock}(w)$ always becomes trivial when the coefficients are in $U(1)$. In other words, the `anomaly' $e^{\frac{2\pi i q}{9} \mathrm{Bock}(w)}$ on a charge-$q$ line is always trivial and does not enforce symmetry breaking. This is unsurprising, because the 0-form symmetry $G^\0$ acts trivially by construction.

\subsubsection{Postnikov Class Implies Symmetry Breaking}
\label{sec:discretebreaking}

Let us now turn to examples where the Postnikov class leads to a non-trivial anomaly and symmetry breaking by lines. We start by considering a generic 3d QFT $\mathcal{T}$ with $\mathbb{Z}_{2,\widetilde{A}}^\0 \times \mathbb{Z}_{2,A}^\0$-symmetry and inflow action
\begin{equation}\label{eq:3dZ2Z2ma}
    i\pi \int \widetilde{A}^{(1)}\cup A^{(1)} \cup A^{(1)} \cup A^{(1)} ~.
\end{equation}
Upon gauging the $\mathbb{Z}_{2,\widetilde{A}}^\0$, we get a dual theory $\mathcal{T}'$ with $\mathbb{Z}_{2,B}^\1$ dual symmetry participating in a 2-group with $\mathbb{Z}_{2,A}^\0$~\cite{Tachikawa:2017gyf}
\begin{equation}
    Z_{\mathcal{T}'}[B^{(2)}, A^{(1)}] =  \sum_{\widetilde{a}^{(1)}\in H^1(\mathcal{M}_3,\mathbb{Z}_2)} Z_{\mathcal{T}}[\widetilde{a}^{(1)},A^{(1)}] \exp\left(i\pi \int \widetilde{a}^{(1)}\cup B^{(2)} \right) ~.
\end{equation}
Gauge invariance of this sum imposes
\begin{equation}
    \delta B^{(2)} = A^{(1)} \cup A^{(1)} \cup A^{(1)} ~.
\end{equation}
Viewing background gauge fields as Poincar\'e dual to insertions of symmetry operators, this relation is equivalent to the statement that a 1-form symmetry operator terminates on the triple junction of 0-form symmetry operators. So there is a non-trivial Postnikov class for the 2-group symmetry $\mathbb{Z}_{2,B}^{\1}\times \mathbb{Z}_{2,A}^{\0}$. Moreover, it does not trivialize when evaluated in $U(1)$ and therefore implies that any simple lines charged under $\mathbb{Z}_{2,B}^{\1}$ must break $\mathbb{Z}_{2,A}^{\0}$. These symmetry breaking constraints can be understood directly in the original theory $\mathcal{T}$. The anomaly~\eqref{eq:3dZ2Z2ma} implies that the 0-form symmetry operator generating $\ZZ_{2,\widetilde{A}}^\0$ carries a $\mathbb{Z}_{2,A}^\0$-anomaly characterized by a 3d inflow action\footnote{Another application of anomalies of symmetry operators under other symmetries is to construct non-invertible symmetries~\cite{Kaidi:2021xfk}.}
\begin{equation}
    i\pi \int A^{(1)} \cup A^{(1)} \cup A^{(1)} ~.
\end{equation}
The lines charged under the dual 1-form symmetry $\mathbb{Z}_{2,B}^{\1}$ after gauging $\ZZ_{2,\widetilde{A}}^\0$ originate from the twist defects for $\mathbb{Z}_{2,\widetilde{A}}^{\0}$ in $\mathcal{T}$, namely the non-genuine line operators that bound the $\mathbb{Z}_{2,\widetilde{A}}^{\0}$-symmetry operator~\cite{Kapustin:2014gua,Copetti:2026ncv}. The above world-sheet anomaly immediately implies any simple boundary condition must break $\mathbb{Z}_{2,A}^{\0}$-symmetry. 

\bigskip

Next, we look at an example in 3d discrete gauge theory with gauge group $\DD_{16} = \langle r,s|r^8 = s^2 = 1, srs = r^{-1}\rangle$. This theory has $\mathbb{Z}_{2,a}^\1 \times \mathbb{Z}_{2,b}^\1 \times \mathbb{Z}_{2,m}^\1$ 1-form symmetries. Here, $\mathbb{Z}_{2,a}^\1 \times \mathbb{Z}_{2,b}^\1$ is generated by four $\DD_{16}$-Wilson lines labeled by four one-dimensional irreducible representations of $\DD_{16}$ while $\mathbb{Z}_{2,m}^\1$ is generated by the pure magnetic line\footnote{Pure magnetic lines in 3d discrete $G^\0$-gauge theory are defined similarly to 4d 't Hooft lines. We excise a tube in spacetime and restrict to configurations, where the discrete gauge field has a specific holonomy around it. The holonomy is labeled by conjugacy classes of $G^\0$, and it is only invertible if the conjugacy class contains a single element.} labeled by the center $\mathbb{Z}_2 = Z(\DD_{16})$. The gauge group $\DD_{16}$ has an order-2 outer automorphism $\rho$, which acts as
\begin{equation}
    \rho: r \mapsto r^5 ~, s \mapsto sr ~. 
\end{equation}
$\rho$ naturally induces a $\mathbb{Z}_2^\rho$ 0-form symmetry in the $\DD_{16}$ gauge theory. Its action exchanges $a$ and $b$, but leaves the pure magnetic line invariant. The twisted group cohomology associated with this action is $H^3_\rho(\mathbb{Z}_{2,\rho}^\0,\mathbb{Z}_{2,a}^\1 \times \mathbb{Z}_{2,b}^\1 \times \mathbb{Z}_{2,m}^\1) \simeq \mathbb{Z}_2$ and the non-trivial generator has a representative
\begin{equation}\label{eq:D16P}
    \beta(\eta,\eta,\eta) = m ~.
\end{equation}
As shown in \cite{Fidkowski:2015oam}, this Postnikov class is realized in the $\DD_{16}$-gauge theory by $\mathbb{Z}_{2,\rho}^\0$ and $\mathbb{Z}_{2,a}^\1 \times \mathbb{Z}_{2,b}^\1 \times \mathbb{Z}_{2,m}^\1$. Although this can directly be seen using the formalism of~\cite{Barkeshli:2014cna}, there is a simpler way to show it. If $\mathbb{Z}_{2,\rho}^\0$ can be gauged, the resulting theory must be a $K$-gauge theory, where $K$ is an order-$32$ group from the group extension
\begin{equation}
    0 \rightarrow \DD_{16} \rightarrow K \rightarrow \mathbb{Z}_{2,\rho} \rightarrow 0 ~.
\end{equation}
However, due to an obstruction valued in $H^3_\rho(\mathbb{Z}_{2,\rho},Z(\DD_{16}))$, no such group extension exists. The proposed gauging cannot be implemented and there must be an obstruction to gauging $\mathbb{Z}_{2,\rho}^\0$. In a 3d TQFT, a 0-form symmetry cannot be gauged, if there is a non-trivial Postnikov class or a 't Hooft anomaly. However, there are no 't Hooft anomalies for $\mathbb{Z}_2^\0$ 0-form symmetry in three dimensions because $H^4(\mathbb{Z}_2,U(1)) = 0$. Thus $\mathbb{Z}_{2,\rho}^\0$ must realize the only non-trivial Postnikov class \eqref{eq:D16P}.\footnote{In 3d discrete $H$-gauge theories invertible 1-form symmetries (abelian anyons) $\mathcal{A}^{\1}$ are labeled by $Z(H) \times \widehat{H}$, where $\widehat{H}$ is the group of one-dimensional irreducible representations of $H$. The obstruction to extending the group, valued in $H^3(G,Z(H))$, naturally embeds into $H^3(G,Z(H)\times \widehat{H})$ and is the Postnikov class of the $H$-gauge theory.} Any line charged under $\mathbb{Z}_{2,m}^\1$ must break the 0-form symmetry $\rho$, because the anomaly in $H^3(\mathbb{Z}_{2,\rho}^\0,U(1))$ is not trivial. This prediction can be verified by computing the $\rho$-action on the line operator spectrum following~\cite{Coste:2000tq,Lu:2025gpt}. Note that this theory provides a counter-example to the lore that spontaneously broken $\calA^\1$ implies spontaneously broken $G^\0$. 

\bigskip

As pointed out above, there are Postnikov classes that lead to non-trivial anomalies for continuous, disconnected $G^\0$. Consider $\text{Spin}(4)=SU(2)_L\times SU(2)_R$ gauge theory with $N_f=2$ Weyl fermions $\psi^{a\Dot{a}}_i$ in the vector representation as discussed in~\cite{Hsin:2020nts} (see also~\cite{Bhardwaj:2022scy}). The ABJ anomaly breaks the naive flavor symmetry to $(SU(2)_f^\0\times \mathbb{Z}_4^\0)/\mathbb{Z}^\0_2$. Furthermore, the gauge-invariant mesons and baryons of this theory are purely bosonic, which leaves us with:
\begin{equation}
    G^\0 = SO(3)_f^\0\times \mathbb{Z}_2^\0 \times \mathbb{Z}_{2,\mathcal{C}}^\0 ~,
\end{equation}
where the last factor is charge conjugation which exchanges the two  $SU(2)$ factors of the gauge group and acts non-trivially on baryons. The 1-form symmetry of the theory is the diagonal center element $\mathbb{Z}_{2,e}^\1$. It acts on Wilson lines $W_{(j_L,j_R)}$ with $j_L, j_R\in \frac{1}{2}\mathbb{Z}_{\geq 0}$ as $(-1)^{2j_L + 2j_R}$. As shown in \cite{Hsin:2020nts}, this theory has Postnikov class:
\begin{equation} \label{eq:disconnected_postnikov}
     \delta B_e^\2 = \beta^\3 = w_2(SO(3)_f)\cup A^\1_{\mathcal{C}}
\end{equation}
which remains non-trivial as a two-dimensional anomaly.\footnote{For instance, activating the subgroup $\ZZ_{2,x}^\0 \times \ZZ_{2,z}^\0 \subset SO(3)^\0_f$ of $\pi$ rotations around the $x$ and $z$ axes leads to the type-III anomaly with inflow $i \pi \int A^\1_x \cup A^\1_z \cup A^\1_{\mathcal{C}}$.  }
In this case, the symmetry breaking is obvious, because any simple Wilson line $W_{(j_L,j_R)}$ charged under $\mathbb{Z}_{2,e}^{(1)}$ will have $j_R\neq j_L$ and is therefore mapped to $W_{(j_R,j_L)} \neq W_{(j_L,j_R)}$ by $\mathcal{C}$. This can be understood more invariantly from the Postnikov class \eqref{eq:disconnected_postnikov} as follows: the charge conjugation action takes $B^\2_e \to B^\2_e + w_2(SO(3)_f)$ and therefore shifts the $SO(3)_f^\0$ fractionalization class of lines charged under $\ZZ_{2,e}^\1$. This is only possible if charged lines come in pairs related by $\mathcal{C}$.

\bigskip 

Finally, we revisit the abelian Goldstone-Maxwell model but explicitly break $U(1)^\0 \to \ZZ_N^\0$ by adding a potential $\sim \cos(N\chi)$ to the Lagrangian. The symmetries are $G^\0 = \ZZ^\0_N, \calA^\1 = \ZZ^\1_N$. The Postnikov class comes from restricting the familiar continuous anomaly $AdA$ to discrete gauge fields, 
\begin{equation} \label{eq:ZN_anomaly} 
\beta^\3(A^\1) \,  =  \, \kappa\, A^\1 \cup  \mathrm{Bock}(A^\1)\,, \quad \kappa \sim \kappa + N\,. 
\end{equation}
This example provides a setting where we can explicitly connect the Wess-Zumino consistency analysis of the family anomaly in Sec.~\ref{sec:abelian} with the symmetry defect-based formalism we use for discrete symmetries. 

First, we can recycle the analysis from Sec.~\ref{sec:abelian_GMM} to see that the $\ZZ_N^\0$ symmetry is explicitly broken by the 't Hooft line. For simplicity, we set $\kappa = 1$ in what follows. Let's start with the (unmodulated) $q=1$ 't Hooft line and apply a global $\ZZ_N^\0$ transformation with $\lambda = \frac{2\pi \widehat{\lambda}}{N} \in \frac{2\pi}{N}\ZZ$. From Eq.~\eqref{eq:abelian_trans}, the 't Hooft line changes by
\begin{equation}
H_\gamma \goesto H_\gamma \, \exp\left( - \frac{2\pi i \widehat{\lambda}}{N} \int_\gamma \frac{d\chi}{2\pi} \right) \,,
\end{equation}
showing that the $\ZZ^\0_N$ symmetry is indeed broken by the line. 

Next, we show how to relate the differential cohomology formalism to the associativity constraint from the triple-junction of symmetry operators. To study discrete 2-groups using differential cohomology we set the local background gauge fields $A^\1_i$ to zero, and consider constant transition functions taking values in $\Phi_{ij}^\0 =\frac{2\pi}{N} \widehat{\Phi}_{ij} \in \frac{2\pi}{N}\ZZ$. In this formalism, there are leftover small gauge transformations with $\lambda^\0_i = \frac{2\pi}{N} \widehat{\lambda}_i \in \frac{2\pi}{N}\ZZ$ and large gauge transformations $K_{ij} \in \ZZ$ which correspond to changes in the integer lifts of the mod-$N$ gauge fields. 

The patch formalism gives a concrete way to compute the anomalous phases $\nu_{\sigma_{\mathbf{g}}(i)}(\mathbf{h},\mathbf{k})$ from Sec.~\ref{sec:discrete}. In the $\ZZ_N$ case we have a set of line operators labeled by group elements $i = a \in \{1,\ldots,N\}$, with $\sigma_{a}(b) = [a+b]$ where $[\cdot]$ denotes mod $N$. We start with the un-modulated charge-1 line through the symmetry operator $a_1$, followed by $a_2$ and $a_3$. Fusing $a_2$ and $a_3$ to $[a_2+a_3]$ using appropriate background gauge transformations will generate a phase from the family anomaly. This phase is nothing but $\nu_{a_1}(a_2,a_3)$, which we use to compute the Postnikov class.  

The computation can be performed using 4 patches. We start with transition functions
\begin{equation}
\Phi_{12} = \frac{2\pi a_1}{N} \,, \ \Phi_{23} = \frac{2\pi a_2}{N} \,,  \ \Phi_{34} =\frac{2\pi a_3}{N}\,, 
\end{equation}
with associated modulation parameters
\begin{equation}
\theta_2 = \frac{2\pi a_1}{N} \,, \ \theta_3 = \frac{2\pi a_2}{N} \,, \ \theta_4 = \frac{2\pi a_3}{N} \,,  \quad W_{12} = a_1\,, \ W_{23} = \{a_1,a_2\}\,,  \ W_{34} = \{a_1+a_2,a_3\}\,,
\end{equation}
where $\{a,b\} \equiv \frac{a+b-[a+b]}{N}$. To fuse $a_2$ and $a_3$ we perform a background gauge transformation with $\widehat{\lambda}_3 = - a_3$. To ensure we are left with the proper defect labeled by $[a_2+a_3]$ we also have to change the integer lift using $K_{23} = \{a_2,a_3\}$. Using the family anomaly from Eq.~\eqref{eq:correct_abelian_anomaly} the fusion holds up to a phase
\begin{equation}
\nu_{a_1}(a_2,a_3) = e^{\frac{2\pi i}{N} \sum_{i<j} W_{ij} \, \widehat{\lambda}^\0_j} = e^{- \frac{2\pi i}{N} \{a_1,a_2\} a_3}\,. 
\end{equation}
The Postnikov class is computed by taking the twisted differential, 
\begin{equation}
\beta(a_1,a_2,a_3) = \frac{\nu_0(a_1,a_2)\, \nu_0([a_1+a_2],a_3) }{\nu_{a_1}(a_2,a_3)\, \nu_0(a_1,[a_2+a_3])} = e^{ \frac{2\pi i}{N} \{a_1,a_2\} a_3} ~.
\end{equation}
This is the group cohomology analog of Eq.~\eqref{eq:ZN_anomaly}. 

\subsection{Discrete Intricacies}
\label{sec:intricacies}

In this section we examine some subtleties of 2-groups with discrete $\mathcal{A}^\1$ that do not occur in the continuous case. First, we ask whether non-simple lines can evade the symmetry breaking derived for simple lines. Using the abelian Goldstone-Maxwell model as an example, we argue that non-simple lines are only symmetric when one extends the global symmetry of the problem. Next, we consider two distinct mechanisms by which the Postnikov class can be trivialized. In the first scenario we extend $G^\0 \to \widehat{G}^\0$ (this is related to the question about non-simple lines). In the second scenario, we start by enlarging $\calA^\1 \to \widetilde{\calA}^\1$. Both mechanisms can be realized along RG flows where either a 0-form symmetry becomes trivially acting or a 1-form symmetry becomes enhanced.

\subsubsection{Non-Simple Lines}\label{sec:nsl}

The abelian Goldstone-Maxwell model gives a setting where we can explore the discrete analog of coupling the line defect to a dynamical spurion as we did in Sec.~\ref{sec:nlsm_on_line}. Recall that in the continuous case, promoting the spurion to a dynamical field results in a new tilt operator coming from the original family anomaly. The discrete analog is to consider \emph{non-simple} line operators as a way to `symmetrize' the defect. We will see that the family anomaly obstructs even non-simple lines from preserving the $\ZZ^\0_N$ symmetry.\footnote{We thank S. Seifnashri for related discussions on non-simple boundary conditions for anomalous theories. } 

A naive first attempt to symmetrize the defect would be to sum over its (global) $\ZZ^\0_N$ orbit. While this makes the 't Hooft line invariant under the global action of $\ZZ^\0_N$ (obtained by wrapping the line with a symmetry operator), summing over orbits is not enough to construct topological junctions between symmetry operators and the line. In order to construct these topological junctions, we have to sum over orbits within each patch, 
\begin{equation}
\widehat{H}_\gamma \, = \,  \sum_{\{\widehat{\theta}_i \in \ZZ\}} H_\gamma \, \exp\left(  \frac{i}{N} \sum_i  \int_{\gamma_i} \widehat{\theta}_i \, d\chi_i  \right)\,.
\end{equation} 
This is clearly invariant under the symmetry action if we shift $\widehat{\theta}_i \to \widehat{\theta}_i + \widehat{\lambda}_i$. The non-simpleness of the above line is reflected in the fact that $e^{i\chi(\x=0,t)}$ evaluated on the line is a topological operator. In fact, this topological operator is generated when we perform a large gauge transformation, or equivalently when we change the integer lift of the $\ZZ^\0_N$ background, 
\begin{equation}
\widehat{H}_\gamma  \goesto \widehat{H}_\gamma \, \exp\left( i \sum_{i<j} \left. K_{ij} \, \chi^\0_j \right|_{\gamma_{ij}} \right)\,. 
\end{equation}
We can summarize the situation as follows: we succeeded in writing down a non-simple 't Hooft line which is symmetric under the bulk $\ZZ^\0_N$ symmetry in the sense that there exist topological junctions between bulk symmetry operators and the line. But if we do parallel fusion of $N$ generators pierced by the line, we are left with the topological point operator $e^{i\chi}$ on the line. In other words, in the presence of the line defect the $\ZZ^\0_N$ symmetry is extended by this topological point operator. We can make this more familiar if we take the deep IR limit where $\chi$ is pinned into $N$ discrete vacua. Then $e^{i\chi}$ is also topological in the bulk, and generates a $\ZZ_N^{(3)}$ symmetry. The global symmetry in the presence of the line defect is not $\ZZ^\0_N$ but rather is extended by $\ZZ_N^\3$ to $\ZZ_{N^2}$. 

Alternatively, we can try to fix the non-invariance under the changes in integer lift by summing over another discrete field on the line, 
\begin{equation}
\sum_{ \{ k_{ij} \in \ZZ \} } \widehat{H}_\gamma \, \exp\left( -i \sum_{i<j}\left. k_{ij}\, \chi^\0_j \right|_{\gamma_{ij}}\right)\,, 
\end{equation}
but this spoils the invariance under ordinary $\ZZ^\0_N$ background gauge transformations, so despite being non-simple, it does not define a symmetric line defect.

\subsubsection{Postnikov Class Resolution: Extending $G^\0$}

In the previous example we argued that non-simple lines can be viewed as symmetric as long as we extend the global symmetry of the problem. In the above discussion, the global symmetry was extended by a symmetry localized to the defect worldline. But we can equally view this as a bulk symmetry, which does not act faithfully. 

To understand this, consider a UV theory $\mathcal{T}_{\text{UV}}$ that has 0-form symmetry $G_{\text{UV}}^\0$ and 1-form symmetry $\mathcal{A}^\1$ without a Postnikov class. In the infrared, a subgroup $G_{\text{triv}}^{\0}$ may become trivially acting, such that the faithfully acting symmetry $G_{\text{IR}}^{\0}$ is the quotient of $G_{\text{UV}}^\0$ by $G_{\text{triv}}^\0$:
\begin{equation}
    0 \rightarrow G_{\text{triv}}^{\0} \rightarrow G_{\text{UV}}^\0 \rightarrow G_{\text{IR}}^\0 \rightarrow 0 ~.
\end{equation}
If the group extension is non-trivial, then some $G_{\text{IR}}^\0$ backgrounds cannot be lifted to a $G_{\text{UV}}^\0$ background. At the IR fixed point $\mathcal{T}_{\text{IR}}$, we have access to more 0-form symmetry background fields $A_{\text{IR}}^{(1)}$. Even if we start with a trivial Postnikov class between $G_{\text{UV}}^\0$ and $\mathcal{A}^\1$ in the UV, coupling to a background field $A_{\text{IR}}^{(1)}$ which cannot be lifted to a $G_{\text{UV}}^\0$-background field, may require $B^{(2)}$ and $A_{\text{IR}}^\1$ to obey 
\begin{equation}
    \delta B^{(2)} = A_{\text{IR}}^{(1)\,*}\beta ~,
\end{equation}
indicating a non-trivial Postnikov class between the faithfully acting quotient $G_{\text{IR}}^\0$ and the 1-form symmetry $\mathcal{A}^\1$.

There are two interesting applications of this mechanism. First, a non-trivial Postnikov class is an obstruction to gauging the $0$-form symmetry $G^\0$ on its own. By enlarging $G^\0$ to $\widetilde{G}^\0$ with a trivially acting kernel $G_{\text{triv}}^{\0}$,\footnote{In fact, this is always possible when $G^\0$ is discrete \cite{meunier2026arbitrary}.} one can often resolve the Postnikov class obstruction and gauge $\widetilde{G}^\0$. As an example, consider the case $G^\0 = \mathbb{Z}_2^\0$ and $\mathcal{A}^\1 = \mathbb{Z}_2^\1$. As mentioned, $H^3(\mathbb{Z}_2^\0,\mathbb{Z}_2^\1) \simeq \mathbb{Z}_2$ and the non-trivial Postnikov class is represented by the relation of the background gauge fields $A^{(1)}, B^{(2)}$ as
\begin{equation}\label{eq:Z22Postnikov}
    \delta B^{(2)} = A^{(1)} \cup A^{(1)} \cup A^{(1)} ~.
\end{equation}
This Postnikov class can be trivialized if we enlarge $\mathbb{Z}_2^\0$ to $\mathbb{Z}_4^\0$ with a trivially acting $\mathbb{Z}_{2,\text{triv}}^{\0}$ kernel. Any $\mathbb{Z}_4^\0$-background gauge field $A_{\mathbb{Z}_4}^{(1)}$ can be written as 
\begin{equation}
    A_{\mathbb{Z}_4}^{(1)} = A^{(1)} +2 C^{(1)} ~, \quad \delta C^{(1)} = \mathrm{Bock}(A^{(1)}) = A^{(1)} \cup A^{(1)} ~, \quad  \delta A^{(1)} = 0 ~,
\end{equation}
where $C^{(1)}$ is a background field for $\mathbb{Z}_{2,\text{triv}}^{\0}$. Then,
\begin{equation}
    A^{(1)} \cup A^{(1)} \cup A^{(1)} = \delta (C^{(1)} \cup A^{(1)}) ~,
\end{equation}
and the Postnikov class can be removed by shifting $B^{(2)} \mapsto B^{(2)} + C^{(1)} \cup A^{(1)}$. In practice the right-hand side of \eqref{eq:Z22Postnikov} is an obstruction to summing over all $\mathbb{Z}_2^{(0)}$ background fields $A^{(1)}$ consistently. However, $A^{(1)} \cup A^{(1)} = \mathrm{Bock}(A^{(1)})$ measures the obstruction of lifting a $\mathbb{Z}_2^{(0)}$-background field to a $\mathbb{Z}_4^{(0)}$-background field. By declaring the symmetry to be $\mathbb{Z}_4^{(0)}$, we only sum over backgrounds $A^{(1)}$ which can be lifted to a $\mathbb{Z}_4^{(0)}$-background and therefore have $\mathrm{Bock}(A^{(1)}) = 0$ which trivializes \eqref{eq:Z22Postnikov}. This process can also be carried out in the 3d $D_{16}$-gauge theory discussed previously. By gauging the enlarged $\mathbb{Z}_4^\0$ symmetry, we get a discrete gauge theory with an order-$64$ gauge group. It is important to emphasize that this operation does not lift the symmetry breaking of simple charged lines. To derive these constraints, we couple the theory to every accessible background field. To avoid an obstruction to gauging, we sum over fewer backgrounds which explains why some constraints are seemingly missing. 

A second application concerns the emergence theorem of 2-group symmetries. If both $G^\0$ and $\mathcal{A}^\1$ emerge in the infrared with a non-trivial Postnikov class, the 0-form symmetry cannot emerge before the 1-form symmetry. If the Postnikov class $\beta$ is non-trivial, there is no UV completion where $\mathcal{A}^\1$ is broken while the faithfully acting symmetry in the infrared $G^\0_{\text{IR}}$ is unbroken. However, the Postnikov class may only exist upon considering the faithfully acting group. If there exists an enlarged symmetry $\widetilde{G}^{\0}$ where
\begin{equation}
    0 \rightarrow G^{\0}_\text{triv} \rightarrow \widetilde{G}^{\0} \rightarrow G_{\text{IR}}^\0 \rightarrow 0 ~,
\end{equation}
such that the Postnikov class $\beta \in H^3(G^\0_{\text{IR}},\mathcal{A}^\1)$ is lifted to a trivial element in $H^3(\widetilde{G}^\0,\mathcal{A}^\1)$, it is possible to find a UV completion where $\calA^\1$ is broken but there is an exact 0-form symmetry $G_{\text{UV}}^{\0} = \widetilde{G}^\0$. Along the RG flow, the subgroup $G^{\0}_\text{triv}$ becomes trivially acting and $\mathcal{A}^\1$ emerges. The emergence theorem prevents $G^{\0}_\text{triv}$ from acting trivially before the 1-form symmetry emerges. This occurs in the 4d abelian Higgs model with two charge-2 scalars. Its Postnikov class
\begin{equation}
    \delta B^{(2)} = \mathrm{Bock}(w_2)
\end{equation}
trivializes if we lift the $SO(3)^{(0)}$ symmetry to $SU(2)^{(0)}$. We can add a heavy flavor-neutral charge-1 scalar, such that $G_{UV}^\0 = SU(2)^{(0)}$ but $\mathbb{Z}_2^{(1)}$ is broken. Upon integrating it out $\mathbb{Z}_2^{(0)} \subset SU(2)^{(0)}$ becomes trivially acting and $\mathbb{Z}_2^{(1)}$ emerges.

\subsubsection{Enlarging $\calA^\1$ with Non-Trivial $\rho$}

We have not yet commented on cases where the symmetry breaking is only due to the action $\rho$ of $G^\0$ on $\calA^\1$. Simple examples can be found when $G^\0$ is charge conjugation $\rho = \mathcal{C}$~\cite{Bhardwaj:2022scy,Jacobson:2024muj}. However, there is a more subtle realization, where a non-trivial $\rho$ action is enforced by a Postnikov class associated to a \emph{subgroup} of the full 1-form symmetry. Consider a 2-group symmetry with non-trivial Postnikov class $\beta \in H^3(G^\0,\mathcal{A}^\1)$ and suppose that the theory actually has a larger 1-form symmetry $\widetilde{\mathcal{A}}^{(1)} \supset \mathcal{A}^{(1)}$. The Postnikov class $\beta$ may become trivializable in $H^3_\rho(G^\0,\widetilde{\mathcal{A}}^\1)$ when the $G^\0$-action $\rho$ on $\widetilde{\mathcal{A}}^{(1)}$ is non-trivial.\footnote{If the Postnikov class does not lead to symmetry breaking on lines, then even if $\rho$ is trivial the Postnikov class may trivialize if we enlarge the 1-form symmetry. For instance take $\beta = \text{Bock}(w_2(SO(3)))$ and enlarge the 1-form symmetry to $U(1)^\1$, or even $\ZZ_4^\1$. This resolves an apparent counter-example to the SSB hierarchy in the 4d abelian Higgs model from Sec.~\ref{sec:nobreaking}. There, we can gap out the scalars to land on pure Maxwell theory, where the $SO(3)$ symmetry acts trivially while the 1-form symmetry is spontaneously broken. The subtle point is that the $\ZZ_2^\1$ 1-form symmetry is enlarged in the IR to $U(1)^\1$, so there is no Postnikov class for the (trivially-acting) $SO(3)$. } If $(G^\0,\mathcal{A}^\1,\beta)$ enforces symmetry breaking, such constraints will however not go away upon considering the enlarged symmetry $\widetilde{\mathcal{A}}^{(1)}$. Instead, they are explained by the non-trivial $G^\0$-action on $\widetilde{\mathcal{A}}^{(1)}$. 

Let us demonstrate this in a concrete example. Previously, we found that gauging $\mathbb{Z}_{2,\widehat{A}}^{(0)}$ in a 3d QFT $\mathcal{T}$ with $\mathbb{Z}_{2,\widehat{A}}^{(0)} \times \mathbb{Z}_2^{(0),A}$ and mixed anomaly
\begin{equation}\label{eq:Z22ma}
    i\pi \int \widehat{A}^{(1)} \cup A^{(1)} \cup A^{(1)} \cup A^{(1)} ~,
\end{equation}
leads to a theory $\mathcal{T}'$ with $\mathbb{Z}_2^{(1)}\times \mathbb{Z}_{2,A}^{(0)}$ and non-trivial Postnikov class in $H^3(\mathbb{Z}_{2,A}^{(0)},\mathbb{Z}_2^{(1)})$. We could also start with a theory $\mathcal{T}$ that has $\mathbb{Z}_{2,C_e}^{(1)} \times \mathbb{Z}_{2,C_m}^{(1)}$ with mixed anomaly
\begin{equation}\label{eq:Z221fma}
    i \pi \int C_e^{(2)} \cup C_m^{(2)} 
\end{equation}
and a non-anomalous $\mathbb{Z}_{2,\widehat{A}}^{(0)} \times \mathbb{Z}_{2,A}^{(0)}$ symmetry with trivial action on $\mathbb{Z}_{2,C_e}^{(1)} \times \mathbb{Z}_{2,C_m}^{(1)}$. We can fractionalize the 0-form symmetry on the 1-form symmetry via
\begin{equation}
    C_e^{(2)} \mapsto C_e^{(2)} + \widehat{A}^{(1)} \cup A^{(1)} ~, \quad C_m^{(2)} \mapsto C_m^{(2)} + A^{(1)} \cup A^{(1)} ~,
\end{equation}
to induce the desired mixed anomaly~\eqref{eq:Z22ma}. In the gauged theory $\mathcal{T}'$, the Postnikov class between $\mathbb{Z}_{2,A}^{(0)}$ and $\mathbb{Z}_{2}^{(1)}$ implies any line charged under $\mathbb{Z}_2^{(1)}$ must break $\mathbb{Z}_{2,A}^{(0)}$ 0-form symmetry. Due to symmetry fractionalization, the symmetry operator $U_A'$ of $\mathbb{Z}_{2,A}^{(0)}$ in the gauged theory $\mathcal{T}'$ is the $\mathbb{Z}_{2,A}^{(0)}$-symmetry operator $U_A$ in $\mathcal{T}$ stacked with the condensation defect 
\begin{equation}
    S = \frac{1}{|H^1(M_2,\mathbb{Z}_2)|} \sum_{\gamma_1,\gamma_2 \in H_1(M_2,\mathbb{Z}_2)} (-1)^{\langle \gamma_1,\gamma_2\rangle} \eta_e(\gamma_1) \, \eta(\gamma_2) ~,
\end{equation}
where $\eta_e,\eta$ generate $\mathbb{Z}_{2,e}^{(1)}, \mathbb{Z}_2^{(1)}$ respectively. The action of $S$ on any simple line $\scrL$ is computed in \cite{Roumpedakis:2022aik} and depends on its charge $(Q_e,Q) \in \mathbb{Z}_2 \times \mathbb{Z}_2$ under $\mathbb{Z}_{2,e}^{(1)} \times \mathbb{Z}_2^{(1)}$:
\begin{equation}\label{eq:stacking}
    S(\scrL) = \eta_e^{Q}\, \eta^{Q_e}\, \scrL ~.
\end{equation}
Due to the mixed anomaly \eqref{eq:Z221fma}, 1-form symmetry generators $\eta_m$ and $\eta_{e} \eta_m$ carry charge $Q_e = 1$. Acting with $U_A' \equiv U_A S$ exchanges them with $\eta_m \eta$ and $\eta_e \eta_m \eta$  respectively. Therefore the 0-form symmetry $\mathbb{Z}_{2,A}^{(0)}$ acts non-trivially on the full 1-form symmetry group $\widetilde{\mathcal{A}}^{(1)} = \mathbb{Z}_{2,e}^{(1)} \times \mathbb{Z}_{2,m}^{(1)} \times \mathbb{Z}_2^{(1)}$ and its action replaces the ordinary differential $\delta$ by a twisted one such that the background field $\widehat{A}$ satisfies
\begin{equation}
    \delta_{A} B^{(2)} = A^{(1)} \cup A^{(1)} \cup A^{(1)} \ \iff \ \delta B^{(2)} = A^{(1)} \cup (C_m^{(2)} + A^{(1)} \cup A^{(1)}) ~.
\end{equation}
If we have access to the \emph{full} 1-form symmetry $\widetilde{\mathcal{A}}^{(1)}$ the Postnikov class can be removed by shifting $C_m^{(2)} \mapsto C_m^{(2)} + A^{(1)} \cup A^{(1)}$. This simply undoes the choice of fractionalization class we made before gauging that led to the 't Hooft anomaly between $\ZZ_{2,\widehat{A}}^\0$ and $\ZZ_{2,A}^\0$. However, the symmetry breaking constraint should not be removed when we perform this shift. Indeed, any simple line $\scrL$ with $Q = 1$ acquires additional charge $\Delta Q_m = 1$ under $U_A'$ due to the $\eta_e$-stacking in \eqref{eq:stacking}. This charge is not invariant under $\mathbb{Z}_2^{(0),A}$ so the line `breaks' this symmetry.

To summarize, when $G^\0$ acts non-trivially on the 1-form symmetry $\widetilde{\mathcal{A}}^\1$, often symmetry fractionalization can allow for a non-trivial Postnikov class of $G^\0$ with a subgroup $\mathcal{A}^\1$. If this Postnikov class leads to non-trivial $H^3(G,U(1))$, it can enforce symmetry breaking even when trivial in $\widetilde{\mathcal{A}}^{(1)}$. The symmetry breaking is naturally explained by the resulting $G^\0$ action on $\widetilde{\mathcal{A}}^\1$.

\section*{Acknowledgements}
We thank A. Cherman, S. Chen, and S. Seifnashri for discussions and especially T. Dumitrescu for invaluable suggestions and conversations, as well as comments on a draft which greatly improved the exposition. We are also grateful to S. Chen and C. Copetti for comments on a draft. T. Jacobson and Z. Sun would like to thank the Simons Center for Geometry and Physics for its hospitality during the workshop ``Paths to Quantum Field Theory," where part of this work was finalized. S. Harder is supported by the Mani L. Bhaumik Institute for Theoretical Physics at UCLA and the Simons Collaboration on Global Categorical Symmetries. Z. Sun is supported by the Simons Collaboration on Global Categorical Symmetries. T. Jacobson acknowledges support from a Schwinger Fellowship at the Mani L. Bhaumik Institute at UCLA.

\appendix

\section{Gauge Variation of the Wess-Zumino-Witten Term}
\label{app:Bvariations}

There are various points in the main text where we need to compute the left $G^\0$-action on the (gauged) 2d Wess-Zumino-Witten term
\begin{equation}
\calB^{(2)}(U,A) = - \kappa\, \calB_{\rm{WZ}}^{(2)}(U) - \frac{i\kappa}{4\pi} \Tr\left( (dU)U^{-1}\wedge A \right)\,. 
\end{equation}
To carry out an explicit computation we need to choose a representative of the 2-form $\calB^\2_{\rm{WZ}}$, which is not globally well-defined. We choose to work with the local tangent space coordinates near the identity of the group manifold, writing $U = e^{i \omega^a T^a}$ and 
\begin{equation}
\calB^\2_{\rm{WZ}}(U) = -\frac{1}{4\pi}\left( \frac{\sinh X(\omega)- X(\omega)}{X^2(\omega)} \right)^{ab} \, d\omega^a \wedge d\omega^b\,,
\end{equation}
where $X^{ab}(\omega) = -f^{abc}\, \omega^c$. Using this trivialization we can in principle compute $\calB^\2_{\rm{WZ}}(g^{-1}\, U) - \calB^\2_{\rm{WZ}}(U)$ for arbitrary $g(x)$. For our purposes, it suffices to compute the variation under local infinitesimal and finite global transformations.

\subsection{Baker-Campbell-Hausdorff Formula}
\label{app:BCH}

The primary tool for these computations is the Baker-Campbell-Hausdorff formula expanded to linear order in one variable and to all orders in the other. When we need an all-orders formula for the matrix on the left, we use 
\begin{equation} \label{eq:BCH_L} 
e^{iC_L} = e^{iA}\, e^{i\delta B} \,, \quad C_L^a = A^a + M^{ab}(A) \, \delta B^b + \calO(\delta B^2) \,,
\end{equation}
where 
\begin{equation}
M^{ab}(A) = \left(\frac{X(A)}{e^{X(A)}-\id}\right)^{ab}\,, \quad X^{ab}(A) = -f^{abc}\, A^c\,. 
\end{equation}
Since $X$ is antisymmetric $M^{ba}(A) = M^{ab}(-A)$. Similarly for a finite matrix acting on the right we have
\begin{equation} \label{eq:BCH_R}
e^{iC_R} =  e^{i\delta B}\, e^{iA} \,, \quad C_R^a = A^a + M^{ab}(-A)\, \delta B^b + \calO(\delta B^2) \,.
\end{equation}

\subsection{Local Infinitesimal Transformations}
\label{app:infinitesimal_trans}

We set $g(x) = e^{i \lambda(x)} \approx \id + i \lambda(x)$ and work to linear order in $\lambda(x)$ (all formulas in this subsection hold only to this leading order). The gauge variation of $\calB^\2[U,A]$ picks up a contribution from both the WZ term and the term involving the gauge field, 
\begin{equation}\label{eq:lambdaB} 
\delta \calB^\2(e^{i \omega},A) = -\kappa  \left(\calB^{(2)}_{\rm{WZ}}(e^{-i \lambda}\, e^{i \omega}) - \calB^{(2)}_{\rm{WZ}}(e^{i \omega})\right) + \frac{i\kappa}{4\pi} \Tr \left( d\lambda \wedge ( (dU)U^{-1} + i A)  \right) \,.
\end{equation}
Recall that $d\, \delta\calB^\2 = d\alpha^\2$. Our goal is to `strip off' a derivative and write
\begin{equation}
\delta\calB^\2(U,A) = \alpha^\2(\lambda,A) + d\nu^\1(\lambda,U,A)\,.
\end{equation} 
Using the convention for the local form of the anomaly in Eq.~\eqref{eq:anomaly_choice}\,,
\begin{equation}
\alpha^\2(\lambda,A) = \frac{\kappa}{4\pi}\Tr\, (\lambda\, dA) \,,
\end{equation}
we get 
\begin{multline} \label{eq:dnu2} 
d\nu^\1(\lambda,U, A) = \delta \calB^\2(e^{i \omega}) - \alpha^\2(\lambda,A)  \\
= - \kappa \left(\calB^{(2)}_{\rm{WZ}}(e^{-i \lambda}\, e^{i\omega})  - \calB^{(2)}_{\rm{WZ}}(e^{i \omega})\right) + \frac{i\kappa}{4\pi} \Tr\, \left[ d\lambda \wedge (dU)U^{-1}   \right] - d\left[\frac{\kappa}{4\pi}\Tr(\lambda A) \right] ~.
\end{multline}
The 1-form $\nu^\1$ is linear in $\lambda$. Matching to the previous equation, it must take the form 
\begin{equation}
\nu^\1(\lambda,U,A) = \lambda^a  f_{ab}(\omega) \, d\omega^b + g_a(\omega)\, d\lambda^a - \frac{\kappa}{4\pi} \lambda^a A^a  \,.
\end{equation}
However, $\nu^\1$ is also only defined up to total derivatives, so without loss of generality we can absorb the second term into the first. Taking the differential,  
\begin{equation}
d\nu^\1(\lambda, U, A) =  f_{ab}(\omega)\, d\lambda^a \wedge d\omega^b + \lambda^a \, \frac{\partial f_{ab}(\omega)}{\partial\omega^c} \, d\omega^c \wedge d\omega^b - d\left[\frac{\kappa}{4\pi}\Tr(\lambda A) \right]  \,. 
\end{equation} 
We can therefore extract $f_{ab}(\omega)$, and hence $\nu^\1$ itself, by computing the coefficient of $d\lambda^a \wedge d\omega^b$ in Eq.~\eqref{eq:dnu2}. Let us first extract the coefficient of $d\lambda^a \wedge d\omega^b$ from the variation of the WZ term: we first use the BCH formula in Eq.~\eqref{eq:BCH_R} and define
\begin{equation}
e^{i \widetilde{\omega}} = e^{-i \lambda}\, e^{i\omega} \ \implies \ \widetilde{\omega}^a = \omega^a - M^{ab}(-\omega)\, \lambda^b + \calO(\lambda^2)\,,
\end{equation}
where $M = \frac{X}{e^X-\id}$. Then, 
\begin{multline}
  -  \kappa\left(\calB^{(2)}_{\rm{WZ}}(e^{-i\lambda}\, e^{i\omega}) - \calB^{(2)}_{\rm{WZ}}(e^{i\omega}) \right) \Bigg|_{d\lambda^a \wedge d\omega^b} =  \kappa \left( M(\omega)\, B(\omega) \right)^{ab} \, d\lambda^a \wedge d\omega^b \\
     =   - \frac{\kappa}{2\pi} \left(\frac{\sinh X(\omega) - X(\omega)}{(e^{X(\omega)}-1)X(\omega)}\right)^{ab} d\lambda^a \wedge d\omega^b ~.
\end{multline}
To compute the other contribution $~\Tr( d\lambda \wedge (dU) U^{-1})$ we write
\begin{multline}
(dU)U^{-1} = i \int_0^1 ds\, e^{i s \omega} d\omega \, e^{-is\omega}  = i \int_0^1 ds\, \left( e^{s X(\omega)} \right)^{ab} d\omega^a \, T^b \\
= i \left(\frac{e^{X(\omega)}- 1}{X(\omega)} \right)^{ab}\, d\omega^a\, T^b = i M^{-1}_{ab}(\omega)\, d\omega^a\, T^b\,. 
\end{multline}
This yields 
\begin{equation}
     \frac{i\kappa}{4\pi} \Tr(d\lambda \wedge (dU)U^{-1}) =  \frac{\kappa}{4\pi} \left(\frac{e^{-X(\omega)}-1}{X(\omega)}\right)^{ab} d\lambda^a \wedge d\omega^b\,.
\end{equation}
Combining the two contributions, we obtain
\begin{equation}
f_{ab}(\omega) = \frac{\kappa}{2\pi} \left(-\frac{\sinh X(\omega) - X(\omega)}{(e^{X(\omega)}-1)X(\omega)} + \frac{e^{-X(\omega)}-1}{2X(\omega)}\right)^{ab} = - \frac{\kappa}{2\pi}\left(\frac{1}{X(\omega)}+\frac{1}{1-e^{X(\omega)}}\right)^{ab}
\end{equation}
and we finally find  
\begin{equation}
\begin{split}
    \nu^{(1)}(\lambda,U,A) &= - \frac{\kappa}{2\pi} \left(\frac{1}{X(\omega)}+\frac{1}{1-e^{X(\omega)}}\right)^{ab} \lambda^a d\omega^b - \frac{\kappa}{4\pi}\lambda^a A^a\,. 
\end{split}
\end{equation}

\subsection{Finite Descent Formula}
\label{app:finite_transgression}

In the previous section, we showed that the variation of $\mathcal{B}^\2(U, A)$ has a nice infinitesimal variation under gauge transformations. We can use the transgression formalism to find a formula for the finite variation. Consider:
\begin{equation}
    U_s = e^{-is\lambda}U = e^{i\omega_s} ~, \quad s\in [0,1] ~.
\end{equation}
Under an infinitesimal change by $ds$, we know that:
\begin{equation}
    \mathcal{B}^\2(U_{s+ds}) - \mathcal{B}^\2(U_s) = d\left(ds\,\nu^\1(\lambda, U_s)\right)+ \mathcal{O}(ds^2)\iff \frac{d}{ds}\mathcal{B}^\2(U_s) = d\nu^\1(\lambda, U_s)
\end{equation}
where we set the gauge field to zero. We can directly integrate to find:
\begin{equation}
    \mathcal{B}^\2(e^{-i\lambda}U) - \mathcal{B}^\2(U) = d\int_0^1 ds\,\nu^\1(\lambda, U_s) ~.
\end{equation}
In practice, it is useful to define
\begin{equation}
    \Theta(\lambda,\omega) = \int_0^1 ds\,\nu^\1(\lambda, \omega_s) = \int_0^1 ds\,\frac{\kappa}{2\pi}\left(\frac{1}{X(\omega_s)}+\frac{1}{1-e^{X(\omega_s)}}\right)^{ab}\lambda^a d\omega_s^b ~.
\end{equation}
We can now use the BCH-formula to get the variation of the form to higher orders.

\subsection{Global Finite Transformations}
\label{app:finite_trans}

Now we consider $g(x) = e^{i\lambda_0}$ and work to all orders in the constant $\lambda_0$. For our purposes it suffices to expand to second order in $\omega(x)$. For global transformations the variation comes just from the WZ term (hence we set $A = 0$ in the following)
\begin{multline}
\delta_{\lambda_0}\, \calB^\2(e^{i\omega(x)},A)  = - \kappa \left( \calB_{\rm{WZ}}^{(2)}(e^{-i\lambda_0} \, e^{i\omega(x)}) - \calB_{\rm{WZ}}^{(2)}(e^{i\omega(x)}) \right) \\
= - \frac{\kappa}{2}\left( B_{ab}(-\lambda_0) \, d\widehat{\omega}^a \wedge d\widehat{\omega}^b - B_{ab}(0)\, d\omega^a \wedge d\omega^b \right) + \calO(\omega^3)\,,
\end{multline}
where $\widehat{\omega}$ is defined using the BCH formula Eq.~\eqref{eq:BCH_L} 
\begin{equation}
e^{i \widehat{\omega}} = e^{-i \lambda_0}\, e^{i \omega}\ \implies \ 
\widehat{\omega}^a = - \lambda_0^a + M^{ab}(-\lambda_0) \, \omega^b + \calO(\omega^2) \,, \quad d\widehat{\omega}^a = M^{ab}(-\lambda_0)\, d\omega^b\,. 
\end{equation}
Plugging this in we find 
\begin{equation}
\delta_{\lambda_0}\,\calB^\2(e^{i\omega(x)}) =  -\frac{\kappa}{2}\left( M(\lambda_0)\, B(-\lambda_0)\, M(-\lambda_0) - B(0) \right)_{ab} \, d\omega^a \wedge d\omega^b   + \calO(\omega^3)\,,
\end{equation}
from which we can read off (again suppressing the $A$-dependence) 
\begin{equation} \label{eq:nu_finite_1}
\nu^\1(e^{i\lambda_0},e^{i\omega(x)}) = - \frac{\kappa}{2}\left( M(\lambda_0)\, B(-\lambda_0)\, M(-\lambda_0) - B(0) \right)_{ab} \, \omega^a \wedge d\omega^b   + \calO(\omega^3)\,. 
\end{equation}
Using the specific trivialization in Eq.~\eqref{eq:BWZ}, which in particular satisfies $B_{ab}(0) = 0$,
\begin{equation} \label{eq:nu_finite_2}
\nu^\1(e^{i\lambda_0},e^{i\omega(x)}) = - \frac{\kappa}{8\pi} \left( \frac{\sinh X(\lambda_0)- X(\lambda_0)}{\cosh X(\lambda_0) - 1} \right)^{ab}\, \omega^a \wedge d\omega^b + \calO(\omega^3)\,.
\end{equation}

\section{Relating the 2-Point and 3-Point Response Functions}
\label{app:2to3pt} 

In this appendix, we show that  
\begin{multline}
\left.\frac{\partial}{\partial u_0^a} \, (e^{X(u_0)})^{a_1b_1}\, (e^{X(u_0)})^{a_2b_2}\, G^{b_1b_2}(U_0; t_1,t_2)\right|_{u_0=0} \\
= -f^{a_1ba}\, G^{ba_2}(\id; t_1,t_2) - f^{a_2ba}\, G^{a_1b}(\id; t_1,t_2) + \left.\frac{\partial}{\partial u_0^a} \, G^{a_1a_2}(e^{iu_0}; t_1,t_2)\right|_{u_0=0} ~.
\end{multline}
We start with the $u_0$ derivative of $\log Z$, 
\begin{multline}
\left.\frac{\partial}{\partial u_0^a} \, \log Z_\scrL[e^{i\omega(t)}\, e^{i u_0}]\right|_{u_0=0} = \left.\int dt\, \frac{\partial \widetilde{\omega}^b(t)}{\partial u_0^a}\frac{\delta}{\delta \widetilde{\omega}^b(t)} \, \log Z_\scrL[e^{i\widetilde{\omega}(t)}]\right|_{u_0=0} \\
 = \int dt\, \left.\frac{\partial \widetilde{\omega}^b(t)}{\partial u_0^a} \right|_{u_0=0}\, \frac{\delta}{\delta \omega^b(t)} \, \log Z_\scrL[e^{i\omega(t)}] \,. 
\end{multline}
Using the BCH formula 
\begin{equation}
\widetilde \omega^b = \omega^b + M^{bc}(\omega)\, u_0^c  \, +\,  O(u_0^2) \  \implies \  \left.\frac{\partial \widetilde{\omega}^b(t)}{\partial u_0^a} \right|_{u_0=0} =   M^{ab}(-\omega)\,,
\end{equation}
where $M^{ab}(\omega) = \left(\frac{X(\omega)}{e^{X(\omega)}-\id}\right)^{ab}$ and $X^{ab}(\omega) = - f^{abc}\, \omega^c$. Then, we obtain an equation describing how a global change on the moduli space is implemented through the local modulation, 
\begin{equation}
\left.\frac{\partial}{\partial u_0^a} \, \log Z_\scrL[e^{i\omega(t)}\, e^{i u_0}]\right|_{u_0=0} =  \int dt\, M^{ab}(-\omega)\, \frac{\delta}{\delta \omega^b(t)} \, \log Z_\scrL[e^{i\omega(t)}]\,.
\end{equation}
Now, we would like to take functional derivatives with respect to $\omega$ to study how the response functions depend infinitesimally on the position on the moduli space. Taking one functional derivative gives two terms 
\begin{multline}
\frac{\delta}{\delta \omega^{a_1}(t_1)}\left.\frac{\partial}{\partial u_0^a} \, \log Z_\scrL[e^{i\omega(t)}\, e^{i u_0}]\right|_{u_0=0} = \int dt\, \frac{\delta M^{ab}(-\omega)}{\delta \omega^{a_1}(t_1)}\, \frac{\delta}{\delta \omega^b(t)} \, \log Z_\scrL[e^{i\omega(t)}] \\
+ \int dt\, M^{ab}(-\omega)\,\frac{\delta}{\delta \omega^{a_1}(t_1)} \frac{\delta}{\delta \omega^b(t)} \, \log Z_\scrL[e^{i\omega(t)}]\,,
\end{multline}
and taking a second functional derivative gives
\begin{multline}
\frac{\delta}{\delta \omega^{a_2}(t_2)}\frac{\delta}{\delta \omega^{a_1}(t_1)}\left.\frac{\partial}{\partial u_0^a} \, \log Z_\scrL[e^{i\omega(t)}\, e^{i u_0}]\right|_{u_0=0} \\
= \int dt\,\frac{\delta^2 M^{ab}(-\omega)}{\delta \omega^{a_2}(t_2)\, \delta \omega^{a_1}(t_1)}\, \frac{\delta}{\delta \omega^b(t)} \, \log Z_\scrL[e^{i\omega(t)}] \\
+ \int dt\, \frac{\delta M^{ab}(-\omega)}{\delta \omega^{a_2}(t_2)}\,\frac{\delta}{\delta \omega^{a_1}(t_1)} \frac{\delta}{\delta \omega^b(t)} \, \log Z_\scrL[e^{i\omega(t)}] \\
+ \int dt\, \frac{\delta M^{ab}(-\omega)}{\delta \omega^{a_1}(t_1)}\, \frac{\delta}{\delta \omega^{a_2}(t_2)} \frac{\delta}{\delta \omega^b(t)} \, \log Z_\scrL[e^{i\omega(t)}] \\
+ \int dt\, M^{ab}(-\omega)\,\frac{\delta}{\delta \omega^{a_2}(t_2)} \frac{\delta}{\delta \omega^{a_1}(t_1)} \frac{\delta}{\delta \omega^b(t)} \, \log Z_\scrL[e^{i\omega(t)}]\,,
\end{multline}
Since we are interested in setting $\omega=0$ at the end, we can simply expand 
\begin{equation}
M^{ab}(-\omega) = \delta^{ab} - \frac{1}{2} f^{abc}\, \omega^c + \frac{1}{12} f^{acd}\, f^{cbe}\, \omega^d \omega^e + \calO(\omega^3) 
\end{equation}
and get
\begin{multline}
\left.\frac{\partial}{\partial u_0^a} \, G^{a_1a_2}(e^{iu_0}; t_1,t_2)\right|_{u_0=0}  = -i \int dt\, G^{a_1a_2a}(\id; t_1,t_2,t) \\
 +\frac{1}{2}  \left( f^{aba_2} \,G^{a_1b}(\id;t_1,t_2) +  f^{aba_1} \,G^{a_2b}(\id;t_2,t_1)\right) \\
 +\frac{i}{12}(f^{aca_1}f^{cba_2}+f^{aca_2}f^{cba_1}) \delta(t_1-t_2)\,  G^b(\id;t_1) 
\,.
\end{multline}
Putting the pieces together, we obtain
\begin{multline} \label{eq:d_of_2pt}
\left.\frac{\partial}{\partial u_0^a} \, (e^{X(u_0)})^{a_1b_1}\, (e^{X(u_0)})^{a_2b_2}\, G^{b_1b_2}(U_0; t_1,t_2)\right|_{u_0=0}= -i \int dt\, G^{a_1a_2a}(\id; t_1,t_2,t) \\
 -\frac{1}{2}  \left( f^{aba_2} \,G^{a_1b}(\id;t_1,t_2) +  f^{aba_1} \,G^{a_2b}(\id;t_2,t_1)\right) \\
  +\frac{i}{12}(f^{aca_1}f^{cba_2}+f^{aca_2}f^{cba_1}) \delta(t_1-t_2)\,  G^b(\id;t_1) 
\end{multline}
which we use in deriving Eq.~\eqref{eq:integrated_3pt}.

\section{Partial Contact Terms} \label{app:partial_contact}

We discuss whether partial contact terms can saturate the anomalous Ward identity in Eq.~\eqref{eq:contracted_ward_id}. To match the anomaly we consider terms $G^{a_1a_2a_3}(\id; t_1,t_2,t_3) = f^{a_1 a_2 a_3} \, \mathsf{F}(t_{12},t_{23})$ where $t_{ij} = t_i-t_j$. Without loss of generality, partial contact terms in $\mathsf{F}$ take the form
\begin{equation}
\mathsf{F}(t_{12},t_{23}) = \sum_n  \delta^{(n)}(t_{12}) \, \mathsf{K}_n(t_{23}) + \text{signed permutations} ~,  
\end{equation}
where $\delta^{(n)}$ is the $n$th derivative of the delta function. For $n=0$ the sum over permutations vanishes, so we just need to consider $n \ge 1$. In momentum space, a term with fixed $n$ translates to
\begin{equation}
\widehat{\mathsf{F}}_n(E_1,E_2,E_3) = (-i)^n \left[ \left( E_3^n-E_2^n \right) \, \widehat{\mathsf{K}}_n(E_1) + \left( E_1^n - E_3^n \right) \, \widehat{\mathsf{K}}_n(E_2) + \left(E_2^n - E_1^n \right) \, \widehat{\mathsf{K}}_n(E_3) \right].
\end{equation} 
Let us assume that $\widehat{\mathsf{K}}_n$ is continuous at $E=0$. Then the Ward identity becomes 
\begin{equation} \label{eq:partial_contact_ward_id} 
(iE)^{n} \left[ \widehat{\mathsf{K}}_n(E) - (-1)^n\, \widehat{\mathsf{K}}_n(-E) - \left(1 - (-1)^n \right) \, \widehat{\mathsf{K}}_n(0) \right] = - \frac{i\kappa}{24\pi}E \,. 
\end{equation}
We reach a contradiction upon taking $E \to 0$. To match the Ward identity we need $\widehat{\mathsf{K}}_n(E) \sim E^{1-n}$, which renders the $E \to 0$ limit ill-defined. This issue can be resolved by interpreting Eq.~\eqref{eq:partial_contact_ward_id} in a distributional sense. We introduce a regulator $\epsilon$,
\begin{equation}
\widehat{\mathsf{K}}_n(E) = \frac{\mathsf{c}_n}{(E - i \epsilon)^{n-1}}
\end{equation}
and perform all integrals before taking $\epsilon \to 0$. Any $n>2$ leads to a position space $\mathsf{K}_n(t)$ with unphysical growing long-distance correlations. We are left with the $n=2$ case, for which the Ward identity reads 
\begin{equation}
(iE)^{2} \left( \widehat{\mathsf{K}}_2(E) - \widehat{\mathsf{K}}_2(-E) \right)  =  (iE)^{2} \frac{2\mathsf{c}_2\, E}{E^2 +\epsilon^2}  = - \frac{i\kappa}{24\pi}E \,. 
\end{equation}
Choosing $\mathsf{c}_2 = \frac{i\kappa}{48\pi}$, we conclude that a partial contact term with 
\begin{equation}
\widehat{\mathsf{K}}_2(E) = \frac{i\kappa}{48\pi} \frac{1}{E - i \epsilon} = \frac{i\kappa}{48\pi}\left[ \text{PV}\left( \frac{1}{E}\right) + i\pi \delta(E) \right] \,,
\end{equation}
can saturate the Ward identity.\footnote{This is simply the propagator for a fermionic oscillator with frequency $\epsilon$. Note that we are free to add any even function of $E$ to this ansatz without spoiling the Ward identity. }  In position space, this ansatz corresponds to 
\begin{equation}
\mathsf{F}(t_{12},t_{23}) = \frac{i\kappa}{48\pi} \delta^{''}\!(t_{12}) \, \Theta(t_{23}) \, e^{-\epsilon\, t_{23}} + \text{signed permutations} ~.
\end{equation}

\section{Relation to BCFT Analysis}\label{app:hbp}

The fact that anomalous Ward identities can lead to a higher Berry curvature in the boundary parameter space was discussed in \cite{Choi:2025ebk, Wen:2025xka, Lo:2026sum}. For instance, in the context of 2d BCFT, the 3-point functions of the lightest boundary-condition-changing (bcc) operators can be used to construct a 2-form connection on the space of conformal boundary conditions \cite{Choi:2025ebk}. Here, we review the necessary ingredients from \cite{Choi:2025ebk} as well as some modifications to show how a similar structure induced by anomaly inflow can be extracted in generic $2d$ QFTs with boundary. The result can be immediately applied to the case of line operators discussed in the main text.

\

Consider a $2d$ BCFT with a boundary conformal manifold $\mathcal{M}$. For any two points $\alpha,\beta \in \mathcal{M}$, there are boundary condition changing operators $O_{\alpha\beta}$ that connect the boundary conditions. Under state-operator correspondence, the space of bcc operators is equivalent to the Hilbert space on an interval with boundary conditions $\alpha,\beta$ on two sides. We denote the lightest bcc operator as $\psi_{\alpha\beta}$. When $\beta$ is sufficiently close to $\alpha$, the operator $\psi_{\alpha\beta}$ is always unique (up to an overall phase). If we place the $2d$ BCFT on a disk, the 2-point function $\langle \psi_{\alpha\beta}(\theta_1) \psi_{\beta \alpha}(\theta_2) \rangle$ of boundary insertions is fixed by conformal symmetry to be
\begin{equation}
    \langle \psi_{\alpha\beta}(\theta_1) \psi_{\beta\alpha}(\theta_2) \rangle = g_M \left[2\sin\left(\frac{\theta_2 - \theta_1}{2}\right)\right]^{-2\Delta_{\alpha\beta}} ~,
\end{equation}
where $g_{M}$ is the $g$-function of the boundary conditions. This normalization does not fix the phase ambiguity in the bcc operators
\begin{equation}\label{eq:psixs}
    \psi_{\alpha\beta}(x) \mapsto e^{i\Lambda(\alpha,\beta)} \psi_{\alpha\beta}(x) ~, \quad e^{i\Lambda(\alpha,\beta)} = e^{-i\Lambda(\beta,\alpha)} \in U(1) ~.
\end{equation}

To define the 2-connection on $\mathcal{M}$, consider the 3-point function of bcc operators
\begin{equation}\label{eq:bcft3pt}
\begin{aligned}
    & \langle \psi_{\alpha\beta}(\theta_1) \psi_{\beta\gamma}(\theta_2) \psi_{\gamma\alpha}(\theta_3) \rangle \\
    & \quad \quad \quad \quad = c(\alpha,\beta,\gamma) g_M \left[2\sin\left(\frac{\theta_{21}}{2}\right)\right]^{-\Delta_{\alpha\beta\gamma}} \left[2\sin\left(\frac{\theta_{32}}{2}\right)\right]^{-\Delta_{\beta\gamma\alpha}} \left[2\sin\left(\frac{\theta_{13}}{2}\right)\right]^{-\Delta_{\gamma\alpha\beta}}
\end{aligned}
\end{equation}
where $\Delta_{\alpha\beta\gamma} = \Delta_{\alpha\beta} + \Delta_{\beta\gamma} - \Delta_{\gamma\alpha}$ and $\theta_{ij} = \theta_i - \theta_j$. The 3-point function is fixed by conformal symmetries up to an overall complex constant $c(\alpha,\beta,\gamma)$, which satisfies
\begin{equation}
    c(\alpha,\beta,\gamma) = c(\beta,\gamma,\alpha) = c(\gamma,\alpha,\beta) ~, \quad c(\alpha,\alpha,\alpha) = 1 ~.
\end{equation}
It is pointed out in \cite{Choi:2025ebk} that the phase $e^{i\phi(\alpha,\beta,\gamma)}$ of $c(\alpha,\beta,\gamma)$ defines a 2-form connection on $\mathcal{M}$
\footnote{Geometrically, this can be understood as taking a function $\mathcal{F}(\alpha,\beta,\gamma)$ on $\mathcal{M}_\alpha\times \mathcal{M}_\beta\times \mathcal{M}_\gamma$, then $d_\beta \wedge d_\gamma \mathcal{F}$ defines a 2-form on $\mathcal{M}_\alpha \times \mathcal{M}_{\beta} \times \mathcal{M}_\gamma$ (where $d_\beta,d_\gamma$ are exterior derivatives on $\mathcal{M}_\beta$ and $\mathcal{M}_\gamma$ respectively). Pulling back this 2-form along the diagonal embedding map $\mathcal{M}_\alpha \hookrightarrow \mathcal{M}_\alpha\times \mathcal{M}_\beta\times \mathcal{M}_\gamma$ leads to a 2-form on $\mathcal{M}_\alpha$.}
\begin{equation}\label{eq:M2f}
    \mathcal{B}^{(2)} = \frac{1}{2} B_{ij}(\alpha) d\alpha^i \wedge d\alpha^j \equiv  \frac{1}{2} \left[\frac{\partial^2 \phi(\alpha,\beta,\gamma)}{\partial \beta^i \partial \gamma^j} - \frac{\partial^2 \phi(\alpha,\beta,\gamma)}{\partial \beta^j \partial \gamma^i}\right]_{\beta = \gamma = \alpha} d\alpha^i \wedge d\alpha^j ~,
\end{equation}
The phase ambiguity \eqref{eq:psixs} implements a gauge transformation of $\mathcal{B}^{(2)}$ via
\begin{equation}
    \mathcal{B}^{(2)} \mapsto \mathcal{B}^{(2)} + d\Lambda^{(1)} ~, 
\end{equation}
where $\Lambda^{(1)} \equiv \frac{\partial \Lambda(\alpha,\beta)}{\partial \beta^i}\Big|_{\beta = \alpha} d\alpha^i$.
And the gauge invariant 3-form curvature on $\mathcal{M}$ is given by:
\begin{equation}
    \mathcal{H}^{(3)} = d\mathcal{B}^{(2)} = \frac{1}{3!} \mathcal {H}_{ijk}(\alpha) d\alpha^i \wedge d\alpha^j \wedge d\alpha^k ~.
\end{equation} 

To connect this to our construction, let us restrict ourselves to the case where the boundary conformal manifold $\mathcal{M}$ is the broken $0$-form symmetry group $G$. Furthermore, the definition of $\mathcal{B}^{(2)}$ in \eqref{eq:M2f} remains unchanged if we replace $\phi(\alpha,\beta,\gamma)$ by $\mathrm{Im} \left(\log \langle \psi_{\alpha\beta}(\theta_1) \psi_{\beta\gamma}(\theta_2) \psi_{\gamma\alpha}(\theta_3) \rangle\right)$. To see this, one only needs to show
\begin{equation}
    \left(\frac{\partial^2 \Delta_{\beta\gamma}}{\partial \beta^a \partial \gamma^b} - \frac{\partial^2 \Delta_{\beta\gamma}}{\partial \beta^b \partial \gamma^ a}\right)\Bigg|_{\beta = \gamma = \alpha} d\alpha^a \wedge d\alpha^b = 0 ~.
\end{equation}
This can be done by first rewriting the above as 
\begin{equation}\label{eq:rewrite}
    \left(\frac{\partial^2 \Delta_{\beta\gamma}}{\partial \beta^a \partial \gamma^b} - \frac{\partial^2 \Delta_{\beta\gamma}}{\partial \beta^b \partial \gamma^ a}\right)\Bigg|_{\beta = \gamma = \alpha} d\alpha^a \wedge d\alpha^b = \left(\mathcal{X}^a_{(\beta)} \mathcal{X}^b_{(\gamma)} \Delta_{\beta\gamma} - \mathcal{X}^b_{(\beta)} \mathcal{X}^a_{(\gamma)} \Delta_{\beta\gamma}\right)\bigg|_{\beta=\gamma = \alpha} \theta_a \wedge \theta_b ~, 
\end{equation}
where $\displaystyle \theta_a = i \left(\frac{e^{X(\alpha)}-1}{X(\alpha)}\right)_{ab} d\alpha^b$ are components of the left-invariant 1-form on $G^\0$ and $\displaystyle \mathcal{X}^a_{(\alpha)}:= -i \left(\frac{X(\alpha)}{1-e^{-X(\alpha)}}\right)_{ab} \frac{\partial}{\partial \alpha^b}$ are components of its dual vector field. Next, notice that $\Delta_{\beta\gamma}$ must be invariant under the $G^\0$-action, which means it only depends on $u$ where $e^{iu} = e^{-i\gamma} e^{i\beta}$. Furthermore, because $\Delta_{\beta\gamma}$ has a global minimum at $u = 0$ with $\Delta_{\beta\gamma}(0) = 0$, its expansion near $u = 0$ is given by
\begin{equation}
    \Delta_{\beta\gamma}(u) = \frac{1}{2}\kappa^{ab} u^a u^b + O(u^3) 
\end{equation}
for some symmetric, positive definite $\kappa_{ab}$. $\mathcal{X}^a_{(\beta)} \mathcal{X}^b_{(\gamma)} \Delta_{\beta\gamma}\big|_{\beta=\gamma = \alpha}$ can therefore be evaluated using the chain rule and the identities  
\begin{equation}
    \mathcal{X}^a_{(\beta)}(u^b) = - i \left(\frac{X(u)}{1- e^{-X(u)}}\right)_{ab} ~, \quad \mathcal{X}^a_{(\gamma)}(u^b) = -i \left(\frac{X(u)}{e^{X(u)}-1}\right)_{ab}~,
\end{equation}
to find
\begin{equation}
    \mathcal{X}^a_{(\beta)} \mathcal{X}^b_{(\gamma)} \Delta_{\beta\gamma}\big|_{\beta=\gamma = \alpha} = -  \kappa_{ab} ~,
\end{equation}
which vanishes upon anti-symmetrizing the $a,b$-indices as in \eqref{eq:rewrite}. This completes the proof for any point $\alpha$ along the space $G$. 

\

In the following, we generalize the above construction to a generic 2d QFT, which can then be directly applied to the line operators discussed in the main text. Without conformal symmetry, the 3-pt function will not take the form \eqref{eq:bcft3pt}. However, motivated by the above discussion, we can directly define a 2-form  $\widetilde{\mathcal{B}}^\2$ from the QFT analog of the BCFT 3-pt function, and use the relation \eqref{eq:2pt_response_final} from the anomaly inflow to separate the anomaly contribution from the non-universal piece (which depends on details of QFT dynamics). We show that the anomaly contribution is topologically non-trivial and leads to a field strength with flux, while the non-universal piece is a globally well-defined 2-form on $G$ and therefore topologically trivial. 

First, the BCFT 3-point function $\langle \psi_{\alpha\beta}\psi_{\beta\gamma}\psi_{\gamma\alpha}\rangle$ is equivalent to the following modulation of the boundary condition $U(t) = e^{i\widetilde{\omega}(t)}$ with
\begin{equation}
    \widetilde{\omega}^a(t) = \alpha^a + (\beta - \alpha)^a \Theta_{[t_1,t_2]}(t) + (\gamma - \alpha)^a \Theta_{[t_2,t_3]}(t) ~,
    \quad 
    \Theta_{[t_i,t_j]}(t) = \begin{cases} 1 \quad t \in [t_i,t_j] ~, \\ 0 \quad \text{otherwise} ~. \end{cases}
\end{equation}
The partition function $Z[e^{i\widetilde{w}(t)}]$ computes the QFT analog of the BCFT 3-point function:
\begin{equation}
Z[e^{i\widetilde{\omega}(t)}] = \begin{tikzpicture}[baseline={(0,-0.25)}]
    \draw[line width = 0.4mm, black] (0,0) circle (1.5);
    \fill[] (1.29904, -0.75) circle (2pt);
    \node[right] at (1.29904, -0.75) {$t_1$};
    \fill[] (-1.29904, -0.75) circle (2pt);
    \node[left] at (-1.29904, -0.75) {$t_3$};
    \fill[] (0,1.5) circle (2pt);
    \node[above] at (0,1.5) {$t_2$};

    \node[left] at (-1.29904, 0.75) {$e^{i\gamma}$};
    \node[right] at (1.29904, 0.75) {$e^{i\beta}$};
    \node[below] at (0,-1.5) {$e^{i\alpha}$};
\end{tikzpicture} \sim \langle \psi_{\alpha\beta}(t_1) \psi_{\beta\gamma}(t_2) \psi_{\gamma\alpha}(t_3) \rangle ~.
\end{equation}
As prescribed above, we want to extract the 2-form 
\begin{equation}\label{eq:2ptfromZ}
    \widetilde{\mathcal{B}}^{(2)} = - \frac{i}{2}\left(\left(\frac{\partial^2}{\partial \beta^a \partial \gamma^b} - \frac{\partial^2}{\partial \beta^b \partial \gamma^a}\right)_{\beta=\gamma = \alpha} \log Z[e^{i\widetilde{\omega}(t)}]\right) d\alpha^a \wedge d\alpha^b ~,
\end{equation}
and we start by rewriting $e^{i\widetilde{\omega}(t)} = e^{-i\omega(t)} e^{i\alpha}$ and find
\begin{equation}\label{eq:Zep}
\begin{aligned}
    \log Z[e^{i\widetilde{\omega}(t)}] = \log Z[e^{-i\omega(t)} e^{i\alpha}] = & \log Z[e^{i\alpha}] - i \int dt_1' \, \omega^{a_1}(t_1') G^{a_1}(e^{i\alpha};t_1') \\ 
    & \quad \quad \quad - \frac{1}{2} \int dt_1' dt_2' \, w^{a_1}(t_1') \omega^{a_2}(t_2') G^{a_1 a_2}(e^{i\alpha};t_1',t_2') + \cdots ~,
\end{aligned}
\end{equation}
where $\omega(t)$ can be expressed in terms of $\alpha, (\beta-\alpha), (\gamma-\alpha)$ using the BCH formula:
\begin{equation}
    \omega^a(t) = -\left(\frac{1 - e^{-X(\alpha)}}{X(\alpha)}\right)_{ab} (\beta - \alpha)^b \Theta_{[t_1,t_2]}(t) - \left(\frac{1 - e^{-X(\alpha)}}{X(\alpha)}\right)_{ab} (\gamma - \alpha)^b \Theta_{[t_2,t_3]}(t) + \cdots  ~.
\end{equation}
It is convenient to compute $\widetilde{\mathcal{B}}^{(2)}$ using $\mathcal{X}^a$ and $\theta_a$ in \eqref{eq:rewrite},
\begin{equation}\label{eq:d2logZ}
\begin{aligned}
    \widetilde{\mathcal{B}}^{\2} = & -\frac{i}{2}\left(\left[\mathcal{X}^a_{(\beta)}\mathcal{X}^b_{(\gamma)} - \mathcal{X}^b_{(\beta)}\mathcal{X}^a_{(\gamma)}\right]_{\beta=\gamma = \alpha} \log Z[e^{i\widetilde{\omega}(t)}] \right) \theta^a \wedge \theta^b\\
    = & -\frac{i}{2}\left((e^{X(\alpha)})_{[a|a_1} (e^{X(\alpha)})_{|b]a_2} \int dt_1' dt_2' \Theta_{[t_1,t_2]}(t_1')\Theta_{[t_2,t_3]}(t_2') G^{a_1 a_2}(e^{i\alpha};t_1',t_2')\right) \theta^a \wedge \theta^b  \\
    = & \underbrace{-\frac{i}{2} \int dt_1' dt_2' \Theta_{[t_1,t_2]}(t_1') \Theta_{[t_2,t_3]}(t_2') G^{[a_1 a_2]}(\id;t_1',t_2')}_{\equiv \Omega_{ab}(t_1,t_2,t_3)} \theta^a \wedge \theta^b \\
    & \quad \quad +\underbrace{\frac{\kappa}{8\pi}\left( \frac{\sinh X(\alpha)- X(\alpha)}{\cosh X(\alpha) - \id } \right)^{a_1 a_2} \int dt_1' dt_2' \, \Theta_{[t_1,t_2]}(t_1') \Theta_{[t_2,t_3]}(t_2') \partial_{t_1'}\delta(t_1'-t_2')}_{\textit{anomaly inflow}} \theta^a \wedge \theta^b ~,
\end{aligned}
\end{equation}
where we used \eqref{eq:2pt_response_final} to separate the anomalous contribution to $\widetilde{\mathcal{B}}^\2$ from non-universal terms. The integral in the second term can be evaluated by choosing a regularization of $\Theta_{[t_1,t_2]}(t)$ that fulfils \footnote{One can construct a regularization of  $\Theta_{[t_1,t_2]}^\epsilon(t) = \theta^\epsilon(t-t_1) - \theta^\epsilon(t-t_2)$ from any regularization $\theta^\epsilon(t)$ of the Heaviside theta function $\theta(t)$. Additionally, we would require $\theta^{\epsilon}(t) = 1 - \theta^\epsilon(-t)$, which ensures that two sides of the jump are treated symmetrically, as well as $\Theta^\epsilon_{[t_1,t_2]}(t) = \Theta^\epsilon_{[t_1,t_2]}(t_1+t_2-t)$. Then,
\begin{equation}
\begin{aligned}
    & \lim_{\epsilon\rightarrow 0^+} \int dt_1' dt_2' \, \Theta_{[t_1,t_2]}^\epsilon(t_1') \Theta^\epsilon_{[t_2,t_3]}(t_2') \partial_{t_1'}\delta(t_1'-t_2') \\
    =& - \lim_{\epsilon\rightarrow 0^+}  \int dt'\, \Theta^\epsilon _{[t_2,t_3]}(t')\partial_{t'}\Theta^\epsilon_{[t_1,t_2]}(t') \\
    =& \lim_{\epsilon\rightarrow 0^+}  \int dt' \, \theta^\epsilon(t'-t_2)\partial_{t'}\theta^\epsilon(t'-t_2) = \frac{1}{2} ~.
\end{aligned}
\end{equation}
}
\begin{equation}
    \int dt_1' dt_2' \, \Theta_{[t_1,t_2]}(t_1') \Theta_{[t_2,t_3]}(t_2') \partial_{t_1'}\delta(t_1'-t_2') = \frac{1}{2} ~.
\end{equation}
We find the 2-form $\widetilde{\mathcal{B}}^\2$ as
\begin{equation}
    \widetilde{\mathcal{B}}^\2 = \Omega_{a b}(t_1,t_2,t_3) \theta^a \wedge \theta^b + \kappa\, \mathcal{B}_{\mathrm{WZ}}^\2 ~.
\end{equation}
The first term is non-universal and depends on details of the QFT dynamics. For fixed and separated $t_i$'s, the coefficient $\Omega_{ab}$ is independent of $G$ and therefore it is a globally well-defined 2-form on $G$ that leads to a topologically trivial 3-form field strength with no flux. The second term is fixed by the anomaly inflow, and leads to a topologically non-trivial 3-form field strength. The flux is determined by the anomaly coefficient. The invariant field strength $\widetilde{\mathcal{H}}^{(3)}$ on the moduli space can be computed using the Maurer–Cartan equation
\begin{equation}
    \widetilde{\mathcal{H}}^{(3)} = (f^{[ab|d}\Omega_{d|c]}) \theta^a \wedge \theta^b \wedge \theta^c + \underbrace{\frac{\kappa}{24\pi} f^{abc} \theta^a \wedge \theta^b \wedge \theta^c}_{\kappa\, \mathcal{H}_{\mathrm{WZ}}^{(3)}} ~,
\end{equation}

and the non-universal piece can be projected out by contracting with $f^{abc}$ via $f_{abc} f^{[ab|d}\Omega_{d|c]} = 0$. 

\section{Symmetry Breaking and Tilts in Massless QED} 
\label{app:SymmetryBreakingMasslessQED}
In this appendix, we follow \cite{vanBeest:2023dbu, Aharony:2023amq} to derive that the 't Hooft line breaks part of the flavor symmetry in massless $\text{QED}_4$. To render the Hamiltonian of the fermion $s$-wave modes in a Dirac monopole background self-adjoint, it is necessary to impose boundary conditions. These break the flavor symmetry explicitly. We implement them by a boundary action for the fermions and use it to derive the tilt operator.

\

Consider a single Weyl fermion of charge $q_e$ in a Dirac monopole background:
\begin{equation}
    \mathscr{L} = -i \Bar{\psi}\Bar{\sigma}^\mu(\partial_\mu -iqa_\mu^M)\psi ~.
\end{equation}
For this analysis, we ignore all gauge field fluctuations.
We use Wess-Bagger conventions with:
\begin{equation}
    \eta_{\mu\nu} = \text{diag}(-,+,+,+), \quad \sigma^\mu = (-\id, \Vec{\sigma}), \quad \Bar{\sigma}^\mu = (-\id, -\Vec{\sigma})
\end{equation}
and use spherical coordinates to leverage the rotation symmetry preserved by the monopole background
\begin{equation}
    a^M = \frac{q_M}{2}(1-\cos(\theta))d\varphi ~,
\end{equation}
with the Pauli matrices in spherical coordinates:
\begingroup
\small
\setlength{\arraycolsep}{3pt}
\renewcommand{\arraystretch}{0.9}

\begin{equation}
\sigma^r =
\begin{pmatrix}
    \cos\theta & e^{-i\varphi}\sin\theta \\
    e^{i\varphi}\sin\theta & -\cos\theta
\end{pmatrix},
\quad
\sigma^\theta =
\begin{pmatrix}
    -\sin\theta & e^{-i\varphi}\cos\theta \\
    e^{i\varphi}\cos\theta & \sin\theta
\end{pmatrix},
\quad
\sigma^\varphi =
\begin{pmatrix}
    0 & -i e^{-i\varphi} \\
    i e^{i\varphi} & 0
\end{pmatrix}.
\end{equation}

\endgroup
The differential operator acting on a left-handed Weyl fermion can be written as:
\begin{equation}
    -i\Bar{\sigma}^\mu D_\mu \psi = i\left(D_t + \vec{\sigma}\cdot \vec{D}\right)\psi =0\iff i\partial_t \psi = -i\vec{\sigma}\cdot \vec{D}\psi ~.
\end{equation}
Due to time-translation invariance, this reduces to an eigenvalue problem of the operator $H = -i\vec{\sigma}\cdot \vec{D}$. As usual, we regulate this by excising a small ball of radius $\epsilon$ around the monopole. The path integral is well-defined if $H$ is self-adjoint with respect to the inner product:
\begin{equation}
    \langle\chi, \psi\rangle = \int_{\mathbb{R}^3/B_{\epsilon}(0)}d^3x\, \Bar{\chi}_{\Dot{\alpha}}\delta^{\Dot{\alpha}\alpha}\psi_\alpha ~.
\end{equation}
The self-adjointness of $H$ requires
\begin{equation}
    \langle\chi, H\psi\rangle - \langle H\chi, \psi\rangle = \lim_{\epsilon \rightarrow 0^+} i \epsilon^2 \int_{S^2}d\Omega\, \Bar{\chi}_{\Dot{\alpha}}(\sigma^r)^{\Dot{\alpha}\alpha}\psi_\alpha =0 ~,
\end{equation}
which holds if the wavefunctions behave like:
\begin{equation}
    \lim_{r\to\epsilon}\psi_\alpha(r) = \frac{c_\alpha}{\epsilon^{1-a}}, \quad a>0 ~.
\end{equation}
The fermion modes can be expanded in terms of monopole harmonics $\Omega_\alpha^{jm}$. For a monopole of charge $q_M$ and a fermion of charge $q_e$, the preserved angular momentum $j$ is given by
\begin{equation}
    j = \frac{|q_eq_M|}{2}-\frac{1}{2} + n \geq 0~,
\end{equation}
where $n\in\mathbb{Z}$ is a mode label, and the solutions take the form
\begin{equation}
    \psi_\alpha = \frac{1}{r}\sum _{j,m}u_j(r) \Omega_\alpha^{jm}(\theta, \varphi)~,
\end{equation}
where the radial function around the origin behaves like:
\begin{equation}
    u_j(r)\sim r^{\pm\lambda_j},\quad \lambda_j = \sqrt{\left(j + \frac{1}{2}\right)^2 - \left(\frac{|q_eq_M|}{2}\right)^2} ~.
\end{equation}
Taking $|q_M| = |q_e|=1$, we see that the radial functions with $n>0$ become a constant as $r\to0$ such that only the $n=j=0$ modes pose a problem to the self-adjointness of the operator. The monopole harmonics for $q_e=\pm1$ are given by:
\begin{equation}
    \Omega_{\alpha, +}=
    \begin{pmatrix}
        \cos\left(\frac{\theta}{2}\right)\\
        e^{i\varphi}\sin\left(\frac{\theta}{2}\right)
    \end{pmatrix} ~,
    \quad
    \Omega_{\alpha, -}=
    \begin{pmatrix}
        -e^{-i\varphi}\sin\left(\frac{\theta}{2}\right)\\
        \cos\left(\frac{\theta}{2}\right)
    \end{pmatrix} ~,
\end{equation}
and we can expand a Weyl-fermion with $q_e=1$ as:
\begin{equation}
    \psi_\alpha = \frac{\eta(t,r)}{\sqrt{4\pi}r}\Omega_{\alpha, +}(\theta, \varphi) + \,\,\left(\, j>0 \text{ modes}\, \right)~.
\end{equation}
Let us look at the simplest example that only has the 2-group and no other 't Hooft or ABJ-anomalies with four Weyl fermions:
\begin{equation}
    \begin{array}{c|cccc}
         & \psi^1 & \psi^2 & \widetilde{\psi}^{\Tilde{1}} & \widetilde{\psi}^{\Tilde{2}} \\
        \cline{1-5}
        U(1)^{\text{Gauge}}_C & +1 & +1 & -1 & -1 \\
        U(1)^\0_{\widetilde{A}} & +1 & -1 & 0 & 0 
    \end{array} ~.
\end{equation}
This can be embedded in $N_f=2$ massless QED where $U(1)_{\widetilde{A}}\subset SU(2)_L$ is the Cartan of the left action.
The Weyl fermions can now be expanded as:
\begin{equation}
    \begin{split}
        \psi^1_{\alpha} &= \frac{\eta^1(t,r)}{\sqrt{4\pi}r}\Omega_{\alpha, +}(\theta, \varphi) +\dots , \quad \psi^2_{\alpha} = \frac{\eta^2(t,r)}{\sqrt{4\pi}r}\Omega_{\alpha, +}(\theta, \varphi) +\dots ~,\\
        \widetilde{\psi}^{\Tilde{1}}_{\alpha} &= \frac{\widetilde{\rho}^{\Tilde{1}}(t,r)}{\sqrt{4\pi}r}\Omega_{\alpha, -}(\theta, \varphi) +\dots , \quad \widetilde{\psi}^{\Tilde{2}}_{\alpha} = \frac{\widetilde{\rho}^{\Tilde{2}}(t,r)}{\sqrt{4\pi}r}\Omega_{\alpha, -}(\theta, \varphi) +\dots ~.
    \end{split}
\end{equation}
We can make the flavor symmetries manifest by writing
\begin{equation}
    \eta^i = 
    \begin{pmatrix}
        \eta^1\\
        \eta^2
    \end{pmatrix}, 
    \quad
    \widetilde{\rho}^{\widetilde{i}} = 
    \begin{pmatrix}
        \widetilde{\rho}^{\Tilde{1}}\\
        \widetilde{\rho}^{\Tilde{2}}
    \end{pmatrix} ~.
\end{equation}
Since $\Omega_\pm$ have opposite eigenvalues under $\sigma^r$, the self-adjointness condition becomes:
\begin{equation}
    0 = i\left(\eta^\dag \eta' - \widetilde{\rho}^\dag \widetilde{\rho}'\right)\bigg\rvert_{r=0} ~.
\end{equation}
This can obviously be fulfilled by:
\begin{equation}
    \eta^i(t,0) = U^i_{\,\,\widetilde{j}} \rho^{\widetilde{j}}(t,0), \quad U\in SU(2) ~.
\end{equation}
However, this condition is not gauge-invariant. The natural gauge-invariant choice is:
\begin{equation}
    \eta^i = g^{i\widetilde{j}} \widetilde{\rho}^\dag_{\widetilde{j}}, \quad g \in U(2) ~,
\end{equation}
which breaks the flavor symmetry. This is a completely generic phenomenon, due to the chiral anomaly of $U(1)_{\widetilde{A}}$ in two dimensions. The boundary condition on the line can be interpreted as a fermion-vertex localized at the boundary. Consider the low-energy effective action of the zero modes:
\begin{equation}
    S_{\text{Bulk}} = \frac{i}{2}\int_0^\infty dr\int dt\,\eta^\dag_i(\partial_t+\partial_r)\eta^i-(\partial_t+\partial_r)\eta^\dag_i\eta^i+\widetilde{\rho}^\dag_{\widetilde{i}}(\partial_t-\partial_r)\widetilde{\rho}^{\widetilde{i}} -(\partial_t-\partial_r)\widetilde{\rho}^\dag_{\widetilde{i}}\widetilde{\rho}^{\widetilde{i}} ~.
\end{equation}
We can add a boundary coupling to make the boundary conditions manifest:
\begin{equation}
    S_{\text{Bdy}} =\frac{i}{2} \int dt\,\eta^\dag g \widetilde{\rho}^\dag -  \widetilde{\rho} g^\dag \eta ~.
\end{equation}
By performing a background gauge transformation, we deduce the tilt operator:
\begin{equation}
    \tau(t) = \frac{1}{2}\left(\widetilde{\rho}g^\dag \sigma^3\eta + \eta^\dag\sigma^3 g \widetilde{\rho}^\dag\right) ~.
\end{equation}

\section{Exactness and Counterterms}
\label{app:counterterms} 

The anomalous variation of a Euclidean effective action is only well-defined up to the variations of local counterterms. In Sec. \ref{sec:3point_res}, we showed that the term
\begin{equation} \label{eq:AnomalySafe}
   \alpha\int dt\,f_{abc}\lambda^a \omega^b \Dot{\omega}^c
\end{equation}
in the family anomaly constrains the three-point response function to include either a partial contact term or a genuine separated points contribution. Here we show that this term is indeed scheme-independent. It cannot be polluted by variations of any local counterterms.

One-dimensional local counterterms are made out of the gauge field \(A^a_0(t)\), the moduli \(\omega^a(t)\), and their time derivatives. Since the anomaly contains one time derivative, there are two natural classes of counterterms to consider:
\[
    \int dt\, C_a(\omega)\dot \omega^a~,
    \qquad
    \int dt\, A^a_0\,G^a(\omega)~,
\]
which we analyze one after the other.

\paragraph{Counterterms containing \(\dot \omega\).}

The most general counterterm of this type includes a function \(C_a(\omega)\):
\begin{equation}
    S_{\text{CT}}^\1
    =
    \int dt\, C_a(\omega)\dot \omega^a ~,
\end{equation}
and its variation, after integrating by parts and dropping boundary terms, is
\begin{align}
    \delta S_{\text{CT}}^\1
    &=
    \int dt \,
    \left[
        \frac{\partial C_b}{\partial \omega^a}
        -
        \frac{\partial C_a}{\partial \omega^b}
    \right]
    \delta \omega^a \dot \omega^b
\end{align}
To produce the anomaly in Eq. \eqref{eq:AnomalySafe}, we need terms of the order \(\lambda \omega\dot \omega\). Therefore, it suffices to study \(C_a(\omega)\) and $\delta_\lambda \omega$ to the following order 
\begin{equation}
    C_a(\omega)
    =
    M_{ba}\omega^b
    +
    \frac12 N_{bca}\omega^b \omega^c
    +
    \dots,\quad \delta_\lambda \omega^a
    =
    -\lambda^a
    +
    \frac12 f^{cda}\lambda^c \omega^d
    +
    \dots~.
\end{equation}
Substituting the gauge variation of \(\omega^a\), the part of
\(\delta S_{\text{CT}}\) of order \(\lambda w\dot w\) is
\begin{equation}
    \left.\delta_\lambda S_{\text{CT}}^\1\right|_{\lambda \omega\dot \omega}
    =
    -\int dt\, N_{b[ac]} \lambda^a  \omega^b \dot{\omega}^c
        +
        \frac12 f^{abe} M_{[ec]} \lambda^a \omega^b \dot{\omega}^c ~.
    \label{eq:variationCcoordinate}
\end{equation}

It is simple to see why neither of these terms can match the anomaly:
\begin{itemize}
    \item The first term requires $N_{b[ac]}\sim f_{bac}$, but $N_{b[ac]}$ is symmetric in $a,b$ indices. 
    \item The second term enforces $f^{abe}M_{[ec]}\sim f^{abc}$ which implies that $M_{[ab]}$ is proportional to the Kronecker delta (after contracting with $f^{abc}$), which is impossible.
\end{itemize}

\paragraph{Counterterms containing \(A_0\).}

The second possible class of counterterms is
\begin{equation}
    S_{\text{CT}}^\2
    =
    \int dt\, A^a_0\,G^a(\omega),
\end{equation}
with an arbitrary function $G^a(\omega)$. The gauge variation of $A^a_0$ is
\begin{equation}
    \delta_\lambda A^a_0
    =
    \dot \lambda^a
    +
    \cdots ~,
\end{equation}
where the omitted terms contain $A_0$. Since the anomaly does not contain $A_0$, the relevant part of the variation is
\begin{equation}
    \delta_\lambda S_{\text{CT}}^\2
    \supset
    -\int dt\, \lambda^a
    \frac{\partial G^a}{\partial \omega^c}
    \dot \omega^c ~.
\end{equation}
This cannot cancel the anomaly, because we need
\begin{equation}
    \frac{\partial G^a}{\partial \omega^c}
    \propto
    f^{abc}\omega^b\implies \frac{\partial^2 G^a}{\partial \omega^d\partial \omega^c}
    \propto
    f^{adc} ~,
    \label{eq:needGgradient}
\end{equation}
which is inconsistent due to the symmetry properties in $d,c$.

\

Combining the two cases, no local one-dimensional counterterm built from
$\omega^a, \Dot{\omega}^a$, and $A^a_0$ can remove the anomaly.

\section{Defect Scattering in the Goldstone-Maxwell Model}
\label{app:rotor}
        
We consider the dynamics due to the rotor degree of freedom on the 't Hooft line in Eq.~\eqref{eq:H_QM} in Lorentzian signature. Without background fields, the bulk Maxwell and Goldstone sectors decouple and the effect of the dynamical moduli on the 't Hooft line is captured by the classical Lagrangian which couples the rotor to the compact scalar in the bulk
\begin{equation}
    \mathcal{L}_{1d} = \frac{I}{2}\Dot{\uptheta}^2 -\frac{q\kappa}{2\pi}\uptheta \Dot{\chi} ~.
\end{equation}
Away from $r=0$, the bulk equations of motion are the unmodified wave equations with solutions $\chi_{\omega,l}$ with the following behavior near $r = 0$:
\begin{equation}
    \lim_{r\to 0}\chi_{\omega,l}\sim
    \begin{cases}
        & r^l ~, \\
        & \frac{1}{r^{l+1}} ~.
    \end{cases}
\end{equation}
Assuming regularity at the origin removes all of the singular modes. However, the constant mode $l=0$ cannot match the delta function in the equations of motion of $\chi$, which would trivialize the rotor. To treat the coupling properly, we regularize the defect by smearing the line defect on a sphere of radius $\epsilon$ around $r = 0$. Then, the Lagrangian becomes
\begin{equation}
    \mathcal{L} =\frac{I_\epsilon}{2}\Dot{\uptheta}^2  +  v^2 \int d^3x\ \left[ (\partial_t \chi)^2 - (\nabla \chi)^2\   +\ \frac{\kappa q}{2\pi}\rho_\epsilon(x)\, \Dot{\uptheta}\, \chi \right]
\end{equation}
where we take the density smearing to be:
\begin{equation}
    \rho_\epsilon(x) = \frac{\delta(r-\epsilon)}{4\pi r^2}
\end{equation}
which makes the integral vanish on higher harmonics of $\chi$ with $l>0$. Note that this smearing is consistent with the periodicity $\chi \to \chi + 2\pi$. We also allow the moment of inertia to depend on $\epsilon$, anticipating singular behavior as $\epsilon \to 0$.

We proceed by solving the wave equation inside and outside of the $\epsilon$-sphere and gluing them together properly. For $r \neq \epsilon$, the wave equation holds and we have solutions
\begin{equation}
    \chi_{\omega, l, m} =\left(A_\ell \frac{J_{l+\frac{1}{2}}(\omega r)}{\sqrt{\omega r}} + B_\ell \frac{Y_{l+\frac{1}{2}}(\omega r)}{\sqrt{\omega r}}\right)Y_{l m}(\theta, \phi) e^{-i\omega t} ~.
\end{equation}
Focusing on the $s$-wave sector $\chi_\omega(r,t)$, the $\delta$-function turns into a discontinuity of $\partial_r \chi_\omega(r,t)$ at $r = \epsilon$
\begin{equation}
    2v^2\left(\partial_t^2\chi_\omega-\frac{1}{r^2}\partial_r (r^2\partial_r)\chi_\omega\right)  = \lambda\, \Dot{\uptheta}\frac{\delta(r-\epsilon)}{4\pi r^2}\iff -2v^2\epsilon^2\partial_r\chi_\omega(r,t)\Big|_{\epsilon_-}^{\epsilon_+} = \lambda\, \Dot{\uptheta}(t) ~,
\end{equation}
where $\lambda \equiv \frac{q\kappa}{2\pi}$. The equation of motion for the rotor $\uptheta(t)$ is given by:
\begin{equation}
    \Ddot{\uptheta} = -\frac{\lambda}{I_\epsilon} \Dot{\chi}\iff \Dot{\uptheta} +\frac{\lambda}{I_\epsilon} \chi = C ~.
\end{equation}
Observe that $C=0$ for $\omega\neq0$. Let us focus on these modes, for which 
\begin{equation}\label{eq:jump}
2v^2\epsilon^2\partial_r\chi_\omega\Big|_{\epsilon_-}^{\epsilon_+} = \frac{\lambda^2}{I_\epsilon}\chi_\omega\Big|_{r=\epsilon} ~.
\end{equation}
Assuming that the field inside the sphere is regular, we expand for small $r$:
\begin{equation}
    \begin{split}
        \chi_{\omega,\text{in}}(r) &= \frac{c}{\omega}\frac{\sin(\omega r)}{r}\approx c+ \mathcal{O}(r^2)\,, \\
        \chi_{\omega,\text{out}}(r) &= a \left( \frac{\sin(\omega r)}{\omega r}  -\ell \, \frac{\cos(\omega r)}{r}\right) \approx a\left(1 - \frac{\ell}{r}\right) + \mathcal{O}(r) \,,
    \end{split}
\end{equation}
where we have introduced the scattering length $\ell$. The continuity of $\chi_\omega$ and the discontinuity \eqref{eq:jump} of $\partial_r\chi_\omega$ imply:
\begin{equation}
    c = a \left(1- \frac{\ell}{\epsilon}\right)\,, \quad I_\epsilon = -\frac{\lambda^2}{2v^2} \left( \frac{1}{\epsilon} -\frac{1}{\ell} \right)\,. 
\end{equation}
Indeed, we find singular behavior for the moment of inertia as $\epsilon \rightarrow 0$. We renormalize the moment of inertia by splitting it into $I_\epsilon = \frac{\iota}{\epsilon} + I_r$ with
\begin{equation}
\iota = -\frac{\lambda^2}{2v^2}\,,\quad I_r = \frac{\lambda^2}{2v^2}\frac{1}{\ell}\,.
\end{equation}
The singular term is fixed, while the constant term determines the scattering length (or phase shift) 
\begin{equation}
\ell = -\frac{\tan(\delta(\omega))}{\omega} = \left(\frac{q\kappa}{2\pi}\right)^2 \frac{1}{2v^2 I_r}\,. 
\end{equation}
The renormalized scattering length $\ell$ depends on the UV completion of the defect, which appears in the low-energy effective theory through the dimensionful parameter $I_r$. The scattering strength of the $s$-wave is proportional to the square of the 2-group structure constant $\kappa$.

\section{Descent Formalism for Anomalous Theories with Boundary}
\label{app:descent}			  

In this appendix, we review the Stora-Zumino descent procedure \cite{Zumino:1983ew,Alvarez-Gaume:1984zlq,Manes:1985df,Thorngren:2020yht} for constructing consistent anomalies. We then generalize the descent procedure for anomalous theories with boundaries and give a complementary proof to the inflow picture in the main text.

\

In the usual Stora-Zumino descent procedure, we add a two-dimensional auxiliary space with coordinates $\theta^\alpha$ to enlarge the total space $\mathcal{M}_{\rm{total}}$ to
\begin{equation}
    \mathcal{M}_{\rm{total}} = \mathcal{M}_{d} \times [0,1]_{\theta^1} \times [0,1]_{\theta^2} ~.
\end{equation}
There is a group valued field $g(x,\theta)$ such that
\begin{equation}
    g(x,0) = 1 \in G ~, \quad g(x,\theta)^{-1}\frac{\partial}{\partial \theta^\alpha} g(x,\theta) \bigg|_{\theta^\alpha = 0} = i \lambda_\alpha \quad ~,
\end{equation}
where the $\lambda_\alpha$'s are organized as a 1-form on $\mathcal{M}_{\rm{total}}$
\begin{equation}
    \widehat{\lambda} = \lambda_\alpha d\theta^\alpha = - i g^{-1} \widehat{d}g ~, \quad \widehat{d} = d\theta^\alpha \frac{\partial}{\partial \theta^\alpha} ~,
\end{equation}
which satisfies the Maurer-Cartan equation
\begin{equation}
    \widehat{d}\ \widehat{\lambda}= - i\, \widehat{\lambda} \wedge \widehat{\lambda} ~.
\end{equation}
We extend any gauge field $A^\1$ on $\mathcal{M}_{d}$ to a gauge field $\overline{A}^\1$ on $\mathcal{M}_{\rm{total}}$ via $g(x,\theta)$:
\begin{equation}
    \overline{A}^\1(x,\theta) = g(x,\theta)^{-1} (A^\1 -i \, d) g(x,\theta) ~,
\end{equation}
where $d$ is the differential on $\mathcal{M}_{d}$ and clearly $\overline{A}^\1(x,0) = A^\1(x)$ with $d\, \widehat{d} = -\widehat{d}\, d$.

Crucially $\widehat{d}$ acts as a gauge transformation on $\overline{A}^\1$ and $F^\2(\overline{A}) = d\overline{A}^\1 + i \overline{A}^\1 \wedge \overline{A}^\1$:
\begin{equation}
    \begin{split}
        \widehat{d}\ \overline{A}^\1 &= - D_{\overline{A}}\widehat{\lambda} = - \delta_{\widehat{\lambda}}\overline{A}^\1 \equiv - \delta_{\widehat{\lambda}_i} \overline{A}^\1\wedge d\theta^i \\
        \widehat{d} F^\2(\overline{A}) &= - i [\widehat{\lambda},F^\2(\overline{A})] = \delta_{\widehat{\lambda}}F^\2(\overline{A}) \equiv  \delta_{\widehat{\lambda}_i} F^\2(\overline{A}) \wedge d\theta^i  ~.
    \end{split}
\end{equation}
This allows us to express the WZ consistency condition as
\begin{equation}
    \widehat{d}\mathcal{A}(\widehat{\lambda},\overline{A}) = 0 ~,
\end{equation}
where the coefficient of $d\theta^1 \wedge d\theta^2$ is $\delta_{\mathrm{WZ}}\mathcal{A}$\footnote{\label{ft:pf1}To see this, note that $\mathcal{A}$ must take the form
\begin{equation}
    \mathcal{A} \equiv \int \alpha^{(2n)}(\widehat{\lambda},\overline{A})= \int \Tr(\widehat{\lambda} \wedge G^{(2n)}(\overline{A})) ~, 
\end{equation}
where $G^{(2n)}(\overline{A})$ is some degree-$2n$ form built from $\overline{A}$ and the spacetime exterior derivative $d$, which can be rewritten as being built from $\overline{A}$ and its field strength $F(\overline{A})$. Then, for the $\widehat{d}$ action on $G^{(2n)}(\overline{A})$, generically there are two possibilities. First, $\widehat{d}$ hits on some $F(\overline{A})$ inside $P_1^{(p)}(\overline{A},F(\overline{A})) \wedge F(\overline{A}) \wedge P_2^{(2n-p-2)}(\overline{A},F(\overline{A}))$, this leads to
\begin{equation}
    (-1)^p P_1^{(p)} \wedge (-\delta_{\widehat{\lambda}_j} F(\overline{A}) \wedge d\theta^j) \wedge P_2^{(2n-p-2)} = P_1^{(p)} \wedge \delta_{\widehat{\lambda}_j} F(\overline{A}) \wedge P_2^{(2n-p-2)} \wedge d\theta^j ~.
\end{equation}
Second, $\widehat{d}$ hits on some $\overline{A}$ inside $\widetilde{P}_1^{(\widetilde{p})}(\overline{A},F(\overline{A})) \wedge \overline{A} \wedge \widetilde{P}_2^{(2n-\widetilde{p}-1)}(\overline{A},F(\overline{A}))$ and leads to
\begin{equation}
    (-1)^{\widetilde{p}} \widetilde{P}_1^{(\widetilde{p})} \wedge (-\delta_{\widehat{\lambda}_j} \overline{A}\wedge d\theta^j) \wedge \widetilde{P}^{(2n-\widetilde{p}-1)} = \widetilde{P}_1^{(\widetilde{p})}\wedge \delta_{\widehat{\lambda}_j}\overline{A} \wedge \widetilde{P}_2^{(2n-\widetilde{p}-1)} \wedge d\theta^j ~.
\end{equation}
Combined, this implies that
\begin{equation}
    \widehat{d} G^{(2n)} = (\delta_{\widehat{\lambda}_j}G^{(2n)})\wedge d\theta^j ~.
\end{equation}
This implies that
\begin{equation}
\begin{aligned}
    \widehat{d} \mathcal{A} & = \int \Tr(\widehat{d}\, \widehat{\lambda} \wedge G^{(2n)} - \widehat{\lambda}\wedge \widehat{d} G^{(2n)}) \\
    & = \underbrace{\left(\delta_{\widehat{\lambda}_1} \mathcal{A}(\widehat{\lambda}_2,\overline{A}) - \delta_{\widehat{\lambda}_2} \mathcal{A}(\widehat{\lambda}_1,\overline{A}) - \mathcal{A}(i[\widehat{\lambda}_1,\widehat{\lambda}_2],\overline{A})\right)}_{(\delta_{\mathrm{WZ}}\mathcal{A})(\widehat{\lambda}_1,\widehat{\lambda}_2,\overline{A})} d\theta^1 \wedge d\theta^2 ~.
\end{aligned}
\end{equation}}.

Since $\mathcal{A}$ can be written as the integral of a local density $\alpha^{(2n)}$, $\widehat{d} \alpha^{(2n)}$ is exact with respect to $d$ and we can locally write:
\begin{equation}\label{eq:WZdensity}
    \widehat{d}\alpha^{(2n)} \ =\ d \alpha^{(2n-1)} ~.
\end{equation}
Consistent anomalies $\alpha^{(2n)}$ can be constructed by starting from a $(2n+2)$-degree anomaly polynomial $\mathcal{I}^{(2n+2)}(\overline{A})$ of the background field $\overline{A}^\1$, satisfying
\begin{equation}
    d \mathcal{I}^{(2n+2)}(\overline{A}) = 0 ~, \quad \widehat{d}\mathcal{I}^{(2n+2)}(\overline{A}) = 0 ~.
\end{equation}
At least locally we can write $\mathcal{I}^{(2n+2)}(\overline{A}) = d\mathcal{I}^{(2n+1)}(\overline{A})$ and furthermore
\begin{equation}
    \widehat{d}\mathcal{I}^{(2n+2)}(\overline{A}) = 0 \ \implies \  d(\widehat{d}\mathcal{I}^{(2n+1)}(\overline{A})) = 0 ~.
\end{equation}
This means $\widehat{d}\mathcal{I}^{(2n+1)}$ is again locally exact, and any $\alpha^{(2n)}$ fulfilling
\begin{equation}
     d\alpha^{(2n)}(\widehat{\lambda},\overline{A}) \ = \  \widehat{d}\mathcal{I}^{(2n+1)}(\widehat{\lambda},\overline{A})
\end{equation}
will satisfy \eqref{eq:WZdensity}. This can be seen by taking $\widehat{d}$ on both sides and using $d\,\widehat{d} = - \widehat{d}\, d$.

Notice that there are ambiguities in the descent procedure.
\begin{itemize}
    \item Given an anomaly polynomial $\mathcal{I}^{(2n+2)}(A)$, the choice of the SPT phase $\mathcal{I}^{(2n+1)}(A)$ is determined up to a (locally) exact term $d\mu^{(2n)}(A)$. This ambiguity corresponds to adding a local, typically non-gauge-invariant counterterm $\mu^{(2n)}(A)$ on $M_{2n}$, which does not affect the WZ consistency condition or change the cohomology class of the anomaly. This simply changes the \textit{presentation} of the anomaly as it shifts $\alpha^{(2n)}(\lambda, A)$ by the gauge variation $\delta_\lambda \mu^{(2n)}(A)$.

    \item After fixing the choice of $\mathcal{I}^{(2n+1)}$, the anomalous variation $\alpha^{(2n)}(\lambda,A)$ is also only determined up to exact terms $d\mu^{(2n-1)}(\lambda,A)$ which vanish in $\mathcal{A}[\lambda,A]$ on a closed manifold (with the appropriately quantized coefficient of the anomaly polynomial).
\end{itemize} 

For theories defined on closed manifolds, the Wess-Zumino consistency condition is guaranteed as long as $\widehat{d}\alpha^{(2n)}(\widehat{\lambda},\overline{A})$ is locally exact. However, this does not suffice if the theory is placed on a manifold $M_{2n}$ \textit{with boundary}, since 
\begin{equation}
    \delta_{\lambda_1}\, \mathcal{A}(\lambda_2,A) - \delta_{\lambda_2}\,\mathcal{A}(\lambda_1,A) - \mathcal{A}(i[\lambda_1,\lambda_2],A) = - \int_{\partial M_{2n}} \alpha^{(2n-1)}(\lambda_1,\lambda_2,A) ~,
\end{equation}
where $d\alpha^{(2n-1)}(\lambda_1,\lambda_2,A)$ is the coefficient of $d\theta^1 \wedge d\theta^2$ in $\widehat{d}\alpha^{(2n)}(\widehat{\lambda},\overline{A})$ at $\theta^\alpha = 0$. As discussed in the main text, this forces the symmetry to be explicitly broken at the boundary. We can introduce a spurion $U(x)$ on the boundary, to render the partition function $Z[U(x),A^\1(x)]$ WZ-consistent, due to an additional boundary anomalous variation $\nu^{(2n-1)}(\lambda,U,A)$ which cancels the WZ inconsistency $\alpha^{(2n-1)}(\lambda_1,\lambda_2,A)$ from the bulk:

\begin{multline}
    \int_{\partial \mathcal{M}_{2n}} \delta_{\lambda_1}\, \nu^{(2n-1)}(\lambda_2, U,A) - \delta_{\lambda_2}\,\nu^{(2n-1)}(\lambda_1, U,A) - \nu^{(2n-1)}(i[\lambda_1,\lambda_2],U,A) \\
    =    -\int_{\partial \mathcal{M}_{2n}} \alpha^{(2n-1)}(\lambda_1,\lambda_2, A) ~.
\end{multline}

Possible $\nu^{(2n-1)}(\lambda,U,A)$ can be acquired by a similar descent formalism. Again, we consider the extended manifold $\mathcal{M}_{\rm{total}} = \mathcal{M}_{2n} \times [0,1]_{\theta^1} \times [0,1]_{\theta^2}$, except now $\mathcal{M}_{2n}$ has a non-trivial boundary. For simplicity, let us consider the case where the symmetry $G^\0$ is completely broken on the boundary such that the boundary conditions are labeled by $G^\0$ itself, i.e., $U \in G^\0$. Let us also choose the symmetry $g \in G^\0$ to act as $U \mapsto g^{-1}U$. We now extend the boundary background field $U$ to a field $\overline{U}$ on $\partial\mathcal{M}_{2n} \times [0,1]^2$ via
\begin{equation}
    \overline{U}(x,\theta) := g(x,\theta)^{-1} U(x) ~.
\end{equation}
Then, we have 
\begin{equation}
    \widehat{d}\,\overline{U}(x,\theta) = - i\widehat{\lambda}(x,\theta)\, \overline{U}(x,\theta) = \delta_{\widehat{\lambda}}\,\overline{U} ~.
\end{equation}
This means that we can express the WZ consistency condition with boundary as \footnote{To show this, one can follow an argument similar to \footref{ft:pf1} to show that the $d\theta^1 \wedge d\theta^2$-component in $\widehat{d}\nu^{(2n-1)}$ is the WZ combination of $\nu^{(2n-1)}$. More specifically, $\nu^{(2n-1)}$ can be written as $\Tr(\widehat{\lambda}\wedge \widetilde{G}^{(2n-1)})$ where $\widetilde{G}^{(2n-1)}$ is a degree-($2n-1$) form built from $\overline{U},\overline{U}^{-1},d\overline{U},\overline{A},F(\overline{A})$. By a similar consideration as the previous case, one can show that
\begin{equation}
    \widehat{d} \widetilde{G}^{(2n-1)} = - (\delta_{\widehat{\lambda}_j}\widetilde{G}^{(2n-1)})\wedge d\theta^j ~.
\end{equation}
Then,
\begin{equation}
    \widehat{d}\nu^{(2n-1)} = \underbrace{\left(\delta_{\widehat{\lambda}_1} \nu^{(2n-1)}(\widehat{\lambda}_2,\overline{U},\overline{A}) - \delta_{\widehat{\lambda}_2} \nu^{(2n-1)}(\widehat{\lambda}_1,\overline{U},\overline{A}) - \nu^{(2n-1)}(i[\widehat{\lambda}_1,\widehat{\lambda}_2],\overline{U},\overline{A})\right)}_{\delta_{\mathrm{WZ}}\nu^{(2n-1)}} d\theta^1 \wedge d\theta^2 ~.
\end{equation}
}
\begin{equation}\label{eq:bdyWZ}
    \widehat{d}\nu^{(2n-1)}(\widehat{\lambda},\overline{U},\overline{A}) \ =\   \omega^{(2n-1)}(\widehat{\lambda},\overline{A}) \ +\  d\nu^{(2n-2)}(\widehat{\lambda},\overline{U},\overline{A}) ~.
\end{equation}
The solution can be acquired by introducing $\mathcal{H}^{(2n+1)}(\overline{U},\overline{A})$ in analogy with the anomaly polynomial $\mathcal{I}^{(2n+2)}$ such that
\begin{equation}\label{eq:bbdesct1}
    d\mathcal{H}^{(2n+1)} = 0 ~, \quad \widehat{d}\left(\mathcal{H}^{(2n+1)} + \mathcal{I}^{(2n+1)}\right)  = 0 ~.
\end{equation}
The first condition allows us to write $\mathcal{H}^{(2n+1)}(\overline{U},\overline{A}) = d\calB^{(2n)}(\overline{U},\overline{A})$ and the second condition implies
\begin{equation}\label{eq:nuinflow}
\begin{aligned}
    & d (\alpha^{(2n)}(\widehat{\lambda},\overline{U},\overline{A}) - \widehat{d} \calB^{(2n)}(\widehat{\lambda},\overline{U},\overline{A}) ) = 0 \\
    &  \quad \quad \quad \quad \quad \quad  \quad \quad  \implies \quad \widehat{d} \calB^{(2n)}(\widehat{\lambda},\overline{U},\overline{A}) = \alpha^{(2n)}(\widehat{\lambda},\overline{U},\overline{A}) + d \nu^{(2n-1)}(\widehat{\lambda},\overline{U},\overline{A}) ~,
\end{aligned}
\end{equation}
for some $\nu^{(2n-1)}(\widehat{\lambda},\overline{U},\overline{A})$. By further taking $\widehat{d}$ on the above equation, we see that any $\nu^{(2n-1)}(\widehat{\lambda},\overline{U},\overline{A})$ automatically satisfies the WZ consistency condition in the presence of the boundary~\eqref{eq:bdyWZ}.

To conclude, let us now discuss various ambiguities that could potentially arise in the descent formalism. 
\begin{enumerate}
    \item Changing the presentation of the 't Hooft anomaly does not affect the boundary anomalous variation $\nu^{(2n-1)}(\lambda, U,A)$: fixing the bulk anomaly polynomial $\mathcal{I}^{(2n+2)}(A)$, recall that $\mathcal{I}^{(2n+1)}(A)$ is determined up to a shift by $d\mu^{(2n)}(A)$ which changes the presentation of the anomaly. This shift will also shift $\calB^{(2n)}(U,A)$ by $-\mu^{(2n)}(A)$, however, it will not affect the inflow equation of $\nu^{(2n-1)}(\lambda,U,A)$ in Eq.~\eqref{eq:nuinflow} as it cancels with the shift of $\widehat{d}\alpha^{(2n)}(\widehat{\lambda},\overline{A})$ from $\mu^{(2n)}(\widehat{\lambda},\overline{A})$.
    
    \item After fixing the SPT phase $\mathcal{I}^{(2n+1)}(A)$, recall that $\alpha^{(2n)}(\lambda,A)$ is determined up to some total derivative $d\mu^{(2n-1)}(\lambda,A)$ (which does not matter in the previous boundary-less case but matters here). This leads to a shift in the boundary anomalous variation $\nu^{(2n-1)}(\lambda,U,A) \to \nu^{(2n-1)}(\lambda,U,A) -  \mu^{(2n-1)}(\lambda,A)$, and the WZ consistency condition is maintained. 
    
    \item After fixing $\mathcal{I}^{(2n+1)}(A)$ and $\alpha^{(2n)}(\lambda,A)$, $\mathcal{H}^{(2n+1)}(U,A)$ is determined up to a shift $d \beta^{(2n)}(U,A)$ satisfying
    \begin{equation}
        \widehat{d} \beta^{(2n)} \ =\  d \beta^{(2n-1)}~.
    \end{equation}
    Physically, this ambiguity means that the boundary may have its own \textit{self-consistent} anomaly involving $U(x)$ and $A(x)$ not induced by the bulk 't Hooft anomaly. This shifts the boundary anomalous variation $\nu^{(2n-1)}(\lambda,U,A)$ by $\beta^{(2n-1)}(\lambda, U,A)$ maintaining the WZ consistency.
    
    \item Finally, after fixing $\mathcal{I}^{(2n+1)}(A), \alpha^{(2n)}(\lambda,A)$, and $\mathcal{H}^{(2n+1)}(U,A)$, $\calB^{(2n)}(U,A)$ is determined up to a total derivative $d \gamma^{(2n-1)}(U,A)$. This can be understood as adding a local counterterm on the boundary, which changes the presentation of the boundary anomaly, but does not affect the WZ consistency condition. 
\end{enumerate}

\linespread{1}\selectfont
\bibliographystyle{utphys}
\bibliography{2group}

\end{document}